\documentclass[extra]{gji}

\usepackage{timet,color}
\usepackage[colorlinks=true,urlcolor=blue,citecolor=blue,linkcolor=blue]{hyperref}
\usepackage{amsmath,amssymb}
\usepackage{graphicx}

\usepackage{subcaption}
\usepackage{lineno}
\usepackage{placeins}
\usepackage{float}

\newcommand{\dd}{\mathrm{d}}
\newcommand{\fdip}{f_{\textrm{dip}}}

\newcommand{\Pra}{\mathit{Pr}}

\newcommand{\Sc}{\mathit{Sc}}
\newcommand{\Rey}{\mathit{Re}} 
\newcommand{\Rm}{\mathit{Rm}}
\newcommand{\Ro}{\mathit{Ro}}
\newcommand{\Pm}{\mathit{Pm}}
\newcommand{\Rol}{\mathit{Ro}_\ell}
\newcommand{\Ek}{\mathit{E}}
\newcommand{\Ra}{\mathit{Ra}}

\newcommand{\Raf}{{\mathit{Ra}_F}}
\newcommand{\lehnert}{\mathit{Le}}
\newcommand{\elsasser}{\Lambda}
\newcommand{\temp}{\theta}

\newcommand{\magfield}{\mitbf{B}}
\newcommand{\vel}{\mitbf{u}}
\newcommand{\cross}{\times}
\newcommand{\grad}{\nabla}
\newcommand{\curl}{\mitbf{\nabla}\times}
\newcommand{\pressure}{P}

\newcommand{\lap}{\nabla^2}
\newcommand{\dt}[1]{\frac{\partial #1}{\partial t}}
\newcommand{\zhat}{\hat{\mitbf{z}}}
\newcommand{\rhat}{\hat{\mitbf{r}}}

\newcommand{\kinvisc}{\nu}

\newcommand{\rotation}{\Omega}
\newcommand{\Vs}{V_{\rm s}}
\newcommand{\shellthick}{D}

\newcommand{\thermexpan}{\alpha}
\newcommand{\thermcond}{k}
\newcommand{\thermdiff}{\kappa}
\newcommand{\compdiff}{\kappa_C}
\newcommand{\magdiff}{\eta}
 
\newcommand{\grav}{g}
\newcommand{\gravity}{\mitbf{g}}
\newcommand{\radius}{r}
\newcommand{\meandensity}{\rho_0} 

\newcommand{\magperm}{\mu_0}

\newcommand{\innerrad}{r_i}
\newcommand{\outerrad}{r_o}

\newcommand{\kinEnergy}{ E_\textrm{kin} }
\newcommand{\magEnergy}{ E_\textrm{mag} }

\newcommand{\charvel}{U}
\newcommand{\charmag}{B}
\newcommand{\chartemp}{\Theta}

\newcommand{\viscdiss}{D_U}
\newcommand{\ohmdiss}{D_B}

\newcommand{\du}{d_{\mathrm{u}}}
\newcommand{\duchrs}{\overline{d}_{\mathrm{u}}}
\newcommand{\dB}{d_{\mathrm{B}}}
\newcommand{\dumin}{d_{\rm u}^{\rm min}}
\newcommand{\dbmin}{d_{\rm B}^{\rm min}}
\newcommand{\dpar}{d_\parallel}
\newcommand{\dupol}{d_u^\text{pol}}
\newcommand{\dbpol}{d_B^\text{pol}}

\newcommand{\ldeg}{\ell}
\newcommand{\lubar}{\overline{\ldeg}_\mathrm{u}}
\newcommand{\lbmin}{\ell_B^{\rm min}}
\newcommand{\lupol}{\ell_u^\text{pol}}
\newcommand{\lbpol}{\ell_B^\text{pol}}
\newcommand{\lmax}{\ell_{\rm max}}
\newcommand{\lh}{\ell_{\rm h}}
\newcommand{\qh}{q_{\rm h}}
\newcommand{\Mratio}{\mathcal{M}}
\newcommand{\tp}{T'}
\newcommand{\fohm}{f_\text{ohm}}
\newcommand{\Nr}{N_{\rm r}}

\title[Accessing the dipole-multipole transition in rapidly rotating spherical shell dynamos]
  {Accessing the dipole-multipole transition in rapidly rotating spherical shell dynamos}
\author[A. T. Clarke$^1$, C. J. Davies$^1$, S. J. Mason$^1$, S. Naskar$^1$]
  {Andrew T. Clarke$^{1*}$, Christopher J. Davies$^1$, Stephen J. Mason$^1$, Souvik Naskar$^1$ \\
  $^1$ School of Earth and Environment, University of Leeds, Leeds LS29JT, UK. E-mail: a.t.clarke@leeds.ac.uk
  }

\begin{document}

\label{firstpage}

\maketitle

\begin{summary}
Earth's magnetic field has exhibited erratic polarity reversals over much of its history; however, the processes that cause polarity transitions are still poorly understood. Dipole reversals have been found in many numerical dynamo simulations and often occur close to the transition between dipole-dominated and multipolar dynamo regimes. However, the physical conditions used in reversing simulations are necessarily far from those in Earth's liquid iron core because of the long runtimes needed to capture polarity transitions and because a systematic exploration of parameter space is needed to find the dipole-multipole transition. Here, we develop a unidimensional path theory in an attempt to simplify the search for the dipole-multipole transition at increasingly realistic physical conditions. We consider three paths that are all built from the requirements of a constant magnetic Reynolds number $\Rm$; one path further attempts to impose balance between Magnetic, Coriolis, and Archimedean forces (a MAC balance) while the other two seek to constrain solutions to an inertia-MAC, or IMAC, balance. The presence of inertia, although not geophysically realistic, allows us to build paths that more closely follow the conditions where simulated reversals have been found to date. Numerical simulations show reasonable agreement with the expected physical conditions along the paths within the accessible parameter space, but also deviate from predicted behaviour for certain diagnostic quantities, particularly the magnetic field strength and the magnetic/kinetic energy ratio. Furthermore, the paths do not follow the dipole-multipole transition; starting from reversing conditions, simulations move into the dipolar non-reversing regime as they are advanced along the path. By increasing the Rayleigh number, a measure of the buoyancy driving convection, above the values predicted by the path theory,  we are able to access the dipole-multipole transition down to an Ekman number $\Ek\sim 10^{-6}$, comparable to the most extreme conditions reported to date. Our results, therefore, demonstrate that the path approach is an efficient method for seeking the dipole-multipole transition in rapidly rotating dynamos. However, the conditions under which we access the dipole-multipole transition become increasingly hard to access numerically and also increasingly unrealistic because $\Rm$ rises beyond plausible bounds inferred from geophysical observations. Future work combining path theory with variations in the core buoyancy distribution, as suggested by recent studies, appears a promising approach to accessing the dipole-multiple transition at extreme physical conditions. 
\end{summary}


\section[]{Introduction}\label{sec:intro}

Paleomagnetic data shows that Earth's magnetic field has reversed polarity many times over it's history \citep{strik_palaeomagnetism_2003, ogg2020geomagnetic}. The present reversal rate is around 4 per Myr, while past reversal rates have dropped to zero for tens of millions of years during superchrons and reached 10 reversals per Myr in ``hyper-reversing'' periods like the Jurassic \citep[e.g.][]{biggin_possible_2012}. The origin of these rare but extreme fluctuations is the dynamo process in Earth's liquid iron core, which converts the kinetic energy of convective fluid motion into magnetic energy. An outstanding problem is to identify the conditions under which dipole-dominated dynamos reverse polarity. Owing to the inherent nonlinearity of the dynamo problem, this question is generally addressed using numerical simulations, and this is the approach taken in the present study. 

Reversals of spherical shell dynamos were first reported by \cite{glatzmaiers_three-dimensional_1995}. Since then, many simulated reversals have been reported at a variety of different physical conditions \citep[e.g.][]{kutzner_stable_2002, wicht_inner-core_2002, driscoll2009polarity, olson_complex_2011, heimpel_testing_2013, olson_magnetic_2014, driscoll2018paleomagnetic, sprain_assessment_2019, menu_magnetic_2020, gwirtz2022can, nakagawa_combined_2022, tassin_geomagnetic_2021, jones2025low}. An important result is that increasing the Rayleigh number $\Ra$, a measure of the buoyancy force driving convection, with all other parameters fixed leads to a transition from a stable dipole-dominated core-mantle boundary (CMB) field at low $\Ra$ to a reversing dipole and then to the loss of dipole dominance and emergence of the so-called multipolar state as $\Ra$ increases \citep{kutzner_stable_2002, christensen_scaling_2006, olson_dipole_2006}. This has led several authors to postulate that the geodynamo lies close to the dipole-multipole transition \citep[e.g.][]{olson_dipole_2006, driscoll_effects_2009}. One interpretation is that reversals represent brief incursions of the dynamo into the multipolar regime \citep{olson_complex_2011}.

The physical processes that determine the multipole-dipole transition have been widely debated, and several possibilities have been proposed. \citet{mcdermott2019physical} argued that the multipole state arises from the suppression of inertial waves since these waves have been hypothesised to sustain the convection columns that maintain the dipole field \citep{ranjan2018internally}. Recently, \citet{majumder_self-similarity_2024} suggested that the transition arises when a specific class of magnetic wave is suppressed in the dynamo, while \citet{frasson2024geomagnetic} argue that lateral heterogeneity in CMB heat flux promotes reversals by suppressing westward zonal flows. Several alternatives focus on the enhanced role of inertia in the force balance \citep{sreenivasan_role_2006}. Hydrodynamic criteria to describe the transition include balances between Inertia (I) and Coriolis (C) forces at the characteristic velocity lengthscale $\du$ \citep[an IC balance measured by the local Rossby number criterion $\Rol \sim 0.12$,][]{christensen_scaling_2006}, inertia and Viscous forces \citep[IV balance,][]{soderlund_influence_2012}, a non-gradient triple VIC balance \citep{oruba2014transition}, and loss of equatorial symmetry due to enhanced inertia \citep{garcia2017equatorial}. However, recent studies have challenged the local Rossby number and equatorial symmetry criteria and instead highlighted the role of the magnetic field \citep{menu_magnetic_2020}. In particular, \citet{tassin_geomagnetic_2021} used a large suite of simulations to show that the transition from dipole-dominated to multipolar solutions arises at a magnetic to kinetic energy ratio $\Mratio \sim 1$ \citep{tassin_geomagnetic_2021}. 

The challenge of describing the dipole-multipole transition arises because of the vast range of spatial and temporal scales inherent to the dynamo problem, which makes direct numerical simulations at Earth's core parameters impossible. The important control parameters are the Ekman number $\Ek$, which compares the importance of rotation with viscosity, the Rayleigh number $\Ra$, measuring the vigour of convection, and the magnetic Prandtl number $\Pm$, the ratio of viscous and magnetic diffusion coefficients. In the most extreme (non-reversing) direct numerical simulations (DNS) $\Ek$ is limited to $\Ek\gtrsim 10^{-7}$ \citep[e.g.][]{schaeffer_turbulent_2017}, much larger than the value of $\Ek \sim 10^{-15}$ in the Earth, while $\Ra$ and $\Pm$ must be set respectively much lower and higher compared to Earth values in order to obtain dipole-dominated dynamo action. In terms of diagnostic quantities, simulations can reach geophysically relevant values of the magnetic Reynolds number (the ratio of magnetic advection to diffusion), $\Rm\sim 1000$, but not the small values of the Rossby number, $\Ro \sim 10^{-6}$, characterising the balance of inertial and Coriolis effects. Reversing simulations are further compromised compared to simulations that attempt to reproduce short-timescale dynamics \citep[e.g.][]{aubert_state_2023} because of the long integration times (usually greater than 1 magnetic diffusion time, or $O(10^5)$~yrs) needed to observe reversals. Indeed, all reversing simulations have been conducted at $\Ek\ge 10^{-5}$ \citep{sheyko_magnetic_2016, tassin_geomagnetic_2021}, except for the DNS study of \citet{sheyko_magnetic_2016} who found periodic reversals at $\Ek = 2.4 \times 10^{-6}$ originating from the propagation of dynamo waves at relatively low $\Rm\sim 100$, and \citet{frasson2024geomagnetic} who used hyperdiffusion (see below) and focused on heterogeneous CMB heat flux. Progress relies on numerical simulations at more realistic physical conditions. 

One promising way forward is to construct a uni-dimensional path that relates the control parameters $\Ek$, $\Ra$ and $\Pm$ as well as the system diagnostics $\Rm$, $\Ro$ and the Lehnert number $\lehnert$ (the ratio of rotation and Alfv\'{e}n timescales) to a single path parameter $\epsilon$ \citep{dormy_strong-field_2016,aubert_spherical_2017}. Paths can be constructed to preserve certain properties of Earth's dynamo that current simulations can already achieve, such as constant $\Rm\sim 1000$, while reducing quantities such as $\Pm$, $\Ro$ and $\Ek$ towards the physical conditions that are thought to represent Earth's core. Construction of the path relies on a theoretical scaling based on an assumed balance of forces in the core. Theoretical considerations suggest that the primary force balance in Earth's core is Quasi-Geostrophic (QG), a balance between pressure gradient and Coriolis effect \citep{aurnou_cross-over_2017,calkins2018quasi}. Dynamo action and convection are not achieved by purely geostrophic flow, so different dynamical balances must prevail at the next order. For Earth's core, theory and high-resolution simulations suggest that this secondary balance involves the Magnetic, Coriolis, and Archimedean (buoyancy) terms, a MAC balance \citep{davidson_scaling_2013, yadav_effect_2016, aubert_spherical_2017, schwaiger_force_2019}. \citet{aubert_spherical_2017}, \citet{aubert_approaching_2019} and \citet{aubert_state_2023} have run simulations along this QG-MAC path down to $\Ek \sim 10^{-13}$ and $\Pm\sim 10^{-3}$ using hyperdiffusion to parameterise the small-scale velocity and density anomalies. They showed that the large-scale structure of the dynamo solutions remains mostly invariant along the path, preserving QG-MAC balance with inertial and viscous terms becoming increasingly subdominant as $\epsilon$ decreases. Importantly, they obtained large-scale solution diagnostics $\Rm$, $\Ro$, and $\lehnert$ in good agreement with the path theory even though these diagnostics are not prescribed. This suggests that it may be possible to build path models that (at least approximately) follow the dipole-multipole transition, e.g. by encoding a constant $\Mratio$ or $\Rol$. 

In QG-MAC theory, inertia is strongly subdominant to the MAC terms. The predicted Rossby number $\Ro = U/\rotation \shellthick$, the ratio of inertial and Coriolis terms, is about $10^{-6}$, which is similar to the value obtained using estimates of the RMS flow speed $U$ in the present core \citep[e.g.][]{holme2015large} together with the present rotation frequency $\Omega$ and shell thickness $\shellthick$. Path theory based on QG-MAC balance predicts $\Mratio\sim 1000$ and $\du \sim 40 \shellthick$ \citep{aubert_spherical_2017} implying $\Rol = U /\rotation \du \ll 0.12$ and hence that Earth lies far from the dipole-multipole transition according to both local Rossby number and energy ratio criteria \citep{tassin_geomagnetic_2021}. 
Therefore, while inertia-based transition parameters do a good job of explaining results from large suites of numerical simulations, they apparently fail to explain reversals of the geomagnetic field.

In this paper we use unidimensional path theory to search for the dipole-multipole transition in low $\Ek$ rapidly rotating dynamo simulations. To do so requires addressing 3 main challenges: 1) the dipole-multipole transition is generally sharp \citep{christensen_scaling_2006, oruba2014transition, olson_complex_2011}, occurring in a narrow range of parameter space; 2) there is currently no agreement on the conditions governing the transition; 3) simulations must be run for long enough to witness reversals. We therefore consider 3 different parameter paths that reflect different assumptions regarding the dipole-multipole transition: one based on the MAC theory of \citet{aubert_spherical_2017} and two based on the IMAC theory summarised in \citet{jones_805_2015}. All three paths encode a constant $\Rm \sim 1000$ and keep inertia weakly (rather than strongly) subdominant in the force balance within the accessible parameter space. We stress that this choice is made for practical rather than geophysical reasons since previous work suggests this is the most likely parameter space where simulated reversals will arise. This choice also allows us to test how well different simulated paths reproduce theoretical diagnostics and to search for changes in the character of the dipole-multipole transition at more extreme physical conditions that may indicate a change in the reversal process. Our simulations employ a scale-dependent hyperdiffusion (HD) \citep{nataf_dynamic_2024, aubert_spherical_2017}, though the physical nature of the parameter paths, together with our use of no-slip velocity boundary conditions, limits the computational gains provided by HD compared to previous studies. Nevertheless, using a combination of ``on-path'' and ``off path'' simulations,  the latter employing an increased value of $\Ra$ compared to that suggested by the path theory, we access the dipole-multipole transition down to $\Ek = 2 \times 10^{-6}$, representing some of the lowest $\Ek$ multipolar solutions obtained to date.

Our simulations are driven by pure basal heating, reflecting the release of latent heat due to inner core growth \citep[e.g.][]{davies_buoyancy_2011, jones_805_2015}. This choice is made for theoretical convenience as explained below; however, it is a simplification because the geodynamo has been driven partly by chemical buoyancy since the formation of the inner core $\sim$0.5-1 Gyrs ago \citep{davies_constraints_2015, labrosse2015thermal, wilson2022powering}. The geomagnetic field has undergone stable fluctuations in reversal frequency over at least the last 2 Gyrs \citep{driscoll2016frequency}, On the other hand, both the semblance between simulated field behaviour and paleomagnetic observations and properties of the dipole-multipole transition do depend on the nature of the buoyancy source \citep{kutzner_stable_2002, meduri_numerical_2021}. In any case, it is worthwhile to understand the nature of the dipole-multipole transition for a range of buoyancy profiles, since the actual ratio of thermal and chemical driving and their changes over time are poorly constrained. 

This paper is organised as follows. In section~\ref{sec:methods} we specify the numerical dynamo simulations conducted for this study, derive the MAC and IMAC scaling relations and from these the three uni-dimensional parameter paths. In section~\ref{sec:results} we validate the HD scheme against selected DNS runs, analyse the force balances and lengthscales output from the simulations and their correspondence to the balances posited in the path theories, and compare the simulation diagnostics to the path predictions. Discussion and conclusions are presented in section~\ref{sec:discussion}.

\section{Methods}\label{sec:methods}

\subsection{Geodynamo Equations and Simulation Details}
\label{subsec:equations}

We consider dynamo action in a spherical shell with outer radius $\outerrad$ and inner radius $\innerrad = 0.35 \outerrad$ rotating with rate $\mathbf{\rotation}=\rotation\zhat$ about the vertical $\zhat$ direction. In spherical polar coordinates $(r, \varphi, \phi)$ the dimensional equations that determine the velocity $\vel$, magnetic field $\magfield$, and temperature perturbation $\temp$, of the basally heated incompressible and electrically conducting Boussinesq fluid are

\begin{linenomath*}
\iftwocol{
\begin{multline}
\label{eq:momentum}
    \meandensity\left( \dt{\vel} + \vel\cdot\grad\vel \right)
    +
    2\meandensity\mitbf{\rotation} \times \vel
    = \\
    -\grad\pressure
    +
    \meandensity \thermexpan \gravity \temp
    +
    \frac{\left( \nabla \times \magfield \right) \times \magfield}{\magperm}
    +
    \meandensity \kinvisc \lap \vel,
\end{multline}
}
{
\begin{equation}
\label{eq:momentum}
    \meandensity\left( \dt{\vel} + \vel\cdot\grad\vel \right)
    +
    2\meandensity\mitbf{\rotation} \times \vel
    =
    -\grad\pressure
    +
    \meandensity \thermexpan \gravity \temp
    +
    \frac{\left( \nabla \times \magfield \right) \times \magfield}{\magperm}
    +
    \meandensity \kinvisc \lap \vel,
\end{equation}
}

\begin{equation}\label{eq:induction}
    \dt{\magfield} 
    = \curl \left(\vel \times \magfield \right) 
    + \magdiff \lap \magfield,
\end{equation}

\begin{equation}\label{eq:temp}
    \dt{\temp} + \vel \cdot \grad \temp = \thermdiff \lap \temp,
\end{equation}

\begin{equation}\label{eq:div_zero}
    \grad \cdot \vel = \grad \cdot \magfield = 0,
\end{equation}
\end{linenomath*}

\citep[e.g.][]{jones_805_2015}. Here $\meandensity$ is the mean density, $\pressure$ is the reduced pressure, $\gravity= -(g_o r/\outerrad)\rhat$ is gravity where $\grav_0$ is reference gravity at the outer boundary, $\kinvisc$ is kinematic viscosity, $\magperm$ is magnetic permeability, $\magdiff$ is magnetic diffusivity and $\thermdiff = \thermcond/(\meandensity c_p)$ is thermal diffusivity where $c_p$ is the specific heat capacity and $\thermcond$ is the thermal conductivity. Parameter definitions and values for Earth and our simulations are detailed in Table~\ref{tab:symbols}. 

The 1D time-independent state with $\vel = 0$ is determined by $\nabla \temp_\text{cond} = -(\beta/r^2)\rhat$. This choice of pure basal heating is a simplification because compositional buoyancy also contributes to driving the geodynamo. The choice of basal heating is motivated partly for theoretical convenience since the radial heat current $F_c = \meandensity c_p u_r \temp - \thermcond\mathrm{d}\temp/\mathrm{d}r$ is constant through each spherical surface, and so the advective flux $\meandensity c_p u_r \temp$, which dominates in the bulk and is unknown \textit{a priori}, is balanced by the conductive flux $d\temp/dr|_{\outerrad}$ at the outer boundary, which is known \citep[e.g.][]{mound_heat_2017}. This simplifies the task of relating the convective power to the prognostic parameters \citep{aubert_spherical_2017}. It is also worth considering different driving modes since the relative proportions of thermal and compositional driving are poorly known and will have changed over time \citep{lister1995strength,davies_buoyancy_2011, labrosse2015thermal}. 

The equations are made dimensionless using the shell thickness $\shellthick=\outerrad-\innerrad$ as a lengthscale, the magnetic diffusion time $\shellthick^2/\magdiff$ as timescale, the Elsasser scaling $(2\rotation\meandensity\magperm\magdiff)^{1/2}$ for the magnetic field, and temperature scale $\beta/\shellthick$. With these definitions, the system is described by the following non-dimensional control parameters:

\begin{linenomath*}
\begin{equation}
\text{Ekman number} : \Ek = \frac{\kinvisc}{\rotation \shellthick^2},
\end{equation}
\begin{equation}
\text{Rayleigh number} : \Ra = \frac{\thermexpan \grav_0 \beta \shellthick^2}{\kinvisc \thermdiff},
\end{equation}
\begin{equation}
\text{Prandtl number} : \Pra = \frac{\kinvisc}{\thermdiff} = 1,
\end{equation}
\begin{equation}
\text{magnetic Prandtl number} : \Pm = \frac{\kinvisc}{\magdiff}.
\end{equation}
\end{linenomath*}

Instead of $\Ra$ it will sometimes prove useful to employ the flux Rayleigh number 

\begin{linenomath*}
\begin{equation}
    \Raf = \frac{\thermexpan \grav_0 \beta \thermdiff}{\rotation^3 \shellthick^4} = \frac{\Ra \Ek^3}{\Pra^2}
    \label{eq:flux_rayleigh}
\end{equation}
\end{linenomath*}

We define the following dimensionless output diagnostics to analyse the simulations. The magnetic Reynolds number $\Rm$ measures the dimensionless kinetic energy $\kinEnergy$ and is given by

\begin{linenomath*}
\begin{equation}
   \Rm = \frac{\charvel \shellthick}{\magdiff} = \sqrt{\frac{2 \kinEnergy}{\Vs}}
\end{equation}
\end{linenomath*}
where $\Vs$ is the shell volume and $\kinEnergy$ is given by
\begin{linenomath*}
\begin{equation}
    \kinEnergy = \frac{1}{2} \int \vel^2 \dd V .
\end{equation}
\end{linenomath*}
The Elsasser number $\elsasser$ measures the dimensionless magnetic energy $\magEnergy$ and is given by 
\begin{linenomath*}
\begin{equation}
    \elsasser = \frac{\charmag^2}{2\rotation\meandensity\magperm\magdiff} = \frac{\magEnergy E}{\Pm \Vs} ,
    \label{eq:elsasser}
\end{equation}
\end{linenomath*}
where
\begin{linenomath*}
\begin{equation}
    \magEnergy = \frac{\Pm}{\Ek} \int \magfield^2 \dd V .
\end{equation}
\end{linenomath*}
From these we obtain the Reynolds number $\Rey = \charvel\shellthick/\kinvisc = \Rm/\Pm$,
the Lehnert number $\lehnert = \charmag/(\meandensity\magperm)^{1/2}\rotation\shellthick = \sqrt{2\elsasser \Ek / \Pm}$, the Rossby number $\Ro = \charvel/\rotation\shellthick = \Rey \Ek$, and the energy ratio $\Mratio=\magEnergy/\kinEnergy$.  

We estimate two characteristic lengthscales for the magnetic field and two for the velocity field. The scales $\dupol$ and $\dbpol$ are based on the spherical harmonic degrees $\lupol$ and $\lbpol$ corresponding to peaks in the poloidal kinetic and magnetic energy spectra respectively \citep{schwaiger_relating_2021}:
\begin{linenomath*}
\begin{equation}
    \frac{\dupol}{\shellthick} = \frac{\pi}{\lupol} , \qquad \frac{\dbpol}{\shellthick} = \frac{\pi}{\lbpol} .
\end{equation}
\end{linenomath*}
These quantities are designed to reflect the dominant scales of the flow and field. 
The second estimate is based on the viscous and magnetic dissipation scales, defined as 
\begin{linenomath*}
\begin{equation}
    \frac{\dumin}{\shellthick} = \left( \frac{\kinEnergy}{\viscdiss} \right)^{1/2} , \qquad \frac{\dbmin}{\shellthick} = \left( \frac{2 \magEnergy}{\ohmdiss} \right)^{1/2},
\end{equation}
\end{linenomath*}
where $\viscdiss$ and $\ohmdiss$ are the non-dimensional viscous and magnetic dissipation, given by 
\begin{linenomath*}
\iftwocol{
\begin{align}
\begin{split}
    \viscdiss &= \Pm \int \left( \nabla \times \vel \right)^2 \mathrm{d} V , \\
    \ohmdiss &= \frac{2 Pm}{E} \int \left( \nabla \times \magfield \right)^2 \mathrm{d} V .
\end{split}
\end{align}
}{
\begin{equation}
    \viscdiss = \Pm \int \left( \nabla \times \vel \right)^2 \mathrm{d} V , \qquad
    \ohmdiss = \frac{2 Pm}{E} \int \left( \nabla \times \magfield \right)^2 \mathrm{d} V .
\end{equation}
}
\end{linenomath*}
The ohmic dissipation fraction is calculated as 
\begin{linenomath*}
\begin{equation}
    f_\text{ohm}=\frac{\ohmdiss}{(\ohmdiss + \viscdiss)} ,
\end{equation}
\end{linenomath*}
and the local Rossby number is calculated according to the original definition of \citet{christensen_scaling_2006}, 
\begin{linenomath*}
\begin{equation}
    \Rol = \Ro \frac{\shellthick}{\duchrs},
    \qquad 
    \duchrs=\frac{\pi}{\lubar}\shellthick, 
    \qquad 
    \lubar=\frac{\sum_\ldeg \ldeg \left\langle\vel_\ldeg \cdot \vel_\ldeg\right\rangle}{2\kinEnergy}
    \label{eq:Rol}
\end{equation}
\end{linenomath*}
where $\duchrs$ is the characteristic velocity lengthscale, $\lubar$ is the mean spherical harmonic degree of the flow, and $\langle \dots \rangle$ denotes a time average.

To estimate the force balances in the simulations we employ the range of tools in \citet{naskar_2025}. \citet{Teed_Dormy_2023} showed that full understanding of the force balance cannot be obtained by looking only at volume integrated or spectral dependent forces, as incompressible flows include dynamically irrelevant gradients which are balanced by pressure. We therefore compute both the fluctuating and fluctuating curled forces. All quantities are integrated over the bulk fluid, which excludes regions of radial thickness 10 times the Ekman layer depth adjacent to the upper and lower boundaries (defined based on the linear intersection method). Forces and curled forces are decomposed into mean (azimuthal average) and fluctuating components \citep{calkins_large-scale_2021}. Here we report the integrated fluctuating forces and fluctuating curled forces and the spectra of these forces as defined in \citet{schwaiger_force_2019}. The force terms are denoted as Inertia ($I$), Coriolis ($C$), Pressure ($P$), buoyancy (i.e. Archimedean, $A$),  Lorentz (i.e. Magnetic, $M$) and Viscous ($V$) and are written in dimensionless form as

\begin{linenomath*}
\iftwocol{
\begin{equation}
\begin{split}
     I = \frac{\Ek}{\Pm} \vel^\prime\cdot\grad\vel^\prime , ~~~
     C =   \left(\zhat \times \vel^\prime\right), ~~~ 
     \pressure = -\grad\pressure^\prime , \\ 
     A = \frac{\Ra~\Ek~\Pm}{\Pra} \frac{r}{\outerrad} \temp^\prime\boldsymbol{\rhat} , ~~~
     M =   \left( \nabla \times \magfield^\prime \right) \times \magfield^\prime ,\\
     V =     \Ek \lap \vel^\prime ,
     \end{split}
\end{equation}
}
{
\begin{equation}
     I = \frac{\Ek}{\Pm} \vel^\prime\cdot\grad\vel^\prime , ~~~
     C =   \left(\zhat \times \vel^\prime\right), ~~~ 
     \pressure = -\grad\pressure^\prime , ~~~  
     A = \frac{\Ra~\Ek~\Pm}{\Pra} \frac{r}{\outerrad} \temp^\prime\boldsymbol{\rhat} , ~~~
     M =   \left( \nabla \times \magfield^\prime \right) \times \magfield^\prime , ~~~
     V =     \Ek \lap \vel^\prime ,
\end{equation}
}
\end{linenomath*}
where $\prime$ here denotes a fluctuating component. The residual between the Coriolis effect and pressure gradient is the ageostrophic Coriolis effect and is denoted $C_{ag}$. A detailed analysis of different methods for analysing dynamical balances in rotating spherical shell convection is given in \citet{naskar_2025}. 

Dipolarity of the magnetic field is calculated  by 
\begin{linenomath*}
\begin{equation}
 f_\text{dip} = \left(
    \frac{\int \magfield_{\ldeg=1}(\radius=\outerrad)\cdot\magfield_{\ldeg=1}(\radius=\outerrad)\dd A}
    {\int \magfield_{\ldeg\leq 12}(\radius=\outerrad)\cdot\magfield_{\ldeg\leq 12}(\radius=\outerrad)\dd A} 
    \right)^{1/2},
\end{equation}
\end{linenomath*}
where $\int \dots \dd A$ indicates integration over a spherical surface, which compares the energy in the dipole at the CMB with the energy in the rest of the magnetic field (up to $\ldeg=12$ in accordance with geomagnetic data resolution). 

We have run 18 dynamo simulations, detailed in the Appendix, which are designed to access 3 different parameter paths described in Section~\ref{sec:imac-theory} below. All simulations use no-slip, electrically insulating and fixed thermal flux boundary conditions on both boundaries. Simulations are run using the Leeds Dynamo Code, which is described in \cite{willis_thermal_2007} and \citet{davies2011scalability}. The code has been validated against other dynamo codes \citep{christensen_numerical_2001}, performs comparably with other dynamo codes \citep{matsui_performance_2016} and has been used in many subsequent studies (see e.g. \cite{davies_influence_2013}, \cite{mound_heat_2017}, \cite{mound_longitudinal_2023}). The variables are expressed in spherical harmonics $Y_\ldeg^m$ of degree $\ldeg$ and order $m$. Radial discretisation uses finite difference. Linear operations are computed in spherical space while non-linear terms and the Coriolis force are computed in real space. A predictor-corrector method is used for time integration.

\begin{table*}
\begin{minipage}{170mm}
\centering
    \caption{Symbols used in this work. }
    \label{tab:symbols}
    \begin{tabular}{|c|c|c|c|c}
    Symbol  & Quantity &~~~~~~~~ Earth Value & Units & Reference \\
    \hline
    $\outerrad$   & Outer radius     &~~~~~~~~ 3480   & km          & \citet{dziewonski1981preliminary} \\
    $\innerrad$   & Inner radius     &~~~~~~~~ 1221   & km          & \citet{dziewonski1981preliminary} \\
    $\meandensity$& mean density     &~~~~~~~~ $10^4$ & kg~m$^{-3}$ & \citet{dziewonski1981preliminary} \\
    $\grav_0$         & gravity          &~~~~~~~~ $9.9$  & m~s$^{-2}$  & \citet{dziewonski1981preliminary} \\
    $\vel$        & velocity         &~~~~~~~~ $10^{-3}-10^{-4}$ & m~s$^{-1}$ & \citet{finlay2011flow,holme2015large} \\ 
    $\magfield$   & magnetic field   &~~~~~~~~ $(2.5-4) \times 10^{-3}$ & T & \citet{gillet2010fast, buffett2010tidal} \\
    $\tp$         & temperature gradient at $\outerrad$ &~~~~~~~~ $10^{-4}$ & K~m$^{-1}$ & \citet{davies_constraints_2015} \\
    $\rotation$      & rotation rate       &~~~~~~~~ $7.272\times 10^{-5}$ & s$^{-1}$ & \citet{gross2007earth} \\
    $\magdiff$    & magnetic diffusivity&~~~~~~~~ 0.7-2 & m$^2$~s$^{-1}$ & \citet{davies_constraints_2015} \\
    $\kinvisc$    & kinematic viscosity &~~~~~~~~ $\sim10^{-6}$ & m$^2$~s$^{-1}$ &\citet{pozzo_transport_2013} \\
    $\thermdiff$  & thermal diffusivity &~~~~~~~~ $(5-15)\times 10^{-6}$ & m$^2$~s$^{-1}$ &\citet{pozzo_transport_2013} \\
    $\thermexpan$ & thermal expansivity &~~~~~~~~ $10^{-5}$     & K$^{-1}$ &  \citet{davies_constraints_2015} \\
    \hline
    Symbol  & Quantity     & \multicolumn{2}{c}{Earth Value} & Simulation Range \\
    \hline    
    $\Ek$     & Ekman Number & \multicolumn{2}{c}{$\sim 10^{-15}$} & $10^{-3} - 2 \times 10^{-6}$ \\
    $\Pm$    & magnetic Prandtl Number & \multicolumn{2}{c}{$\sim 10^{-6}$} & $10^{1} - 10^{0}$ \\
    $\Raf$  & Flux Rayleigh Number & \multicolumn{2}{c}{$\sim 10^{-15}$} & $10^{-3} - 2 \times 10^{-6}$ \\
    $\elsasser$ & Elsasser number & \multicolumn{2}{c}{$3-30$} & $15 - 100$\\
    $\lehnert$  & Lehnert  number & \multicolumn{2}{c}{$\sim 10^{-4}$} & $(0.9 - 6) \times 10^{-2}$ \\
    $\Ro$        & Rossby   number & \multicolumn{2}{c}{$\sim 10^{-5}-10^{-6}$} & $(2 - 35 ) \times 10^{-3}$ \\
    $\Rey$        & Reynolds number & \multicolumn{2}{c}{$\sim 10^{9}-10^{10}$} & $30 - 3000$ \\
    \hline
    \end{tabular}
    \end{minipage}
\end{table*}

\subsection{MAC and IMAC Scaling Theory}
\label{sec:imac-theory}

In this paper, we consider 3 different uni-dimensional parameter paths whereby all input and output quantities of the dynamo system are related to a single path parameter denoted by $\epsilon$. All three paths assume that the magnetic Reynolds number remains invariant and differ by the assumed dynamics and variations in core material properties. The first path is based on the Magneto-Archimedian-Coriolis (MAC) theory of \citet{davidson_scaling_2013} and \citet{aubert_spherical_2017} and is called a ``MAC'' path. The second two are based on Inertia-Magneto-Archimedian-Coriolis (IMAC) theory inspired by \citet{jones_805_2015} and are called ``IMAC'' paths. The MAC path and the two IMAC paths differ by the assumed variation of $\Pm$ with $\epsilon$. 

The path theories we consider are derived from time-averaged conservation of vorticity and total magnetic plus kinetic energy, which in the absence of viscosity, are
\begin{linenomath*}
\iftwocol{
\begin{equation}
\begin{split}
    \curl \left( \vel \cdot \grad \right) \vel 
    +
    2 \rotation \frac{\partial \vel}{\partial z}
    =
    -\curl \left( \frac{\thermexpan \temp \grav_0 r}{\outerrad} \rhat\right)
    + \\
    \frac{1}{\meandensity \magperm} \curl \left[ \left(\nabla \times \magfield \right) \cross \magfield)\right]
    \label{eq:vorticity}
    \end{split}
\end{equation}
}{
\begin{equation}
    \curl \left( \vel \cdot \grad \right) \vel 
    +
    2 \rotation \frac{\partial \vel}{\partial z}
    =
    -\curl \left( \frac{\thermexpan \temp \grav_0 r}{\outerrad} \rhat\right)
    +
    \frac{1}{\meandensity \magperm} \curl \left[ \left(\nabla \times \magfield \right) \cross \magfield)\right]
    \label{eq:vorticity}
\end{equation}
}
\end{linenomath*}
and
\begin{linenomath*}
\begin{equation}
    \fohm \int \thermexpan \grav_0 u_r \temp \mathrm{d} V = \frac{\eta}{\rho_0 \mu_0} \int  \left( \nabla \times \magfield \right)^2 \mathrm{d} V , 
    \label{eq:energy}
\end{equation}
\end{linenomath*}
where the factor $\fohm$ accounts for the fact that buoyant power is not all dissipated ohmically in numerical simulations. We reiterate that the inertial term is likely to be strongly subdominant in Earth's core compared to the remaining terms. However, in order to observe and understand the dipole-multipole transition in rapidly rotating dynamos, which generally reverse when the inertia is non-negligible, we retain the inertial force in the first-order vorticity balance.

To write the scaling estimates of equations~(\ref{eq:vorticity}) and (\ref{eq:energy}), we use four length scales. Following \cite{davidson_scaling_2013}, $\du$ denotes the width of convective columns in a plane perpendicular to the rotation axis, $\dpar$ denotes the length of the convective columns in the plane parallel to the rotation axis, which we take to be the system lengthscale $\dpar=\shellthick$, and $\dbmin$ denotes the ohmic dissipation length scale at which magnetic energy is converted to heat. We differ from \cite{davidson_scaling_2013} and \cite{aubert_spherical_2017} by assigning the magnetic field a lengthscale $\dB$ that can differ from  $\du$, based on the assumed scaling rules. We denote the scaling estimate of velocity, magnetic field, and temperature anomaly by $\charvel$, $\charmag$ and $\chartemp$, respectively. Assuming that $u_r \temp \sim \charvel\chartemp \sim F_c / (\meandensity c_p) \sim \thermdiff \tp \sim \kappa \beta / D^2$, where $\tp$ is the temperature gradient at the outer boundary, the scaling estimate of equations~(\ref{eq:vorticity}) and (\ref{eq:energy}) are 
\begin{linenomath*}
\begin{equation}
    \frac{\charvel^2}{\du^2} 
    \sim
    \frac{\rotation\charvel}{\shellthick} 
    \sim 
    \frac{\thermexpan \grav_o \thermdiff \tp}{\charvel \du}
    \sim 
    \frac{\charmag^2}{\meandensity\magperm\dB^2} ,
    \label{eq:IMAC-balance}
\end{equation}
\end{linenomath*}
\begin{linenomath*}
\begin{equation}
    \fohm \thermexpan \grav_0 \thermdiff \tp
    \sim 
    \frac{\magdiff \charmag^2}{\rho_0 \magperm (\dbmin)^2},
    \label{eq:power_balance}
\end{equation}
\end{linenomath*}
Consistency between equations~(\ref{eq:IMAC-balance}) and (\ref{eq:power_balance}) implies that 
\begin{linenomath*}
\begin{equation}
    \fohm \charvel\du \sim \magdiff \frac{\dB^2}{(\dbmin)^2} ,
    \label{eq:vort_eq}
\end{equation}
\end{linenomath*}
For the IMAC paths, the first three terms in equation~(\ref{eq:IMAC-balance}) immediately yield scaling predictions for $\Ro$ and $\du$: 
\begin{linenomath*}
\iftwocol{
\begin{align}
\begin{split}
    \Ro_{\rm IMAC} = \frac{\charvel}{\rotation \shellthick} \sim 
    \left( \frac{\thermexpan \grav_0 \thermdiff \tp }{\shellthick^2 \rotation^3} \right) ^{2/5} = \left( \Raf \right)^{2/5} , 
    \\
    \frac{d_{\mathrm{u}, {\rm IMAC}}}{\shellthick} \sim 
    \left( \frac{\thermexpan \grav_0 \thermdiff \tp }{\shellthick^2 \rotation^3} \right) ^{1/5} = \left( \Raf \right)^{1/5} 
    \label{eq:Ro_lp_IMAC}
    \end{split}
\end{align}
}{
\begin{equation}
    \Ro_{\rm IMAC} = \frac{\charvel}{\rotation \shellthick} \sim 
    \left( \frac{\thermexpan \grav_0 \thermdiff \tp }{\shellthick^2 \rotation^3} \right) ^{2/5} = \left( \Raf \right)^{2/5} , \qquad
    \frac{d_{\mathrm{u}, {\rm IMAC}}}{\shellthick} \sim 
    \left( \frac{\thermexpan \grav_0 \thermdiff \tp }{\shellthick^2 \rotation^3} \right) ^{1/5} = \left( \Raf \right)^{1/5} 
    \label{eq:Ro_lp_IMAC}
\end{equation}
}
\end{linenomath*}
\citep{aubert2001systematic}, where again we have used $\tp \sim \beta / D^2$ and the definition of $\Raf$ from equation~(\ref{eq:flux_rayleigh}). 

The magnetic field scaling depends on $\dB$. For the IMAC paths we follow \citet[][c.f. \citet{christensen_power_2004, moffatt_magnetic_1978}]{jones_805_2015} and assume that 
\begin{linenomath*}
\begin{equation}
    \dbmin \sim \Rm^{-1/2} \shellthick ,
    \label{eq:dbmin1}
\end{equation}
\end{linenomath*}
from which equation~(\ref{eq:power_balance}) implies that $B^2/(\meandensity\magperm) \sim \fohm \thermexpan \grav_0 \thermdiff \tp \shellthick / \charvel$ and hence
\begin{linenomath*}
\iftwocol{
\begin{align}
\begin{split}
    \lehnert_{\rm IMAC} &= \frac{\charmag}{\sqrt{\meandensity \magperm}\rotation \shellthick}
     \\ 
     &\sim \fohm^{1/2}
    \left( \frac{\thermexpan \grav_0 \thermdiff \tp }{\shellthick^2 \rotation^3} \right) ^{3/10} = \fohm^{1/2} \left( \Raf \right)^{3/10} .
    \label{eq:le1}
\end{split}
\end{align}
}{
\begin{equation}
    \lehnert_{\rm IMAC} = \frac{\charmag}{\sqrt{\meandensity \magperm}\rotation \shellthick}
    \sim \fohm^{1/2}
    \left( \frac{\thermexpan \grav_0 \thermdiff \tp }{\shellthick^2 \rotation^3} \right) ^{3/10} = \fohm^{1/2} \left( \Raf \right)^{3/10} .
    \label{eq:le1}
\end{equation}
}
\end{linenomath*}
Equation~(\ref{eq:dbmin1}) together with equation~(\ref{eq:vort_eq}) imply that
\begin{equation}
    d_{\mathrm{B}, {\rm IMAC}}^2 \sim \fohm d_{\mathrm{u}, {\rm IMAC}} \shellthick ,
    \label{eq:dB1}
\end{equation}
i.e. the field adopts a scale $d_{\mathrm{B}, {\rm IMAC}} \sim \left( \Raf \right)^{1/10}$ that is intermediate between the system scale and the flow scale. These results imply that 
\begin{linenomath*}
\begin{equation}
    \Mratio_{\rm IMAC} \sim  \fohm \left( \Raf \right)^{-1/5} .
    \label{eq:Mratio1}
\end{equation}
\end{linenomath*}

We note that the lengthscales $\dB$ and $\du$ are often assumed to scale in the same manner, i.e.
\begin{linenomath*}
\begin{equation}
    \dB \sim \du 
    \label{eq:dB2}
\end{equation} 
\end{linenomath*}
\citep{davidson_scaling_2013, aubert_spherical_2017}. 
In this case, the magnetic field strength scaling can be obtained directly from equation~(\ref{eq:IMAC-balance}) as 
\begin{linenomath*}
\begin{equation}
    \lehnert_{\rm IMAC} \sim \left( \Raf \right)^{2/5} .
    \label{eq:le2}
\end{equation}
\end{linenomath*}
It follows that 
\begin{linenomath*}
\begin{equation}
    \Mratio \sim  O(1),
    \label{eq:Mratio2}
\end{equation} 
\end{linenomath*}
while consistency with equation~(\ref{eq:power_balance}) requires that 
\begin{linenomath*}
\begin{equation}
\frac{\dbmin}{\shellthick} \sim \Rm^{-1/2} \left( \du/\shellthick \right)^{1/2} \sim \Rm^{-1/2} (\Raf)^{1/10} .
\label{eq:dbmin2}
\end{equation}
\end{linenomath*}
We have based our numerical calculations on the predictions from equations~(\ref{eq:dbmin1})--(\ref{eq:Mratio1}). However, since the properties of the field are determined as part of the numerical solution rather than being imposed by the path theory it is possible that the simulations can follow the predictions given by equations~(\ref{eq:dB2})--(\ref{eq:dbmin2}), or fall between the two sets of predictions. Therefore, while we adopt equations~(\ref{eq:dbmin1})--(\ref{eq:Mratio1}) in the following, we will also check the simulation outputs against the predictions from equations~(\ref{eq:dB2})--(\ref{eq:dbmin2}). 
 
In the MAC theory of \citet{davidson_scaling_2013} the inertial term is neglected in equation~(\ref{eq:vorticity}) and $\du = \dB$. By further assuming that the field strength is independent of the rotation rate and diffusion coefficients and setting $\fohm = 1$, \citet{davidson_scaling_2013} obtained the scalings $\Ro \sim \left( \Raf \right)^{4/9}$, $\du/\shellthick \sim \left( \Raf \right)^{1/9} $, and $\lehnert \sim \left( \Raf \right)^{1/3}$. \citet{aubert_spherical_2017} noted that the weak scaling of $\du$ with $\Raf$ implies a weak scaling of $\dbmin$ with $\Raf$ through the vorticity equivalence relation (\ref{eq:vort_eq}) and suggested a scale-invariant form of the \citet{davidson_scaling_2013} theory with $\du$ and $\dbmin$ independent of $\Raf$. In this case the MAC balance in equation~(\ref{eq:vorticity}) implies that
\begin{linenomath*}
\iftwocol{
\begin{equation}
\begin{split}
    \Ro_{\rm MAC} &\sim \left( \Raf \right)^{1/2} , \qquad
    d_{\mathrm{u}, {\rm MAC}} \sim  \shellthick , \\
    \lehnert_{\rm MAC} &\sim \left( \Raf \right)^{1/4}, \qquad \frac{\dbmin}{\shellthick} \sim \Rm^{-1/2}, \\ \Mratio &\sim \left( \Raf \right)^{-1/2}.
    \label{eq:Ro_lp_MAC}
\end{split}
\end{equation}
}{
\begin{equation}
    \Ro_{\rm MAC} \sim \left( \Raf \right)^{1/2} , \qquad
    d_{\mathrm{u}, {\rm MAC}} \sim  \shellthick , \qquad 
    \lehnert_{\rm MAC} \sim \left( \Raf \right)^{1/4}, \qquad \frac{\dbmin}{\shellthick} \sim \Rm^{-1/2}, \qquad \Mratio \sim \left( \Raf \right)^{-1/2}.
    \label{eq:Ro_lp_MAC}
\end{equation}
}
\end{linenomath*}
Note that if $\fohm$ is retained in the vorticity equivalence relation, these scalings are modified such that $\lehnert_{\rm MAC} \sim \fohm^{1/2} \left( \Raf \right)^{1/4}$ and $\Mratio \sim \fohm \left( \Raf \right)^{-1/2}$.

In the following section, we will use equations~(\ref{eq:Ro_lp_IMAC})--(\ref{eq:Ro_lp_MAC}) to derive uni-dimensional parameter paths based on IMAC and MAC theories.

\subsection{MAC and IMAC Path Theory}

The goal of the path theory is to express the input parameters $\Ek$, $\Pm$, $\Pra$ and $\Raf$ alongside the main output parameters $\Ro$, $\lehnert$, $\du$, $\dB$ and $\dbmin$ in terms of a single path parameter $\epsilon$. Formally, for a parameter $x$ we seek relations of the form 
\begin{linenomath*}
\begin{equation}
  x \sim \epsilon^{\alpha_x} x_0 ,
\end{equation}
\end{linenomath*}
where the exponent $\alpha_x$ describes the variation of $x$ along the path and subscript $0$ denotes quantities obtained from a simulation at the start of the path. Once this task is completed, other diagnostics such as $\Rey$, $\elsasser$ and $\Mratio$ can be obtained directly. Moving along the path (in this case by decreasing $\epsilon$) should lead to parameter values that approach those of Earth's core, and the level of agreement can be used to validate and compare different path theories. Here we keep $\Pr=1$, so it will be neglected in further discussion. 

The path parameter is defined following \citet{aubert_spherical_2017} as
\begin{linenomath*}
\begin{equation}
   \Raf = \epsilon (\Raf)_0 .
\end{equation}
\end{linenomath*}
As explained by \citet{aubert_spherical_2017}, a reasonable path should minimally preserve the value of $\Rm$, since current simulations can already access values of $\Rm \sim 1000$ that are inferred from geomagnetic secular variation \citep[e.g.][]{holme2015large}. This condition is written
\begin{linenomath*}
\begin{equation}
    \Rm = \Rm_0 
    \label{eq:rm0}
\end{equation}
\end{linenomath*}
along all paths. From the definition of $\Rm$ we obtain
\begin{linenomath*}
\iftwocol{
\begin{equation}
\begin{split}
    \Rm = \Ro \frac{\Pm}{\Ek} 
    = 
    (\Ro)_0 \epsilon^{\alpha_{\Ro}} \frac{(\Pm)_0\epsilon^{\alpha_{\Pm}}}{(\Ek)_0\epsilon^{\alpha_{\Ek}}} 
    \\ =
    \Rm_0  \epsilon^{\alpha_{\Ro} + \alpha_{\Pm} - \alpha_{\Ek}},
    \label{eq:Rm}
\end{split}
\end{equation}
}{
\begin{equation}
    \Rm = \Ro \frac{\Pm}{\Ek} 
    = 
    (\Ro)_0 \epsilon^{\alpha_{\Ro}} \frac{(\Pm)_0\epsilon^{\alpha_{\Pm}}}{(\Ek)_0\epsilon^{\alpha_{\Ek}}} 
    =
    \Rm_0  \epsilon^{\alpha_{\Ro} + \alpha_{\Pm} - \alpha_{\Ek}},
    \label{eq:Rm}
\end{equation}
}
\end{linenomath*}
where $\alpha_{\Ro} \in \{ 0.4, 0.5\}$ is the exponent obtained from the theoretical $\Ro \sim (\Raf)^{\alpha_{\Ro}}$ scaling. This implies that $\Ek/\Pm$ must also scale as $\epsilon^{\alpha_{\Ro}}$ in order to maintain an approximately constant $\Rm$ along the path \eqref{eq:rm0}. This condition is written
\begin{linenomath*}
\begin{equation}
    \frac{\Ek}{\Pm} = \epsilon^{\alpha_{\Ro}} \frac{\Ek_0}{\Pm_0} .
    \label{eq:EPm}
\end{equation}
\end{linenomath*}

The constraint of invariant $\Rm$ constrains the ratio of $\Ek/\Pm$ but not $\Ek$ or $\Pm$ individually. Clearly, $\Pm$ must decrease along the path from the $O(1)$ values used in current simulations to the values of $O(10^{-6})$ that characterise liquid metals (Table~\ref{tab:symbols}). \citet{aubert_spherical_2017} chose the scaling $\Pm = \sqrt{\epsilon} \Pm_0$, which we denote the Pm0.5 path. We also consider the scaling $\Pm = \epsilon \Pm_0$, which enforces a faster decrease of $\Pm$ along the path. Based on the study of \citet{dormy_strong-field_2016} we expect that this Pm1 path will lead to weaker magnetic forces, potentially placing the simulations closer to the dipole-multipole transition \citep{tassin_geomagnetic_2021}. Denoting the $\Pm$ exponent $\alpha_{\Pm} \in \{ 0.5, 1\}$, equation~(\ref{eq:EPm}) gives the Ekman number exponent $\alpha_\Ek$ as 
\begin{linenomath*}
\begin{equation}
    \alpha_E = \alpha_{Ro} + \alpha_{Pm} .
\end{equation}
\end{linenomath*}

The MAC and IMAC scaling theories give the exponents for the Lehnert number  $\alpha_{\lehnert} \in \{ 0.25, 0.3\}$ and  flow lengthscale $\alpha_{\du} \in \{ 0, 0.2\}$. From these the exponents for $\Rey$, $\elsasser$, $\Rol$ and $\Mratio$ are
\begin{linenomath*}
\begin{align}
    \alpha_{\Rey}        = & \alpha_{\Ro}   - \alpha_\Ek        , \\
    \alpha_{\elsasser} = & 2\alpha_{\lehnert}  + \alpha_{\Pm} - \alpha_{\Ek}  , \\ 
    \alpha_{\Rol}      = & \alpha_{\Ro}   - \alpha_{\du} , \\
    \alpha_\Mratio           = & 2\alpha_{\lehnert}  - 2\alpha_{\Ro}   .
\end{align}
\end{linenomath*}

The starting simulation for the IMAC paths is LEDT002 from \cite{nakagawa_combined_2022}, which was chosen because it has $\Mratio>1$, $\Rm \sim 1000$, and a dipolar field with long periods (on the order of a magnetic diffusion time) of stable dipole tilt, whilst still exhibiting reversals of the magnetic polarity. These properties are all known to be present in the geodynamo, and we should attempt to preserve them on any path deemed to be approaching the Earth. The MAC path starts at slightly different conditions to better match existing MAC paths eg. \cite{aubert_spherical_2017}.

The properties of the 3 paths are summarised in Table~\ref{tab:paths} together with the MAC-Pm0.5 path of \citet{aubert_spherical_2017}. To compare the path predictions to the parameter values inferred for Earth's present-day core (see Table~\ref{tab:symbols}) a value of $\epsilon$ is required. This value is in some sense arbitrary; \citet{aubert_spherical_2017} found that $\epsilon = 10^{-7}$ gave good agreement between the path predictions and parameter values inferred for Earth's core. The paths considered here show optimal agreement with Earth's core at different values of $\epsilon$, so for direct comparison, we use the value $\epsilon = 10^{-8}$, which is a good compromise between the results for the different paths. Taking the predictions for this value of $\epsilon$ at face value, the Pm1 paths provide an excellent match to Earth's values of $\Pm$,  $\Ek$, and $\Rey$; the Pm0.5 paths fall short of Earth's values, but may still have reached asymptotically small values \citep{aubert_spherical_2017}. All paths provide acceptable matches to the value of $\Ro$ and $\lehnert$ given current uncertainties in $\charvel$ and $\charmag$. The IMAC paths give plausible values for $\elsasser$ and $\Mratio$; they are somewhat lower than values generally quoted in the literature, but within uncertainties on the parameters. It is worth noting that IMAC paths based on the assumption $\du \sim \dB$ predict values of $\elsasser$ and $\Mratio$ that are too small to be geophysically relevant. 

\begin{table*}
\begin{minipage}{170mm}
    \centering
     \caption{Predictions from the original MAC-Pm0.5 path of \citet{aubert_spherical_2017} and the 3 different unidimensional parameter paths considered in this study: MAC-Pm1, IMAC-Pm0.5, IMAC-Pm1. For each path the exponent of quantity $x$, $\alpha_x$, is provided alongside the predicted value of $x$ at Earth's core conditions (assumed to be at $\epsilon = 10^{-8}$), according to $x = \epsilon^{\alpha_x} x_0$ where $x_0$ denotes the starting point on the path. Predictions assume $\fohm = 1$. The starting point for IMAC paths is the simulation LEDT002, which has $\Rm_0 = 1185$, $\Ek_0 = 10^{-3}$, $\Pm_0 = 35.3$, $\lehnert_0 = 7.158 \times 10^{-2} \times \sqrt(0.33)$, $\Ra_{F,0} = 4.8 \times 10^{-4}$, $\Mratio_0 = 5.0$. Starting point for MAC-Pm1 path is sim 15 of this work, with $\Rm_0=1417.5$, $\Ek_0 = 3\times 10^{-4}$, $\Pm_0 = 14.96$, $\lehnert = 4.39\times 10^{-2}$, $\Ra_{F,0}=2.031\times10^{-4}$, $\Mratio=2.388$. }
    \begin{tabular}{|c|cc|cc|cc|cc|cc|}
    \hline
         Quantity & \multicolumn{2}{c}{MAC-Pm0.5} & \multicolumn{2}{c}{MAC-Pm1} &  \multicolumn{2}{c}{IMAC-Pm0.5} & \multicolumn{2}{c}{IMAC-Pm1} \\
         & $\epsilon^{\alpha_i}$ & $\epsilon = 10^{-8}$ 
         & $\epsilon^{\alpha_i}$ & $\epsilon = 10^{-8}$ 
         & $\epsilon^{\alpha_i}$ & $\epsilon = 10^{-8}$ 
         & $\epsilon^{\alpha_i}$ & $\epsilon = 10^{-8}$  \\
         \hline
         $\Pm$      & $1/2$ & $3\times 10^{-3}$  & $1$   & $3\times 10^{-7}$  
                   & $1/2$ & $3\times 10^{-3}$  & $1$   & $3\times 10^{-7}$ \\
         $\Ro$      & $1/2$ & $3\times 10^{-6}$  & $1/2$ & $3\times 10^{-6}$  
                   & $2/5$ & $2\times 10^{-5}$  & $2/5$ & $2\times 10^{-5}$ \\
         $\Ek$       & $1$   & $1\times 10^{-11}$ & $3/2$ & $1\times 10^{-15}$ 
                   & $9/10$& $6\times 10^{-11}$ & $7/5$ & $6\times 10^{-15}$\\
         $\Rey$      & $-1/2$& $3\times 10^{5}$      & $-1$  & $3\times 10^{9}$  
                   & $-1/2$& $3\times 10^{5}$      & $-1$  & $3\times 10^{9}$     \\
         $\du/\shellthick$   & $0$   & $0.1-1$ & $0$   & $0.1-1$  
                   & $1/5$ & $0.01-0.1$ & $1/5$ & $0.01-0.1$ \\
         $\Rol$    & $1/2$ & $7 \times 10^{-6}$ & $1/2$ & $7\times 10^{-6}$  
                   & $1/5$ & $1 \times 10^{-3}$ & $1/5$ & $1 \times 10^{-3}$    \\
        \hline
         $\dB/\shellthick$   & $0$   & - & $0$    & -    
                   & $1/10$& - & $1/10$ & - \\       
         $\dbmin/\shellthick$ & $0$   & - & $0$ &  -    
                   & $0$   & - & $0$ & - \\  
         $\lehnert$      & $1/4$ & $4\times 10^{-4}$  & $1/4$ & $4\times 10^{-4}$  
                   & $3/10$& $2\times 10^{-4}$  & $3/10$& $2\times 10^{-4}$ \\
         $\elsasser$ & $0$   & $60$               & $0$   & $60$ 
                   & $1/5$ & $1.5$              & $1/5$ & $1.5$               \\
         $\Mratio$ & $-1/2$& $5\times 10^{4}$   & $-1/2$& $5\times 10^{4}$  
                   & $-1/5$& $200$              & $-1/5$& $200$        \\
         \hline
    \end{tabular}
    \label{tab:paths}
    \end{minipage}
\end{table*}

Table~\ref{tab:paths} shows that the four paths make similar predictions for several gross properties of Earth's present-day core despite being built from different dynamical assumptions. For the present core the inferred values of $U$ and $B$ present sound evidence to favour the MAC paths, while the paleointensity record has been advocated as a discriminator between different scaling theories on longer timescales \citep{davies_dynamo_2022}. However, in the numerically accessible parameter space, these results suggest that it will be difficult to discriminate between the path theories based on their scaling predictions alone. We will therefore conduct a detailed anaylsis of the dynamical balances in the force and vorticity equations to attempt to distinguish the behaviour along the numerically accessible sections of the paths.  

\subsection{Hyperdiffusion}

In order to reach lower Ekman numbers, we adopt the hyperdiffusion approach of \citet[][following \citet{nataf_806_2015}]{aubert_spherical_2017}, whereby computations are restricted to spherical harmonic degrees $\ell < \lbmin  = \pi / \dbmin$ and enhanced diffusion is applied to the velocity and temperature field for degrees $\ell > \lh$. The diffusion operators for the velocity and temperature equations are replaced by 
\begin{linenomath*}
\begin{equation}
\lap ~~ \text{for} ~~\ldeg<\lh, \qquad \qh^{\ldeg-\lh} \lap ~~ \text{for}~~ \ldeg\geq \lh
\label{eq:HD}
\end{equation}
\end{linenomath*}
where $\qh = 1.0325$ is set close to unity to ensure a smooth increase of the hyperdiffusion with wavenumber. 

The magnetic dissipation scale is $\dbmin \sim \Rm^{-1/2}\shellthick$ for the MAC paths and $\dbmin \sim \Rm^{-1/2} (\Raf)^{1/10}\shellthick$ for the IMAC paths, which translate into spherical harmonic truncation degrees of $\lmax \approx 100-260$ for the simulations considered here and so we use this to define the values of $\lmax$ in Table~\ref{tab:sim_outputs} in the Appendix. We also calculate $\dbmin$ directly from the simulations to check for consistency. The value of $\lh$ must be sufficiently larger than the characterisitc degree of the flow, which we estimate as $\lupol$. \citet{aubert_spherical_2017} used $\lh \approx 3\lubar$ and found little effect of the hyperdiffusion on the large-scale solution. Here we take $\lh \approx 4-5\lupol$. For the MAC path $\lupol$ is relatively constant at $\dupol \sim 10$ \citep{aubert_spherical_2017}. For the IMAC paths $\lupol$ varies from $4-15$ as $\Raf$ increases. We therefore set $\lh = 50$ for the simulations with the lowest $\Ra$, increasing to $\lh = 192$ for the simulations with the highest $\Ra$. 

\begin{figure}
    \centering
    \includegraphics[height=4cm]{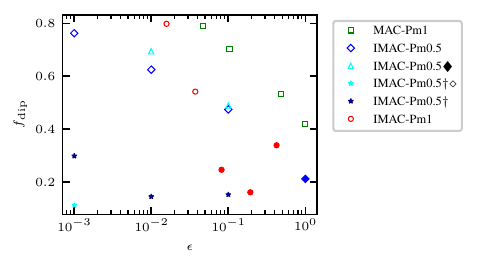}
    \caption{Dipolarity $\fdip$ as a function of path parameter $\epsilon$ for all simulations. Filled symbols indicate reversing/multipolar dynamo simulations, whilst open symbols indicate dipolar. Symbol shape indicates path as described in text. For IMAC-Pm0.5 path, $\dagger$ indicates off-path high Rayleigh number, $\blacklozenge$ indicates DNS (no hyper-diffusion used), and $\diamond$ indicates higher resolution sim used for numerical convergence testing.}
    \label{fig:fodip_scaling}
\end{figure}

\section{Results} \label{sec:results}

Table~\ref{tab:sim_outputs} summarises the numerical simulations performed for this study (A complete list of inputs and outputs can be found in the accompanying spreadsheet). Simulations on the IMAC-Pm0.5, IMAC-Pm1 and MAC-Pm1 path have been conducted down to an Ekman number $\Ek = 2\times 10^{-6}$. Additionally, several simulations have been run by taking the parameters from the IMAC-Pm0.5 path and increasing $\Ra$. These ``off path'' simulations are denoted by a \textdagger. 

Figure~\ref{fig:fodip_scaling} summarises the dynamo regime for all simulations by plotting the dipolarity $\fdip$ as a function of $\epsilon$. In this plot, solid symbols denote reversing/multipolar solutions while open symbols denote dipole-dominated non-reversing simulations. $\fdip$ increases with decreasing $\epsilon$ in all cases, reaching values of $0.7$ at the lowest values of $\epsilon$ considered. Numerical dynamos generally reverse when $\fdip < 0.4-0.5$ \citep{christensen_convection-driven_2006,oruba2014transition, tassin_geomagnetic_2021} and so the increase in $\fdip$ is consistent with the observation that the dynamo transitions from a reversing to non-reversing state along each path. 

In our suite of simulations, all except 2 use the hyperdiffusion defined by equation~(\ref{eq:HD}) and so in the following we first demonstrate that the HD treatment does not significantly perturb the DNS solution (Section~\ref{sec:HDval}). In Sections~\ref{sec:forces} and \ref{sec:scalings} we respectively analyse the dynamical balances and scaling behaviour along the 3 considered paths as a function of $\epsilon$. In Section \ref{sec:ekman} we briefly consider the scaling behaviour as a function of $\Ek$. Finally, in Section~\ref{sec:reversals} we  analyse the behaviour of the multipolar solutions.

\subsection{Validation of the HD Scheme} 
\label{sec:HDval}

\begin{figure*}
    \centering
         \includegraphics[scale=0.5]{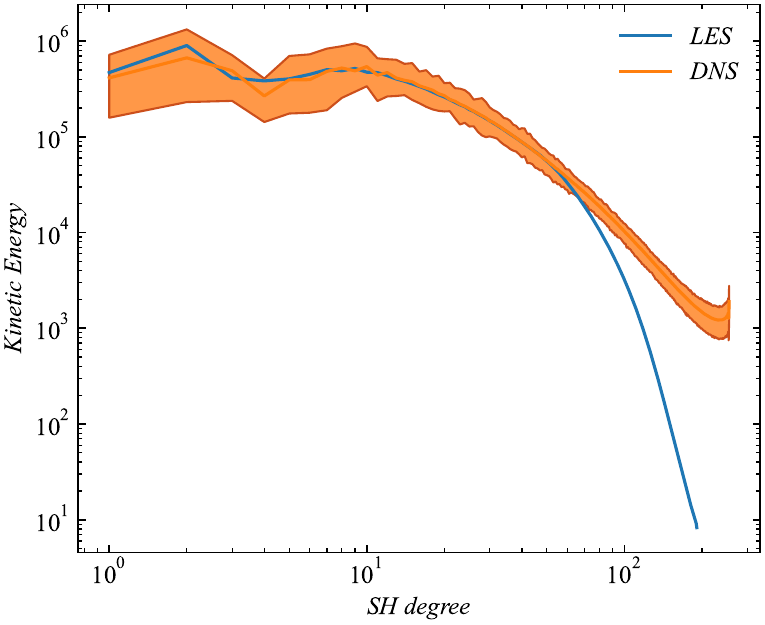}%
        \includegraphics[scale=0.5]{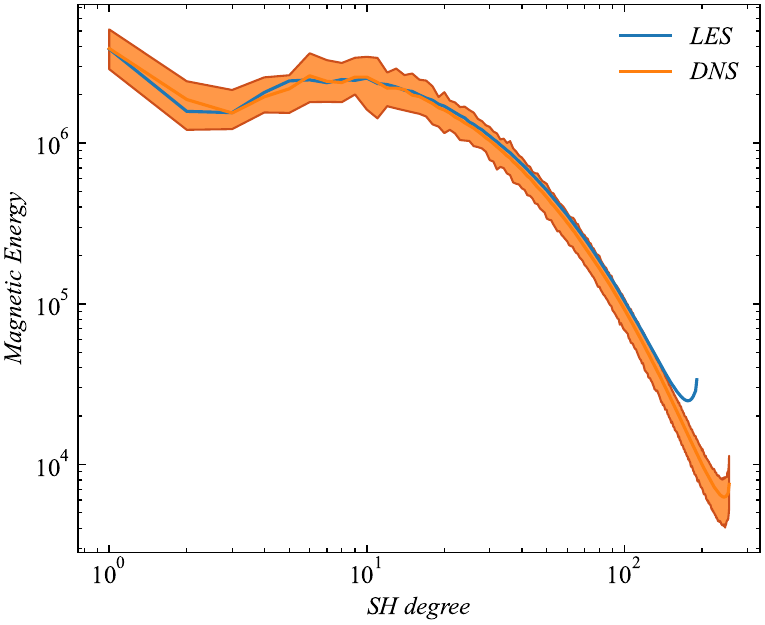} \\
        \includegraphics[scale=0.5]{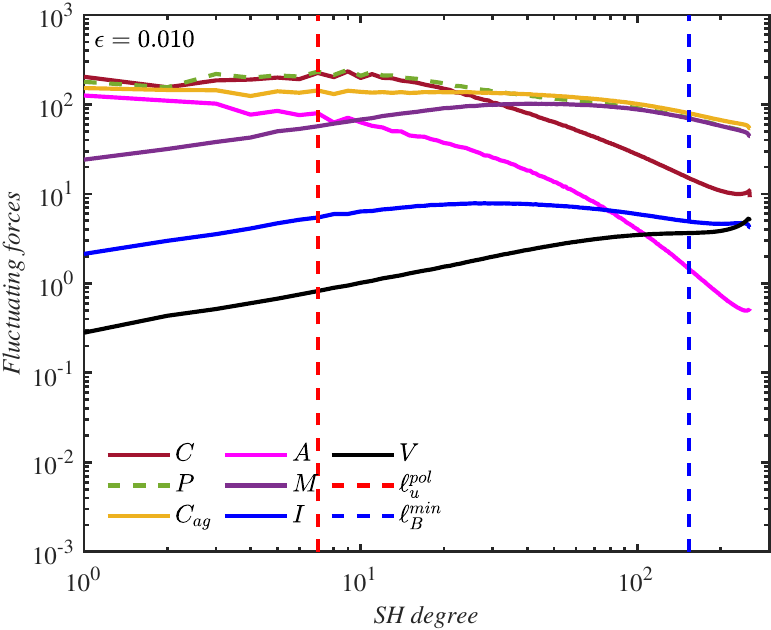} 
        \includegraphics[scale=0.5]{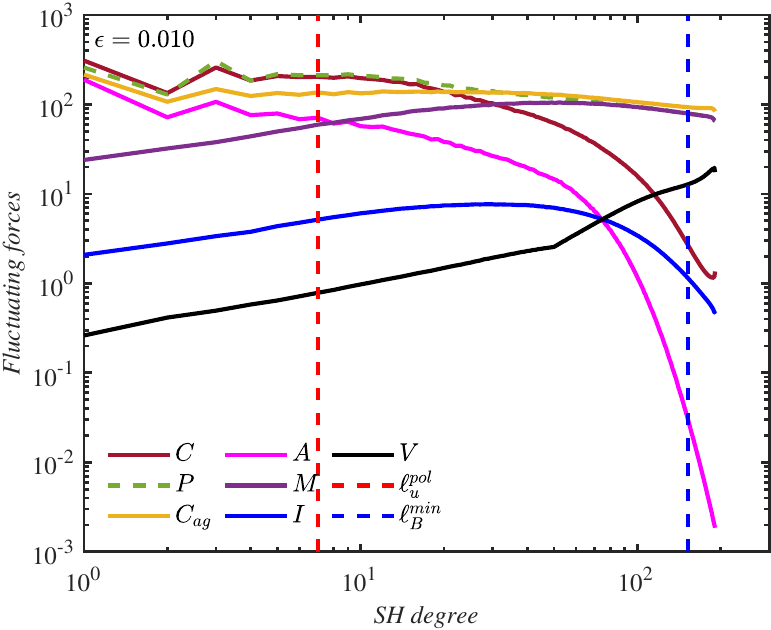}
    \caption{Effect of HD on energy spectra (top row) and force spectra (bottom row), for a simulation on the IMAC-Pm0.5 path with Ekman number $1.58\times 10^{-5}$. In the top row, kinetic energy is shown on the left and magnetic energy on the right. The blue and orange lines correspond to the HD and DNS simulations, respectively, whilst the orange shaded region highlights the maximum and minimum energy values calculated over the DNS simulation time series for each SH degree. In the bottom row, the DNS simulation is shown on the left and the HD simulation on the right. Fluctuating forces $C$ (brown), $P$ (dashed green), $C_{ag}$ (yellow), $A$ (pink), $M$ (purple), $I$ (solid blue) and $V$ (black) are defined in Section \ref{subsec:equations}. The red and blue dashed lines show the spherical harmonic degree corresponding to the peak in poloidal kinetic energy spectra $\lupol$ and the ohmic dissipation $\lbmin$, respectively.}
    \label{fig:dns_vs_les_ek_1e-5}
\end{figure*}

Figure~\ref{fig:dns_vs_les_ek_1e-5} compares HD and DNS simulations on the IMAC-Pm0.5 path with $\Ek = 1.58 \times 10^{-5}$. The DNS simulations use a maximum spherical harmonic truncation of $\lmax = 256$, while the HD simulation uses $\lmax=192$ and $\lh=50$, which is a factor of 5 greater than the dominant energy-containing  wavenumber $\lupol \approx 10 $ of the DNS. Magnetic and kinetic energy spectra for the HD cases closely approximate both the amplitude and wavenumber dependence of the corresponding DNS case down to wavenumbers well past the spectral peaks. Both HD spectra slightly over-estimate the energy at intermediate wavenumbers; however, the difference is within the max/min of the DNS and is expected to reduce further were longer temporal averaging available. The effect of the HD is clear in the kinetic energy spectrum (note that HD is not applied to the magnetic field) where energy in the HD case is suppressed for wavenumbers above the cutoff $\lh=50$. Figures~\ref{fig:dns_vs_les_ek_1e-5}c,d show that force balance spectra are also very similar between the HD and DNS cases. The effect of the HD is evident in the viscous term, which rises above its DNS counterpart above $\lh$. This causes a stronger decrease of the Coriolis, buoyancy and inertial terms with wavenumber above $\lh$ compared to the DNS. It is apparent from Table~\ref{tab:sim_outputs} that these effects produce differences of at most 3\% in the global field and flow diagnostics (compare simulations 6 and 10). 

\begin{figure*}
    \centering
    \includegraphics[width=0.35\linewidth]{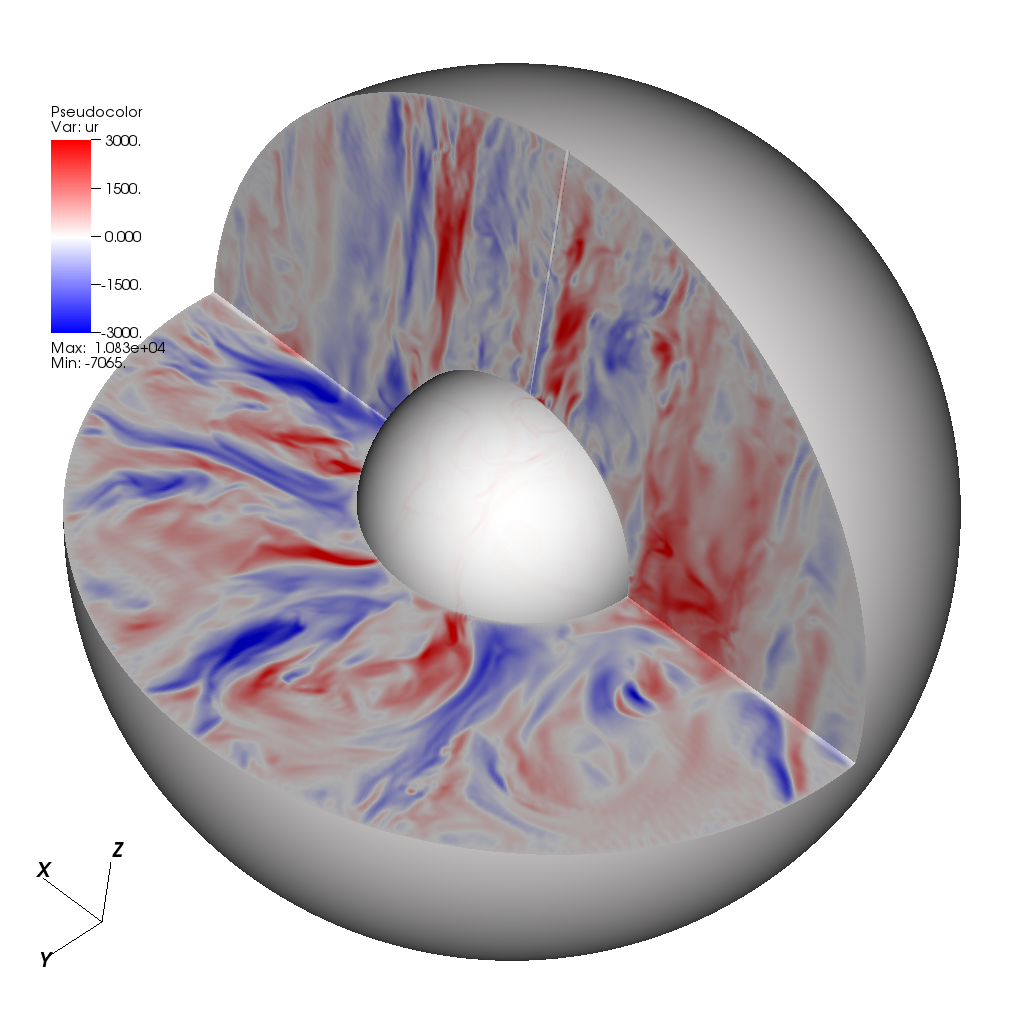}%
    \includegraphics[width=0.35\linewidth]{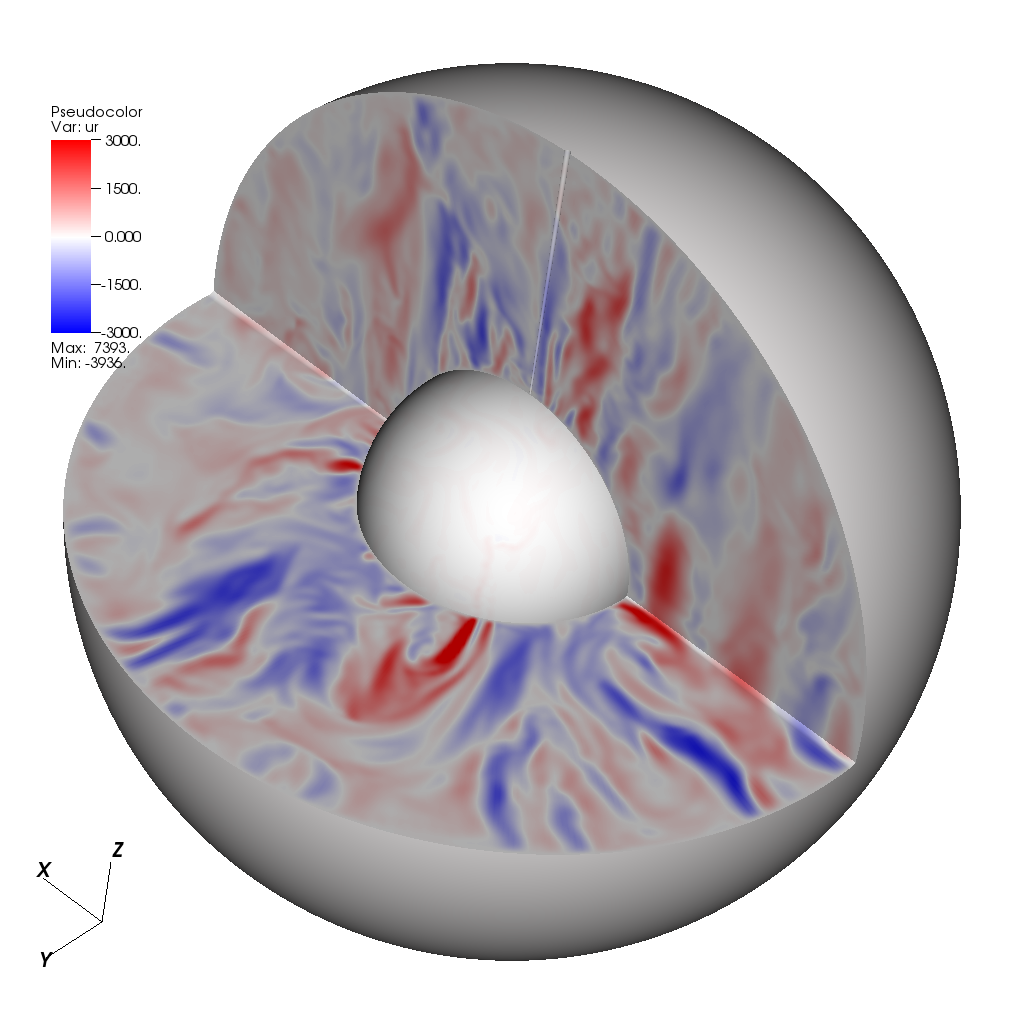}\\
    \includegraphics[width=0.35\linewidth]{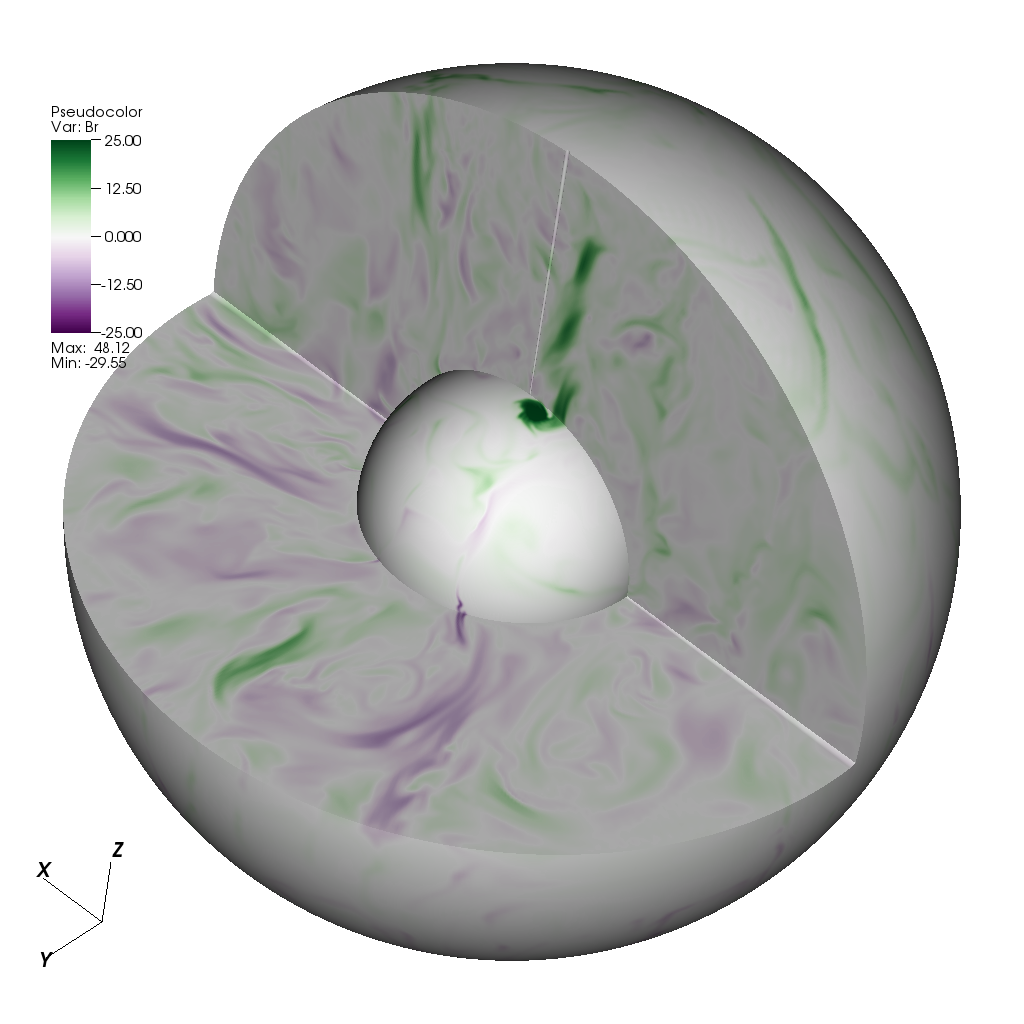}%
    \includegraphics[width=0.35\linewidth]{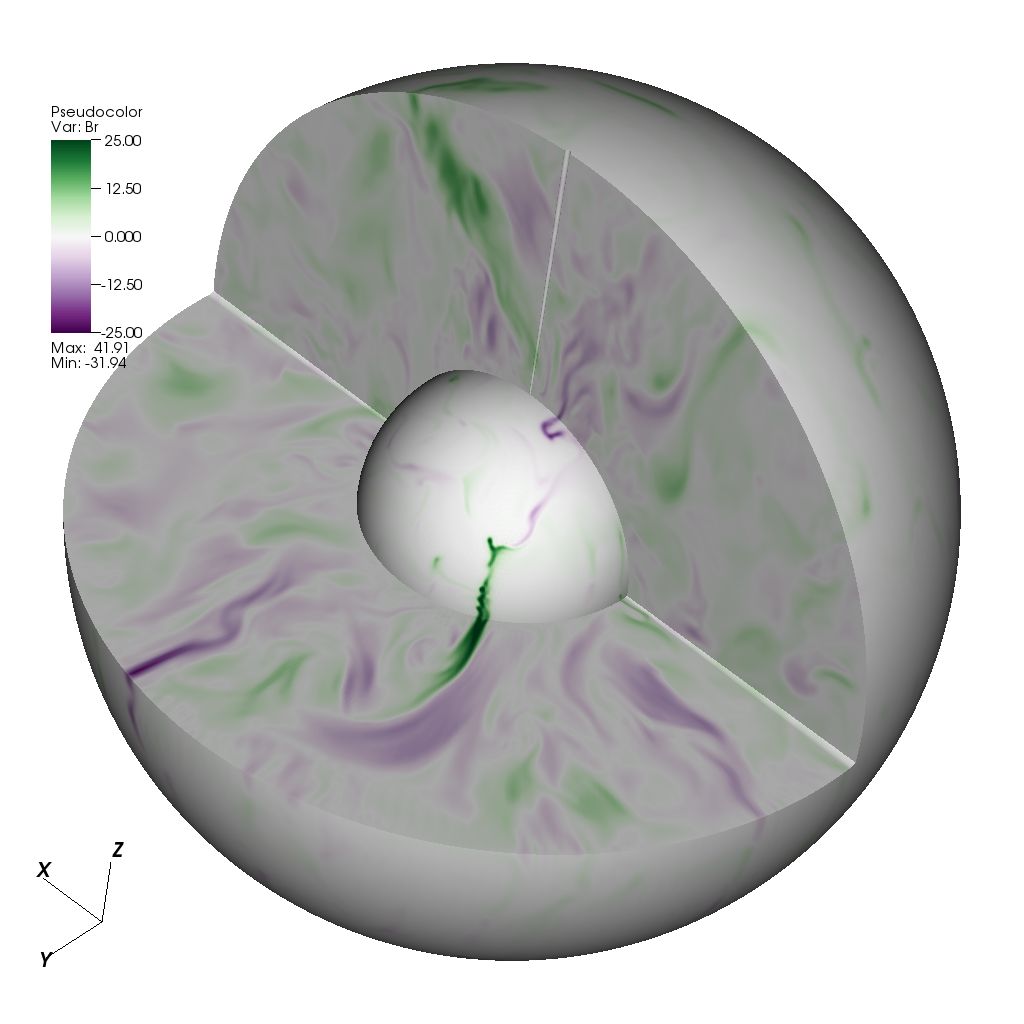}
    \caption{3D isometric projections of the radial component of velocity (top) and of the radial component of magnetic field (bottom), for equivalent DNS (left) and HD (right) simulations at $\Ek=1.58 \times 10^{-5}$.}
    \label{fig:dns_vs_les_vol}
\end{figure*}
To further check the effect of HD, Figure~\ref{fig:dns_vs_les_vol} shows snapshots of the radial flow and magnetic field in a 3D isometric projection. It is apparent that the general alignment of flow structures parallel to the rotation axis as well as the lateral scale of field and flow features, are comparable between the two cases. The outer boundary magnetic fields are dipole-dominated, with high-latitude normal polarity flux patches situated outside the tangent cylinder. This confirms the structural similarities apparent from the spectral plots in Figure~\ref{fig:dns_vs_les_ek_1e-5}.

We made a further comparison between HD and DNS cases with $\Ek = 1.26\times 10^{-4}$ where an HD cutoff of $\lh=50$ is again used; the results are shown in the Supplementary Figure~\ref{fig:dns_vs_les_ek_1e-4}, and diagnostics are reported in Table~\ref{tab:sim_outputs} (see cases 7 and 9). This test confirms the behaviour seen in Figure~\ref{fig:dns_vs_les_ek_1e-5} and leads us to expect that the HD treatment has a minimal effect on the large-scale behaviour of our simulations. This conclusion is reinforced by comparing to the results of \citet{aubert_spherical_2017}, who came to the same conclusion when analysing simulations where HD was enforced from a lower wavenumber, $\lh=30$. 

\subsection{Dynamical Balances along the Paths} 
\label{sec:forces}

When assessing the dynamical balances in our simulations, and the variation of scaling predictions in section~\ref{sec:scalings} below, it must be borne in mind that the computationally accessible range of parameters is set by the Ekman number, which reaches $\Ek = 2-3\times 10^{-6}$ along each path. However, this $\Ek$ corresponds to very different values of $\epsilon$, with the MAC-Pm1 path reaching $\epsilon = 5 \times 10^{-2}$, the IMAC-Pm1 path reaching $\epsilon = 10^{-2}$ and the IMAC-Pm0.5 path reaching $\epsilon = 10^{-3}$. We therefore have limited capacity to test the $\epsilon$-dependence of the dynamical balances that we have attempted to impose through the path theory. However, we can establish whether the balances used to derive the different paths are consistent with outputs from the simulations and this is what we do here. 

\begin{figure*}
    \centering
    \includegraphics[width=0.32\linewidth]{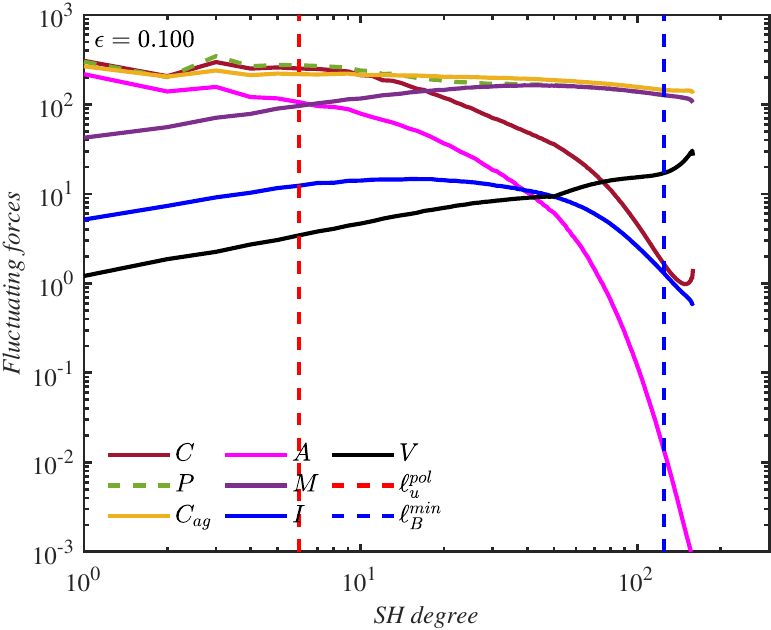}%
    \includegraphics[width=0.32\linewidth]{figs/fluc_force_spec_dyn_IMAC_PM0p5_eps=0.01.pdf}%
    \includegraphics[width=0.328\linewidth]{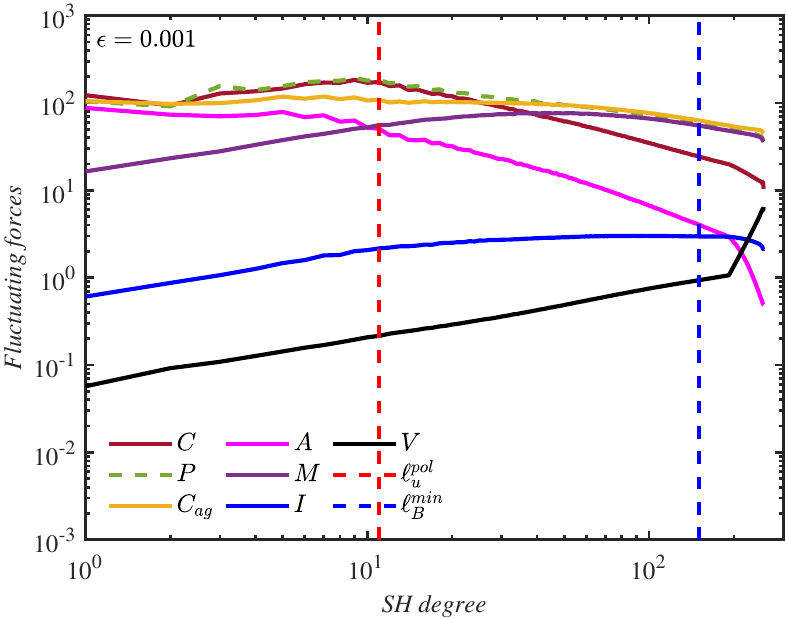}\\
    \includegraphics[width=0.32\linewidth]{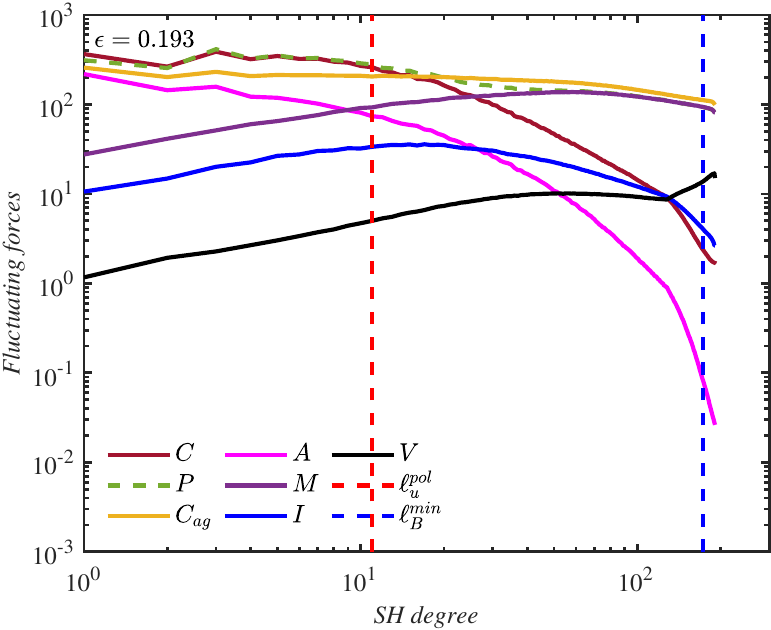}%
    \includegraphics[width=0.32\linewidth]{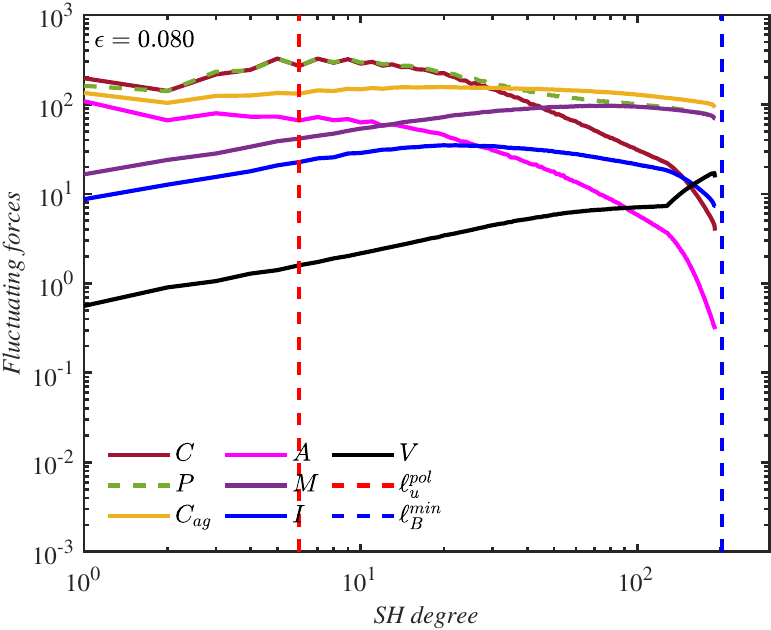}%
    \includegraphics[width=0.32\linewidth]{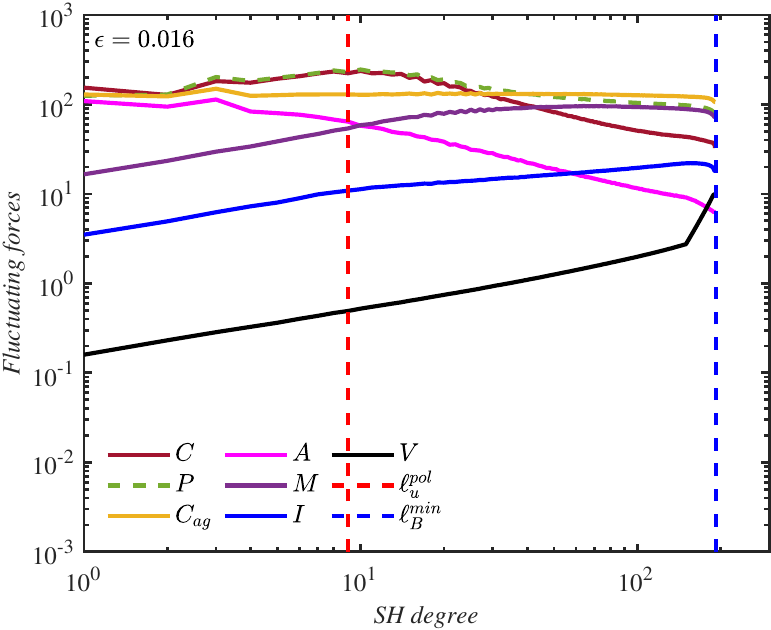}%
    \caption{Force spectra for simulations along the IMAC-Pm0.5 (top) and IMAC-Pm1 (bottom) paths The three plots in each row show the lowest values of $\epsilon$ for each path. Forces are and dashed lines are labelled in the same way as Figure \ref{fig:dns_vs_les_ek_1e-5}.}
    \label{fig:force_spectra_imacpm0.5}
\end{figure*}

Force spectra for the three lowest $\Ek$ cases along the IMAC-Pm0.5 and IMAC-Pm1 paths are shown in Figure~\ref{fig:force_spectra_imacpm0.5}. The balance of forces appears broadly consistent with the theoretical QG-IMAC balance, but differs in detail. For both paths, the dominant balance at the largest scales is essentially QGA (the buoyancy term is within a factor of 2 of the Coriolis and pressure gradient), with the Lorentz force replacing the Coriolis effect at small scales. For the IMAC-Pm0.5 path, the secondary balance is MAC around the dominant scale, shown by a vertical red dotted line, with inertia subdominant by up to an order of magnitude as $\epsilon$ decreases.  The secondary balance that determines the flow and field strength is therefore somewhere between MAC and IMAC. This is expected given the moderate parameter values employed and the design of the path to ensure that inertia remains comparable to the MAC terms. For the IMAC-Pm1 path a secondary IMAC balance is robust in the range of $\epsilon$ considered, though the inertial term does seem to be gradually dropping further below the Lorentz force. Along both paths $\dupol$ is close to the crossing between buoyancy and Lorentz terms, as has been found in previous studies \citep{schwaiger_force_2019}. 

\begin{figure*}
    \centering
    \includegraphics[width=0.32\linewidth]{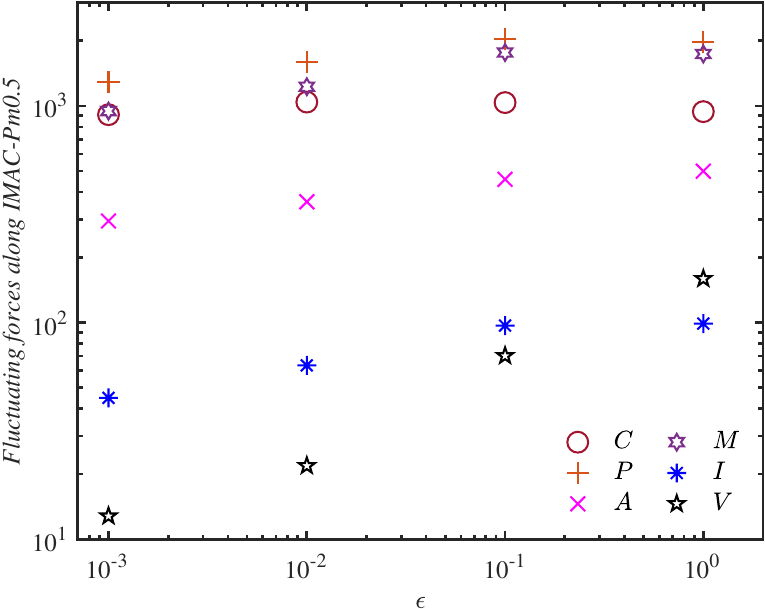}%
    \includegraphics[width=0.34\linewidth]{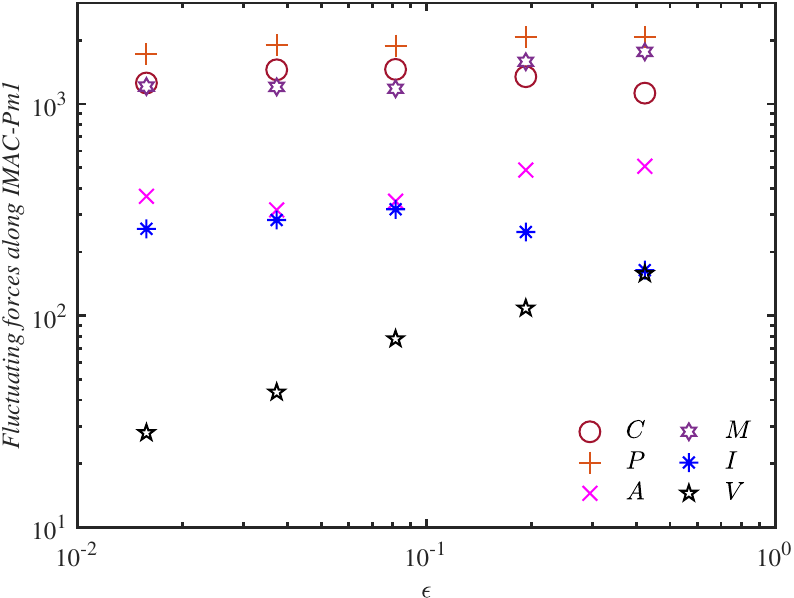}
    \includegraphics[width=0.32\linewidth]{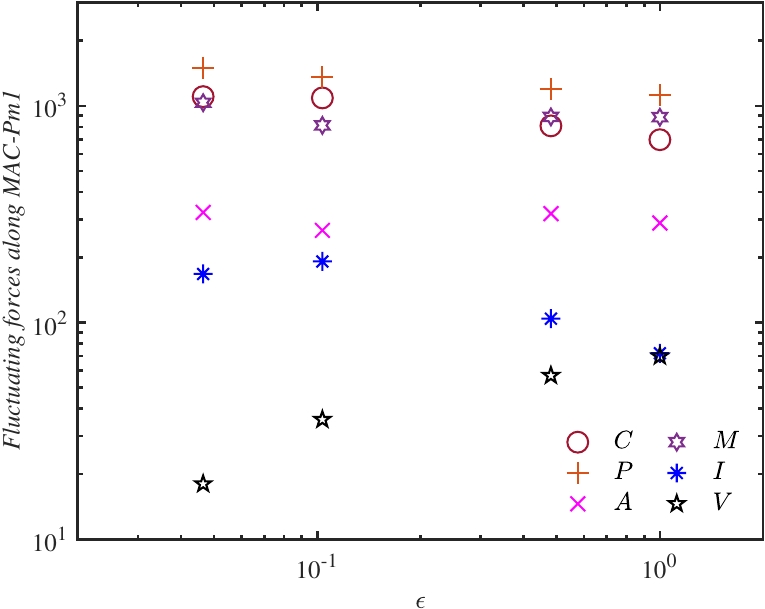} \\
    \includegraphics[width=0.32\linewidth]{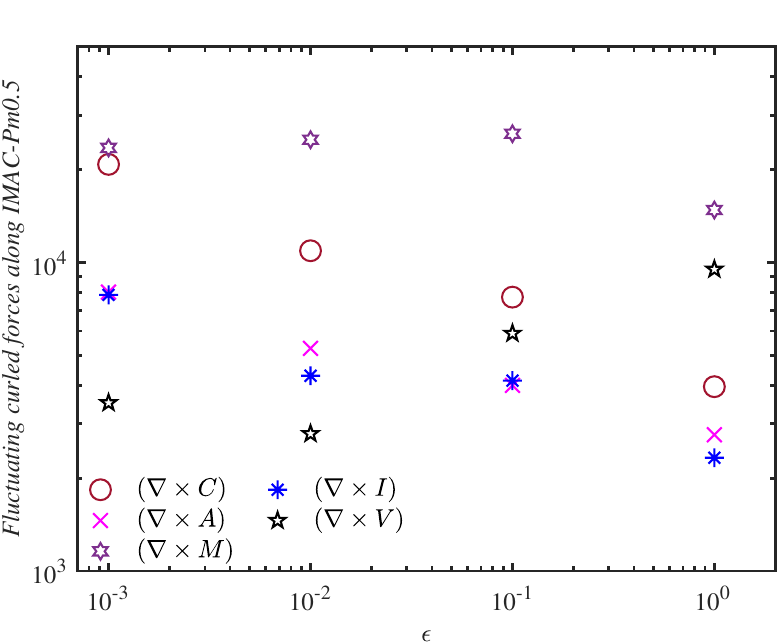}%
    \includegraphics[width=0.327\linewidth]{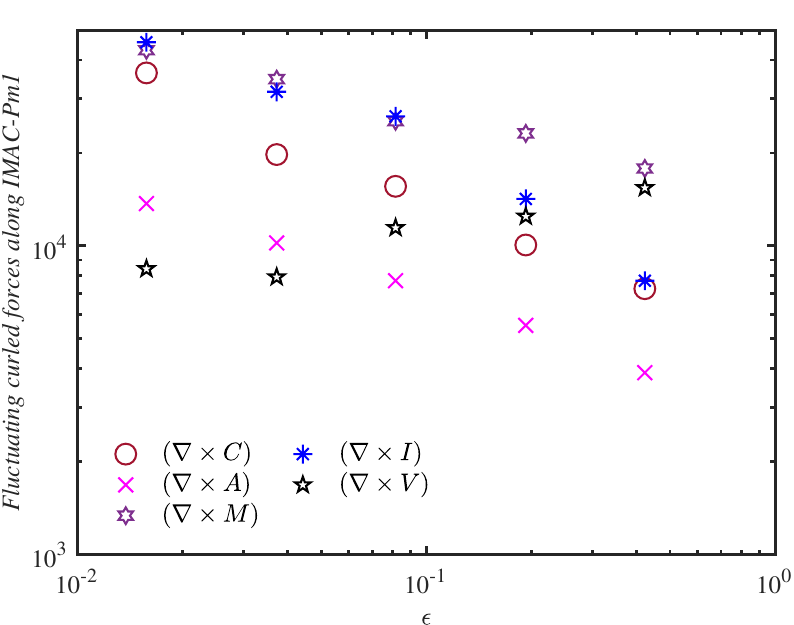}
    \includegraphics[width=0.32\linewidth]{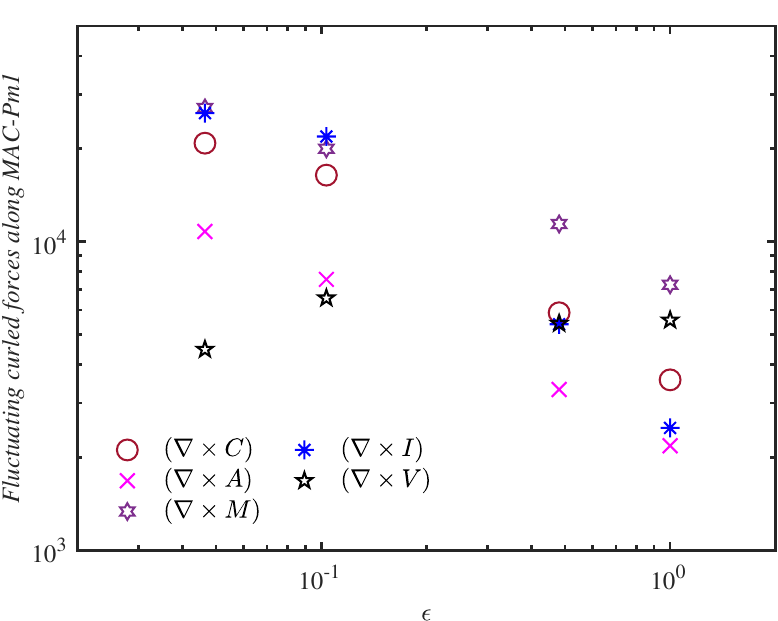} \\
    \caption{Volume-integrated forces (top row) and curled forces (bottom row) along the IMAC-Pm0.5 (left), IMAC-Pm1 (middle) and MAC-Pm1 (right) paths. The forces are summed over all spherical harmonic degrees.}
    \label{fig:force_curl_int_avg}
\end{figure*}

Figure~\ref{fig:force_curl_int_avg} (top row) shows the volume-integrated forces along the three paths. The leading order balance is between the pressure gradient, the Coriolis effect and the Lorentz force. The apparent difference compared to the force spectra in Figure~\ref{fig:force_spectra_imacpm0.5} arises because the Lorentz force has a fairly white spectrum, while the buoyancy term falls off at high $\ldeg$. Therefore, the QG(A) balance appears only at the large scales and turns into a magnetostrophic balance upon integration, as has been noted previously \citep{schwaiger_force_2019}. This is confirmed by evaluating the integrated forces at the scale $\lupol$ (Supplementary Figure~\ref{fig:force_int_lpol}), which reveals the leading QG balance and a greater separation of the terms at large scales, with inertia being more subdominant than in the volume-integrated representation.

In Figure~\ref{fig:force_curl_int_avg}, the integrated magnetostrophic balance arises at approximately the same amplitude along all three paths, and it is therefore insensitive to $\Pm$ and the assumed dynamical balance, at least at these modest values of $\epsilon$. The buoyancy term is subdominant compared to the magnetostrophic terms, though it is comparable to these terms at the scale $\lupol$ (Figure~\ref{fig:force_spectra_imacpm0.5}). Viscosity is strongly subdominant on all paths, as expected. The main difference in integrated balances between paths arises from the effect of $\Pm$ on the size of the inertial term. In the IMAC-Pm0.5 path, inertia remains below the leading order balance by at least an order of magnitude and decreases steadily with $\epsilon$, while for the two Pm1 paths, it is within an order of magnitude of the leading order balance and is relatively constant as $\epsilon$ decreases. Therefore, as expected, the Pm1 paths do a better job of preserving the IMAC balance. The integrated curled forces (Figure~\ref{fig:force_curl_int_avg}, bottom row) demonstrate a comparable level of invariance along the paths and furthermore show the IMAC balance suggested by the forces, though with a larger contribution from viscosity. The appearance of the Lorentz force at leading order in the curled balance suggests that the magnetic pressure, which is removed on taking the curl, is not the dominant contribution to the Lorentz force in these simulations.

\begin{figure*}
    \centering
    \includegraphics[width=0.335\linewidth]{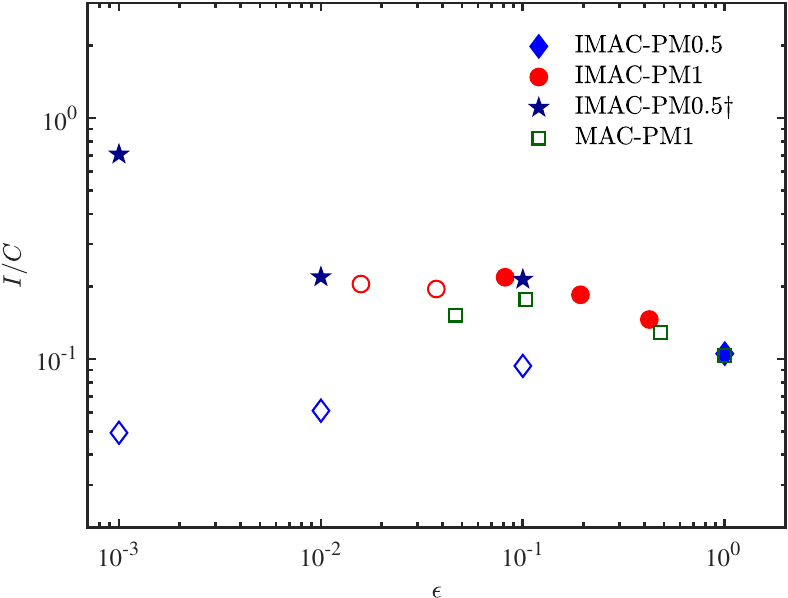}
    \includegraphics[width=0.325\linewidth]{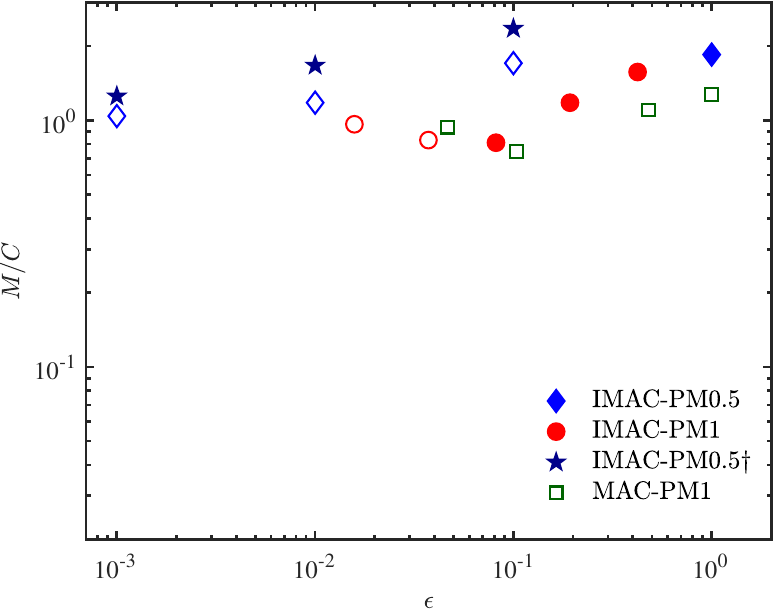}
    \includegraphics[width=0.325\linewidth]{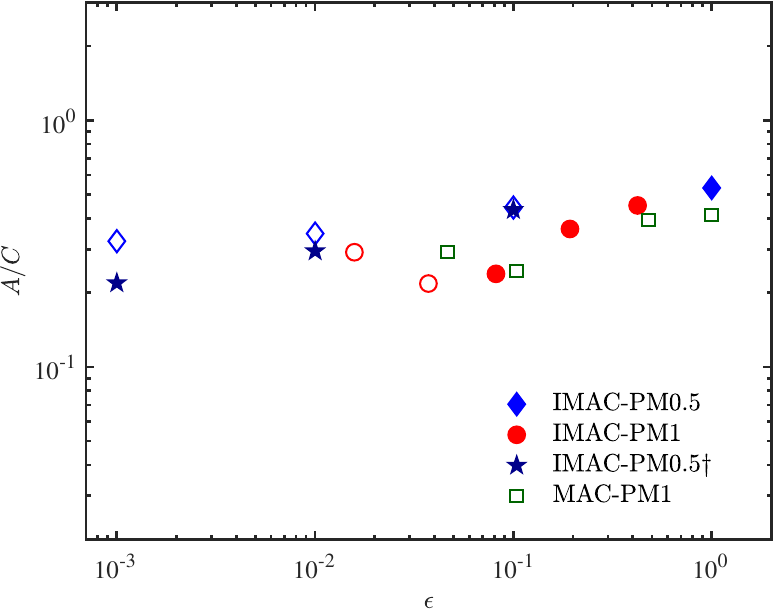}\\
    \includegraphics[width=0.33\linewidth]{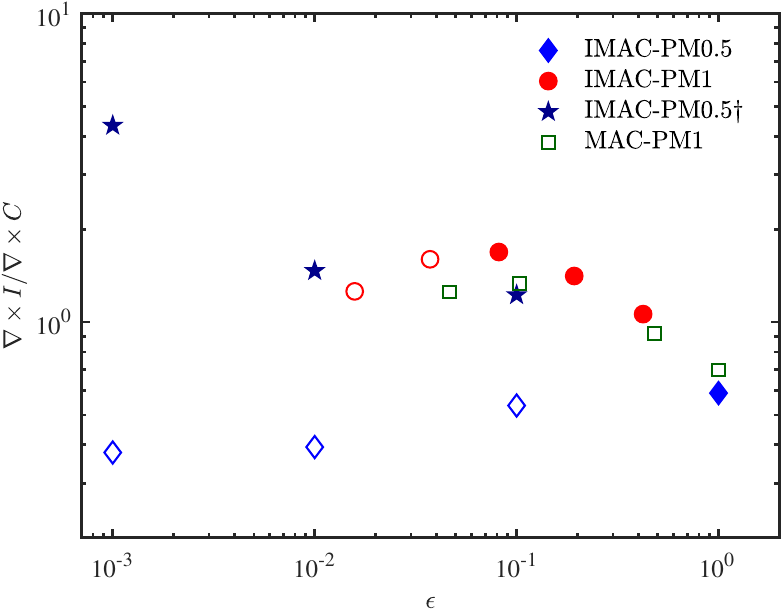}
    \includegraphics[width=0.32\linewidth]{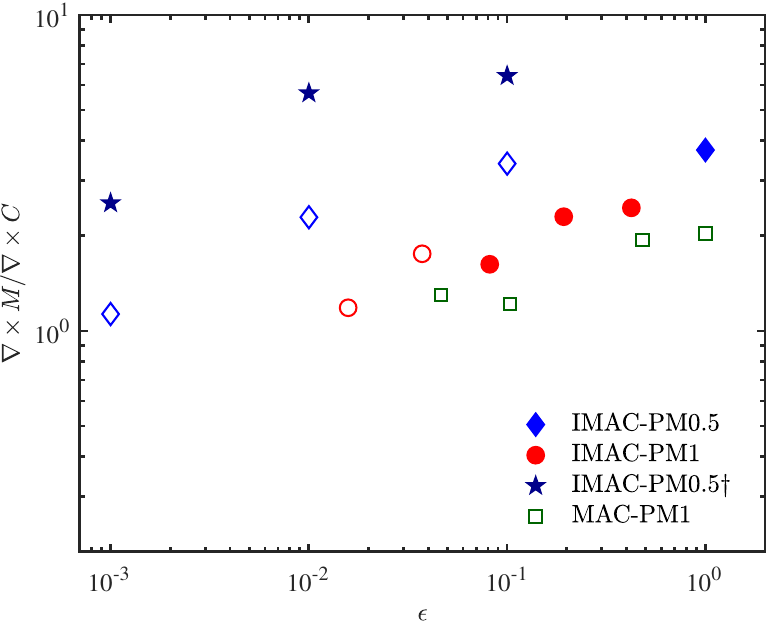}
    \includegraphics[width=0.32\linewidth]{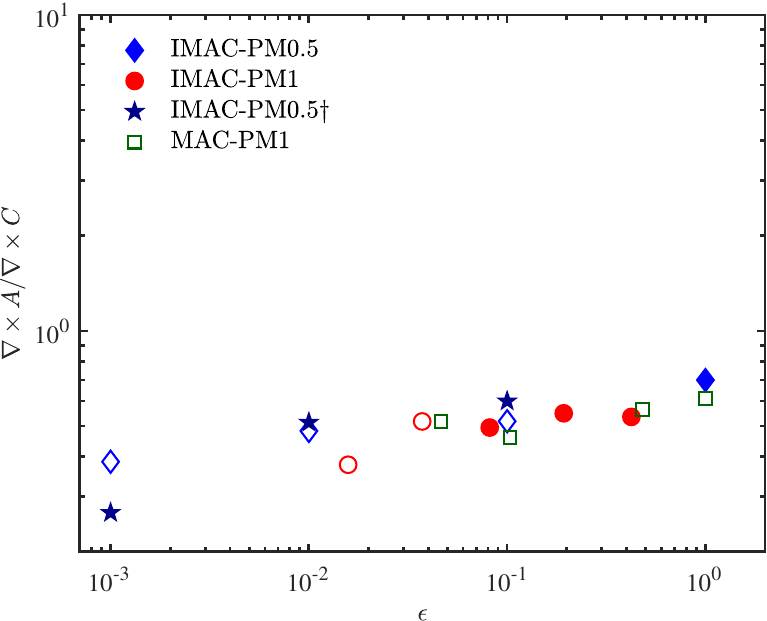}
    \caption{Ratio of rms forces (top row) and rms curled forces (bottom row) integrated over all spherical harmonic degree as a function of $\epsilon$. Ratios are Inertia/Coriolis (left), Lorentz/Coriolis (middle) and Buoyancy/Coriolis (right). Symbol shape indicates path as described in the text. Filled symbols indicate reversing multipolar dynamos, while open symbols are dipolar.}
    \label{fig:fr_vs_eps}
\end{figure*}

Figure~\ref{fig:fr_vs_eps} shows volume-integrated force ratios of inertia/Coriolis ($I/C$), Lorentz/Coriolis ($M/C$), and buoyancy/Coriolis ($A/C$) terms and the corresponding ratios of curled terms. In these plots, solid symbols show reversing/multipolar simulations, and empty symbols show dipolar dynamos. Here, it is clear that on the IMAC-Pm0.5 path, the inertial term gradually falls below the Coriolis and Lorentz terms as $\epsilon$ decreases, while inertia and buoyancy remain of comparable amplitude. On the other two paths, the approximate IMAC balance varies little with $\epsilon$. On these paths, the trends in the various ratios generally change at the point where the dynamo state changes from multipole-dominated to dipole-dominated. Over the range of $\epsilon$ covered by the simulations, the three ratios each vary only by a factor of 2-3, which is consistent with the path assumptions. However, the expected behaviour at lower $\epsilon$ is unclear and is not obviously consistent between force and curled force ratios. For example, on the IMAC-Pm0.5 path, the force ratios suggest a trend towards a MAC balance as $\epsilon$ decreases, while the curled force ratios suggest that the Lorentz term is falling faster than inertial and buoyancy terms compared to the Coriolis effect. The distinction is important because the forces convey the dynamical balance, but the scaling predictions of the path theory are based on the vorticity equation.  

\begin{figure*}
    \centering
    \includegraphics[width=0.32\linewidth]{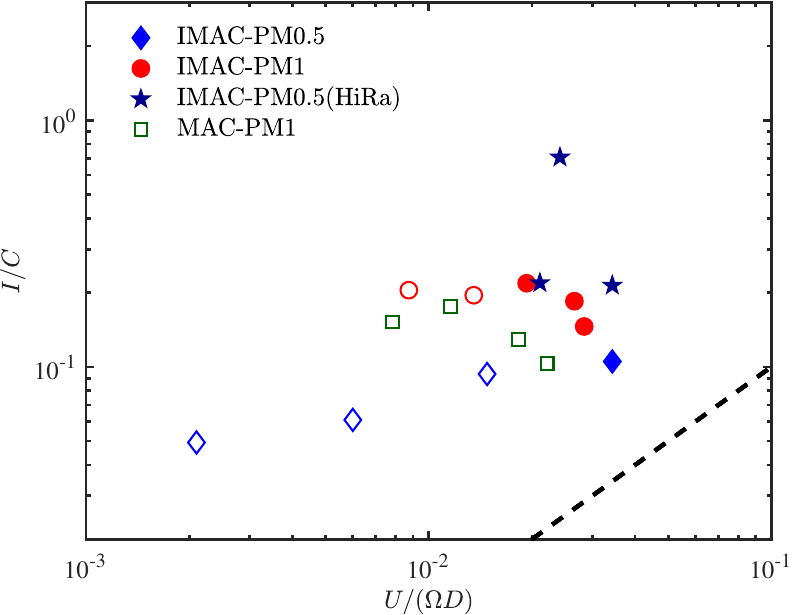}
    \includegraphics[width=0.32\linewidth]{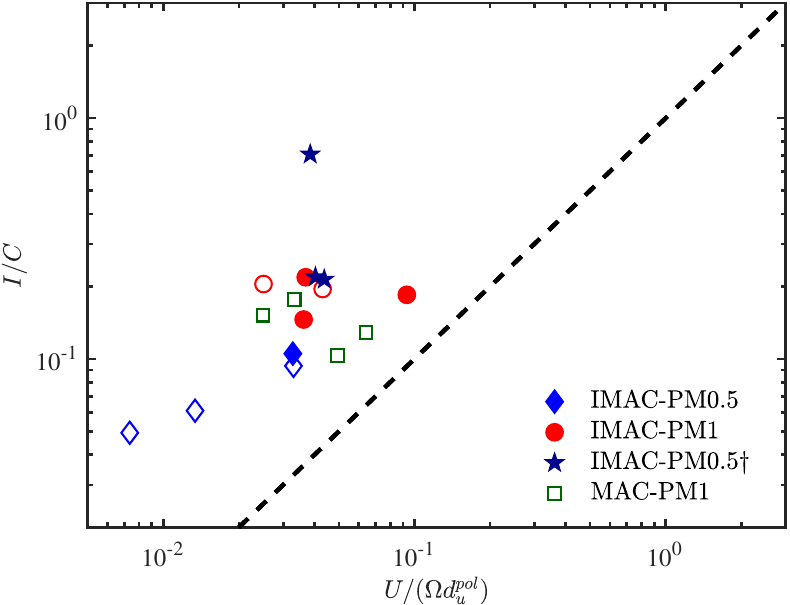}
    \includegraphics[width=0.32\linewidth]{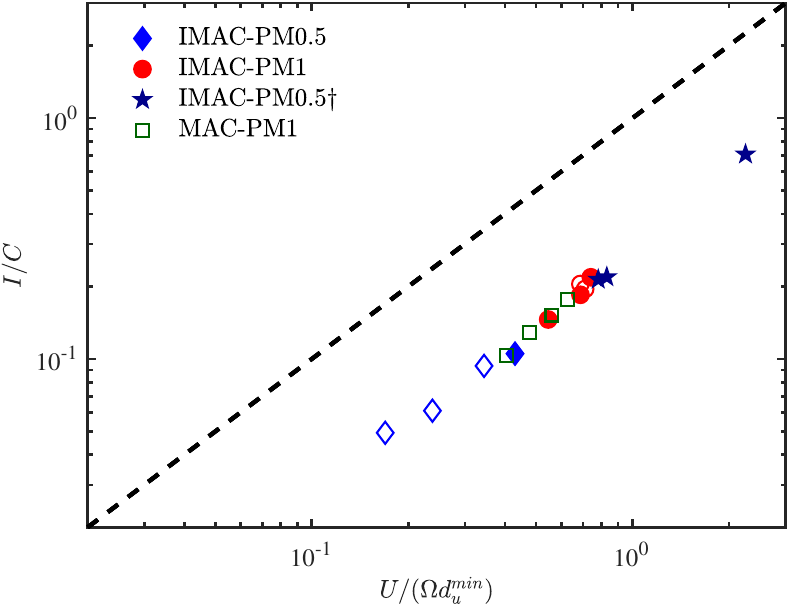}\\  
    \includegraphics[width=0.32\linewidth]{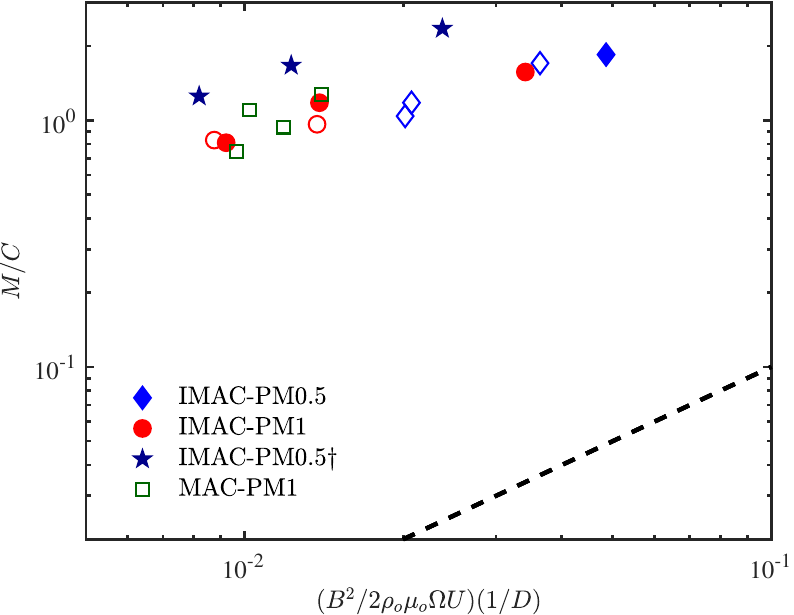}
    \includegraphics[width=0.32\linewidth]{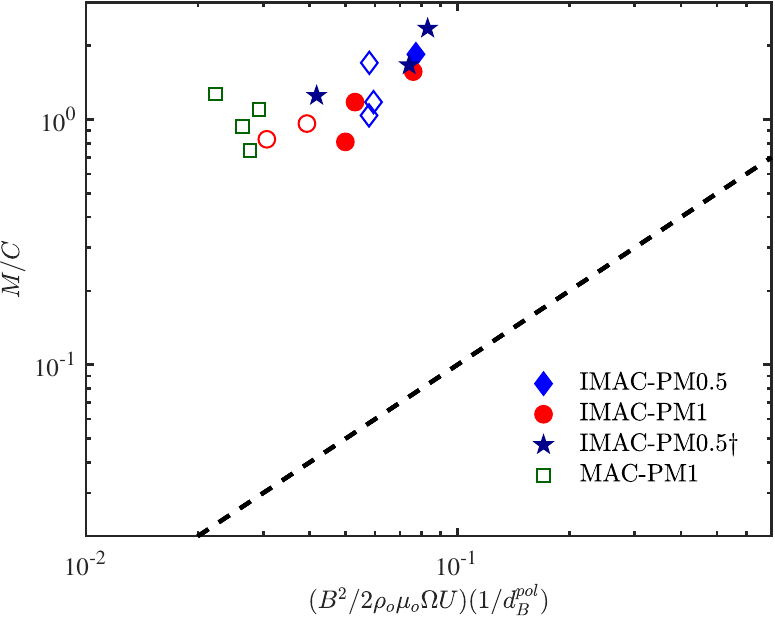}
    \includegraphics[width=0.32\linewidth]{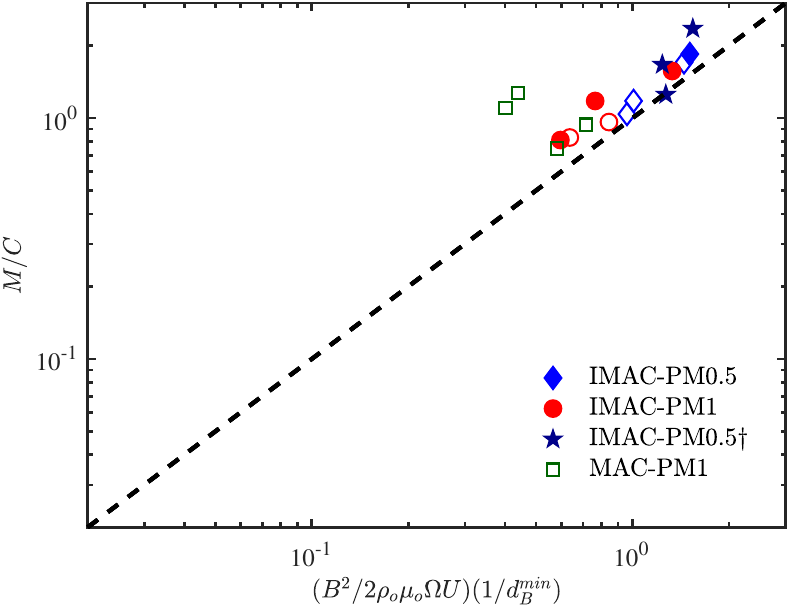}
    \caption{(Top row) The ratio of Inertia to Coriolis force as a function of Rossby number $\Ro$ based on various length scales $D$, $\dupol$ and $\dumin$ from left to right, respectively. (Bottom row) The ratio of Lorentz to Coriolis force vs $\elsasser$ based on various length scales $D$, $\dbpol$ and $\dbmin$ from left to right, respectively. Filled symbols indicate reversing multipolar dynamos, while open symbols are dipolar. The dashed black 1:1 line indicates the perfect representation of the force ratio in the $y$-axis by the corresponding estimate in the $x$-axis.}
    \label{fig:fr_vs_Ro_Els}
\end{figure*}

We end this section by looking for dimensionless numbers that provide accurate measures of the force and curled force balances that have been 
calculated directly in Figure~\ref{fig:fr_vs_eps}. The force ratios can be written in terms of generic flow and field lengthscales, $\du$ and $\dB$ respectively, as 
\begin{linenomath*}
\begin{align}
    \frac{I}{C} & \sim \frac{U}{\rotation \du} = \Ro \frac{\shellthick}{\du}, 
     \\
    \frac{M}{C} & \sim \frac{B^2}{2\rho \mu_0 \dB \rotation U } = \frac{\elsasser}{2\Rm} \frac{\shellthick}{\dB}. 
\end{align}
\end{linenomath*}
In Figure~\ref{fig:fr_vs_Ro_Els} we show the estimates of $I/C$ and $M/C$ using three different estimates for the scale $\du$ ---$\shellthick$, $\dupol$ and $\dumin$--- and the analogous estimates for the scale $\dB$. For the ratio $I/C$, the estimate $I/C \sim \Ro$ does not provide a good fit to the simulation data: $\Ro$ consistently underestimates $I/C$ and does not follow the same trend. The estimate $I/C \sim \Ro \shellthick/\dumin$ follows the trend of $I/C$ with a constant offset that overestimates the simulation data, while the estimate $I/C \sim \Ro \shellthick/\dupol$ underestimates the simulation data, does not follow its general trend, and shows greater variance. This may suggest that a scale that is intermediate between $\dumin$ and $\dupol$ would provide a better estimate of $I/C$, or that the discrepancy could be accounted for by a constant prefactor. We also calculated the lengthscale $\duchrs$ (not shown) based on the weighted poloidal kinetic energy spectrum \citep[][; see equation~(\ref{eq:Rol})]{christensen_convection-driven_2006}, which plots essentially on the 1:1 line for all simulations, and provides excellent agreement with $I/C$. \citet{oruba_predictive_2014} have previously noted that $\dumin$ and $\duchrs$ follow similar scalings. 

For the ratio $M/C$, the estimates $M/C \sim \elsasser$ and $M/C \sim \elsasser/\Rm$ underestimate the simulation data, does not follow the same trend, and are quite scattered. The assumption $\elsasser/\Rm (\shellthick/\dupol)$ underestimates $M/C$, which probably reflects the fact that the spectrum of the Lorentz force peaks at high spherical harmonic degree. Indeed, as found by \citet{dormy2018three}, the assumption $\elsasser/\Rm(\shellthick/\dumin)$, provides an excellent match to the simulated data.

\begin{figure*}
    \centering
    \includegraphics[height=4.5cm]{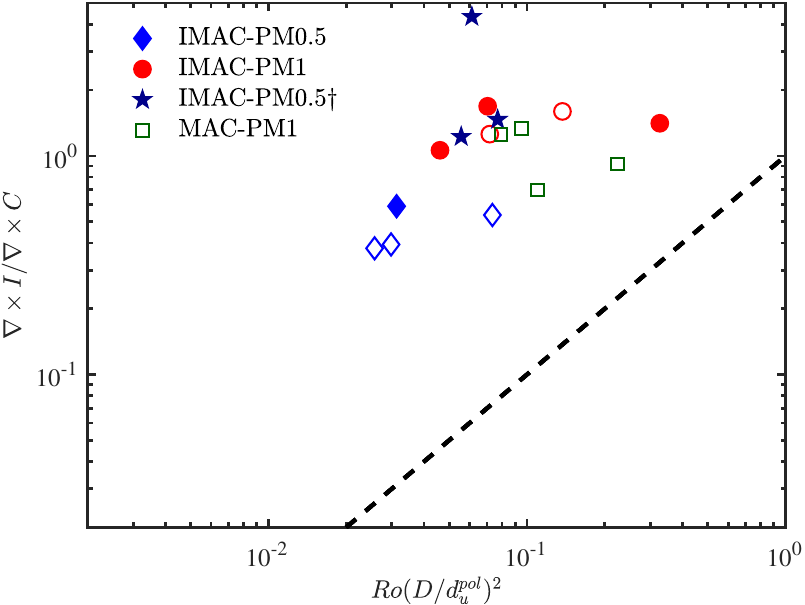}
    \includegraphics[height=4.5cm]{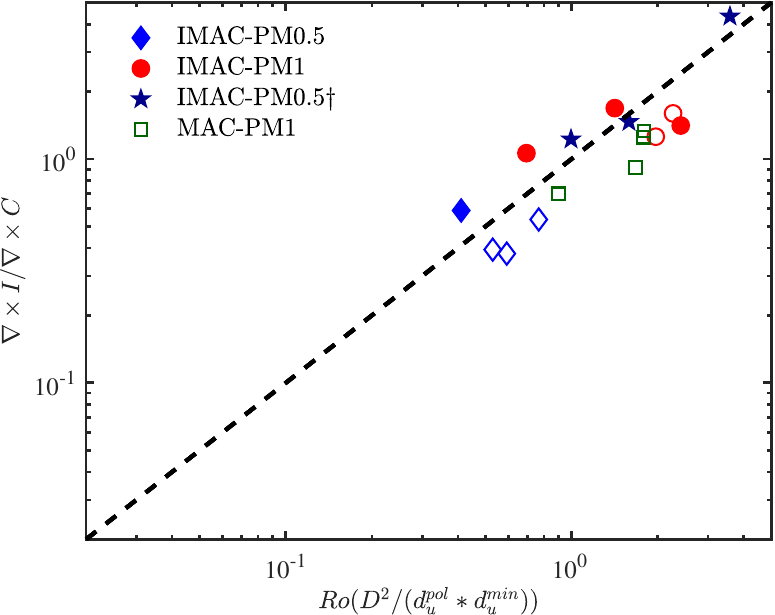}\\
    \includegraphics[height=4.5cm]{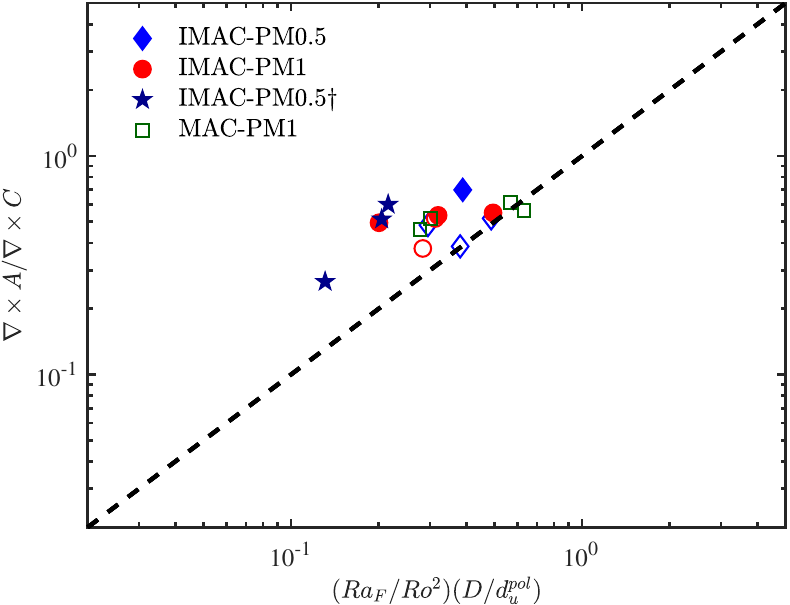}
    \includegraphics[height=4.5cm]{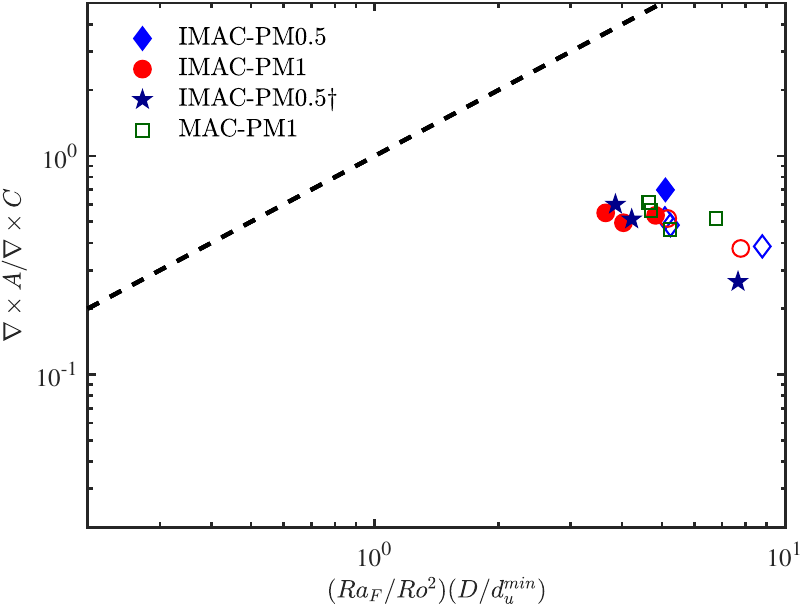} \\
    \includegraphics[height=4.5cm]{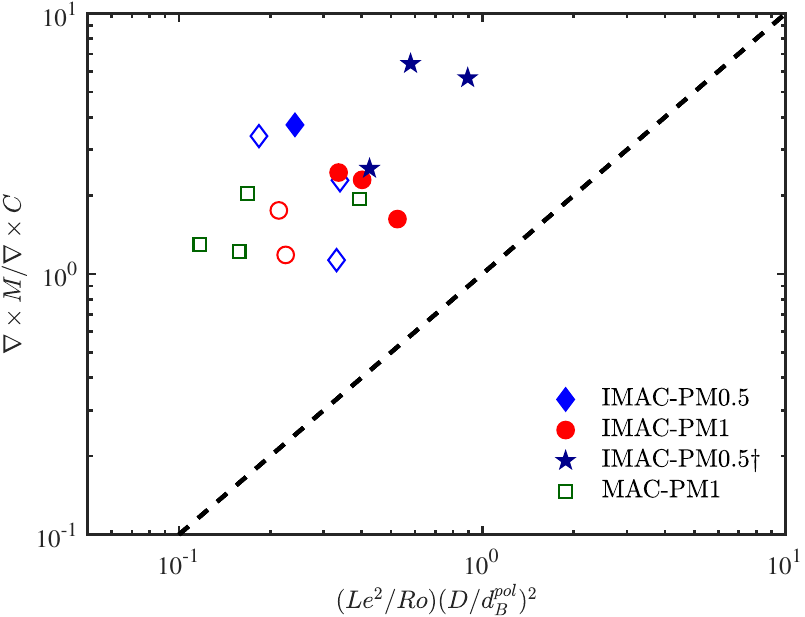}
    \includegraphics[height=4.5cm]{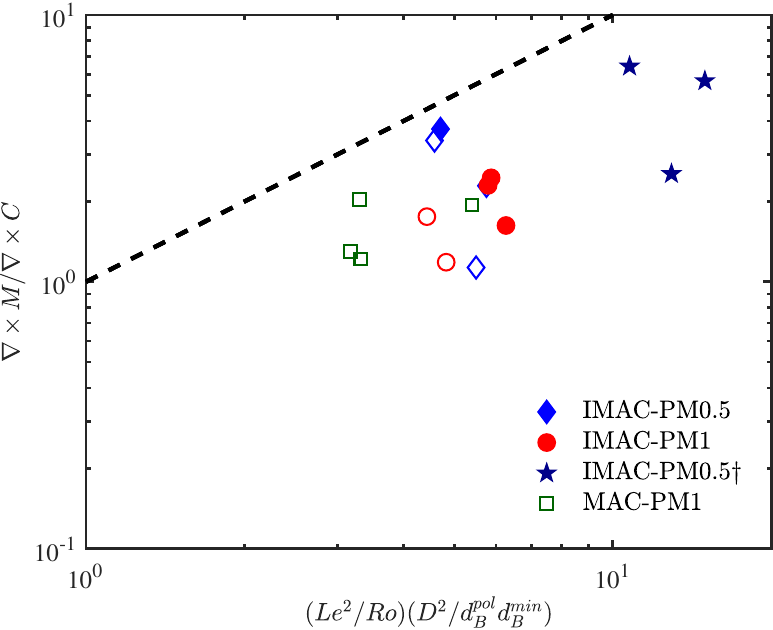}  
    \caption{(Top row) Ratio of curled Inertia to Coriolis force as a function of (left) and (right) (Middle row) Ratio of curled Buoyancy to Coriolis force as a function of (left) and (right). (Bottom row) Ratio of curled Lorentz to Coriolis force as a function of (left) and (right). Filled symbols indicate reversing multipolar dynamos, while open symbols are dipolar. Paths and the dashed line are defined in the same way as Figure \ref{fig:fr_vs_Ro_Els}.}
    \label{fig:cfr_vs_Ro_Els}
\end{figure*}

The curled force ratios are most relevant for diagnosing dynamical balances since they represent ratios of terms in the vorticity equation. In Figure~\ref{fig:cfr_vs_Ro_Els} we consider the ratios
\begin{linenomath*}
\begin{align}
    \frac{\nabla\times I}{\nabla\times C} & \sim \frac{I}{C} \frac{\shellthick}{\du}, 
     \\
    \frac{\nabla\times M}{\nabla\times C} & \sim \frac{M}{C} \frac{\shellthick}{\dB}, 
     \\
    \frac{\nabla\times A}{\nabla\times C} & \sim \frac{A}{C} \frac{\shellthick}{\du} = 
    \frac{\Raf}{Ro^2} \frac{D}{\du}.
\end{align}
\end{linenomath*}
For the inertia/Coriolis vorticity ratio, the assumption $\left(\nabla\times I\right) /\left( \nabla\times C\right) \sim Ro D/(\dupol)^2$ consistently underestimates the data, in some cases by up to an order of magnitude. By contrast, the assumption $\left(\nabla\times I\right) / \left(\nabla\times C\right) \sim Ro D/(\dupol\dumin)$ shows excellent agreement with both the amplitude and trend of the data. For the buoyancy/Coriolis vorticity ratio, the estimate $\left(\nabla\times A\right) /\left( \nabla\times C\right) \sim (\Raf/\Ro^2) D/\dupol$ provides a reasonable fit to the data, and agrees much better than the estimate
$\left(\nabla\times A\right) / \left(\nabla\times C\right) \sim (\Raf/\Ro^2 D/\dumin$. For the Lorentz/Coriolis vorticity ratio the estimate $(\elsasser/\Rm) D^2/(\dupol)^2$ gives a marginally worse match to the data than the estimate $(\elsasser/\Rm) D^2/(\dupol\dumin)$. This latter estimate is also consistent with the estimate $\nabla \times \sim 1/\dupol$ for all terms in the vorticity equation. 

Figure~\ref{fig:cfr_vs_Ro_Els} suggest that the vorticity scaling relation appropriate to describe our simulations is
\begin{linenomath*}
\begin{equation}
    \frac{\charvel^2}{\dupol \dumin} 
    \sim
    \frac{\rotation\charvel}{D} 
    \sim 
    \frac{\thermexpan \grav_o \beta \thermdiff }{U \dupol D^2}
    \sim 
    \frac{\charmag^2}{\meandensity\magperm\dbpol \dbmin} ,
    \label{eq:IMAC-balance-estimate}
\end{equation}
\end{linenomath*}
The scaling relations for the various lengthscales (Section~\ref{sec:imac-theory}) suggest that $\dupol \sim \dumin$ and $\dbpol \sim \dbmin$, at least at moderate values of $\epsilon$, in which case equation~(\ref{eq:IMAC-balance-estimate}) essentially becomes equation~(\ref{eq:IMAC-balance}), which has been used in several previous studies. Nevertheless, we expect that the scaling behaviour of the simulations will deviate from that predicted from equation~(\ref{eq:IMAC-balance}). 

Overall the large-scale dynamics of our simulations are broadly consistent with theoretical predictions when viewed through the lens of scale-dependent force spectra. The integrated balances show a somewhat different picture, with a leading order magnetostrophic balance, while the integrated vorticity balance does not suggest obvious asymptotic behaviour.

\subsection{Scaling Predictions along the Paths} 
\label{sec:scalings}

\begin{figure*}
\centering
    \includegraphics[width=0.33\linewidth]{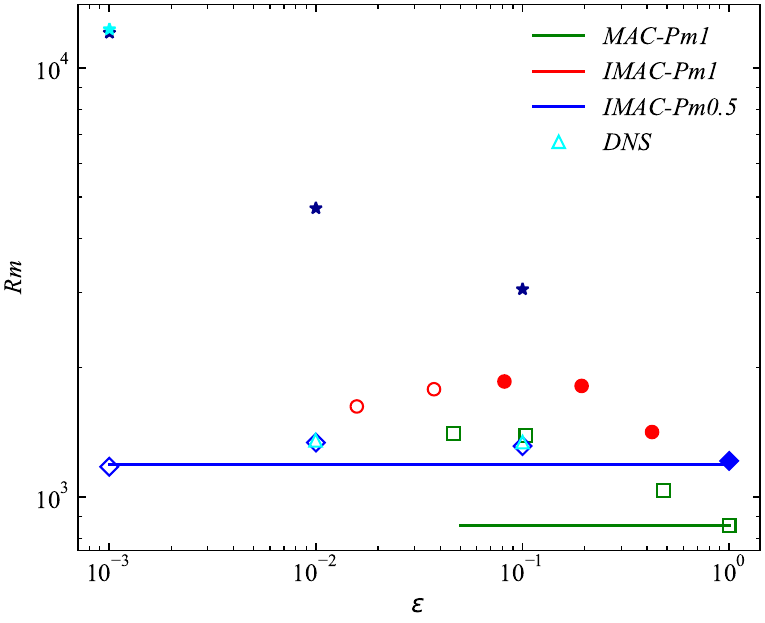}
    \includegraphics[width=0.33\linewidth]{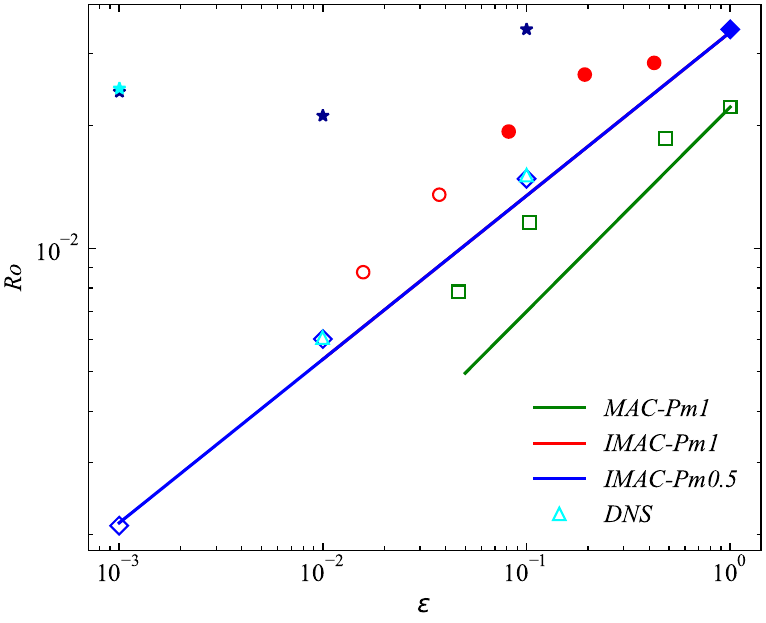}
\caption{Magnetic Reynolds number $\Rm$ (left) and Rossby number $\Ro$ (right)}
\label{fig:Rm_scaling}
\end{figure*}

\begin{figure*}
    \centering
    \includegraphics[height=4cm]{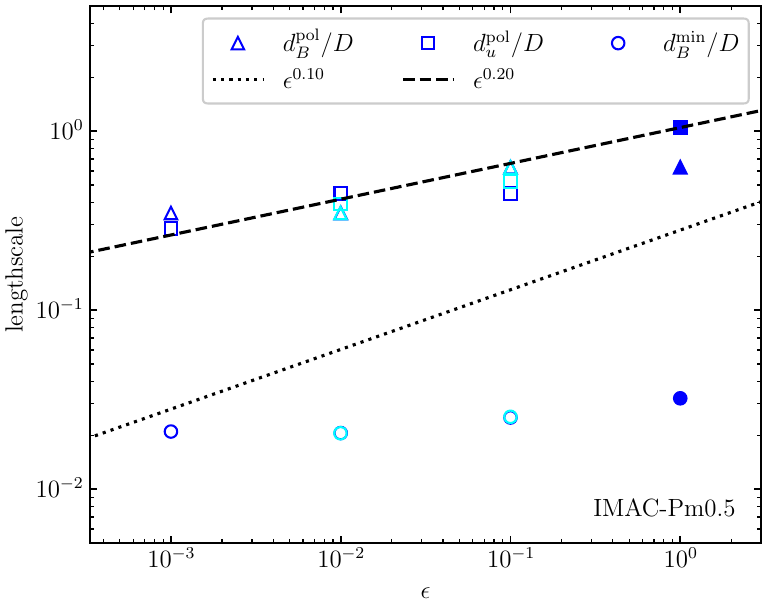}%
    \includegraphics[height=4cm]{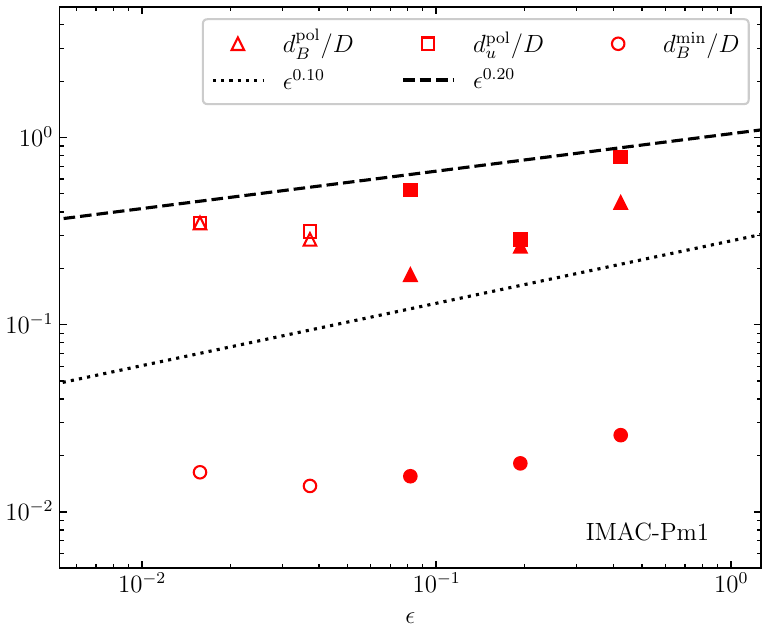}%
    \includegraphics[height=4cm]{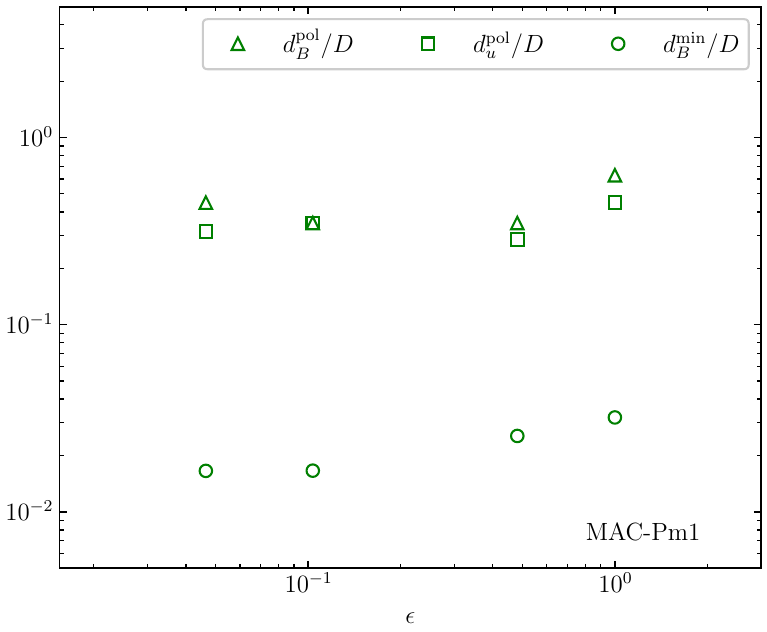}
    \caption{Lengthscales of magnetic field, velocity, and magnetic dissipation, plotted for IMAC Pm0.5 (left), IMAC Pm1 (middle), and MAC Pm1 (right). Light blue symbols in left plot indicate DNS values. Dotted and dashed lines show indicative scalings for $\epsilon$}
    \label{fig:lengthscales}
\end{figure*}

\begin{figure*}
    \centering
    \includegraphics[height=4.1cm]{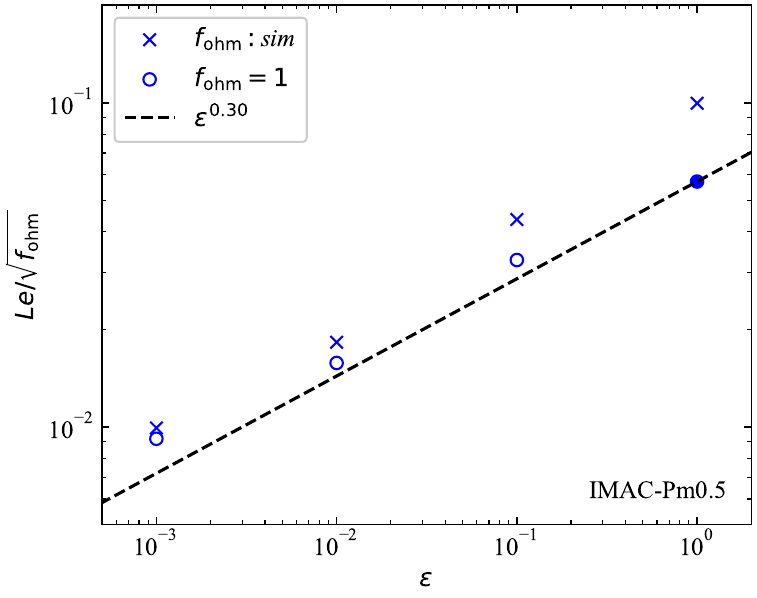}
    \includegraphics[height=4.1cm]{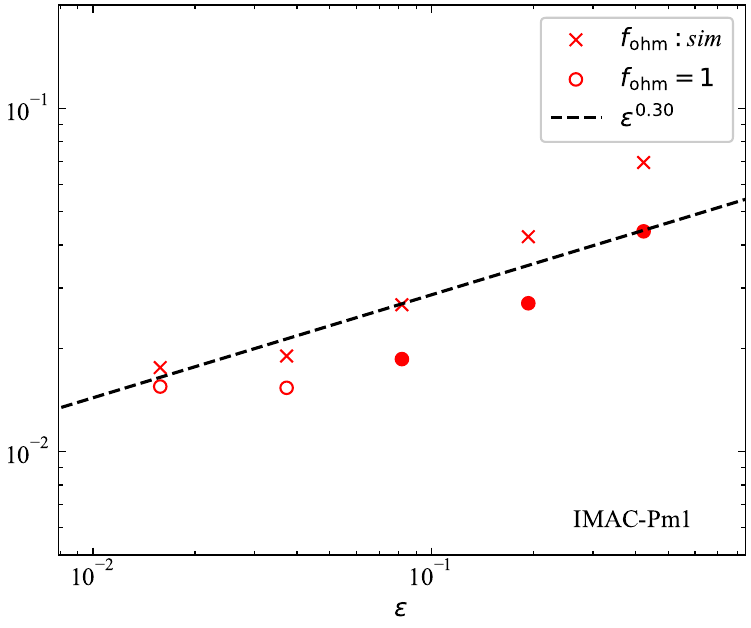}
    \includegraphics[height=4.1cm]{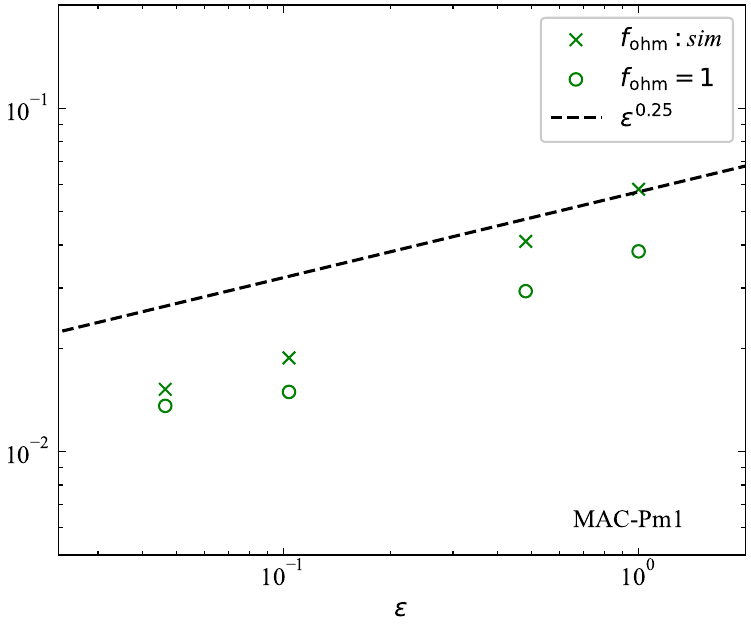}
    \caption{Lehnert number, showing effect of normalising with $\sqrt{\fohm}$. Left plot shows IMAC\_Pm10.5, middle shows IMAC\_Pm1, and right plot shows MAC\_Pm1 rules, defined in the text.}
    \label{fig:Le_sqrt_fohm}
\end{figure*}

\begin{figure*}
    \centering
    \includegraphics[height=4.5cm]{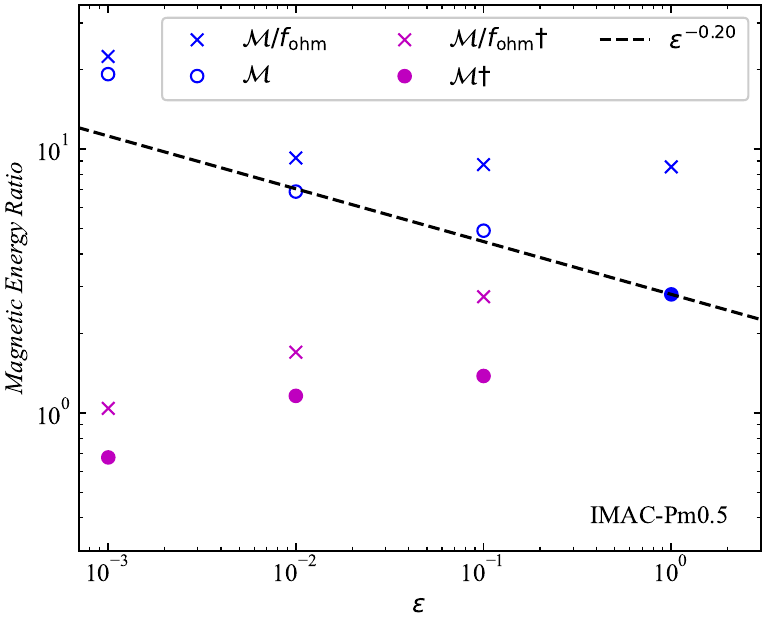}
    \includegraphics[height=4.5cm]{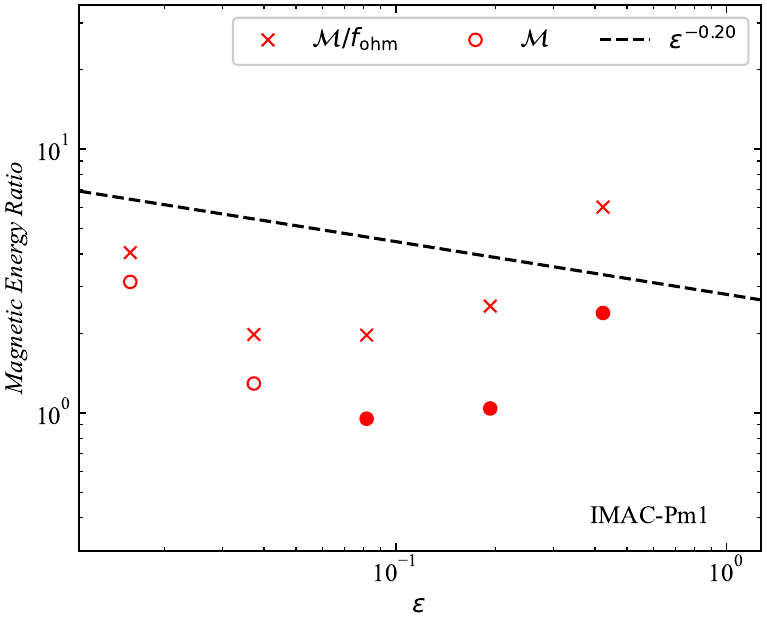} \\
    \includegraphics[height=4.5cm]{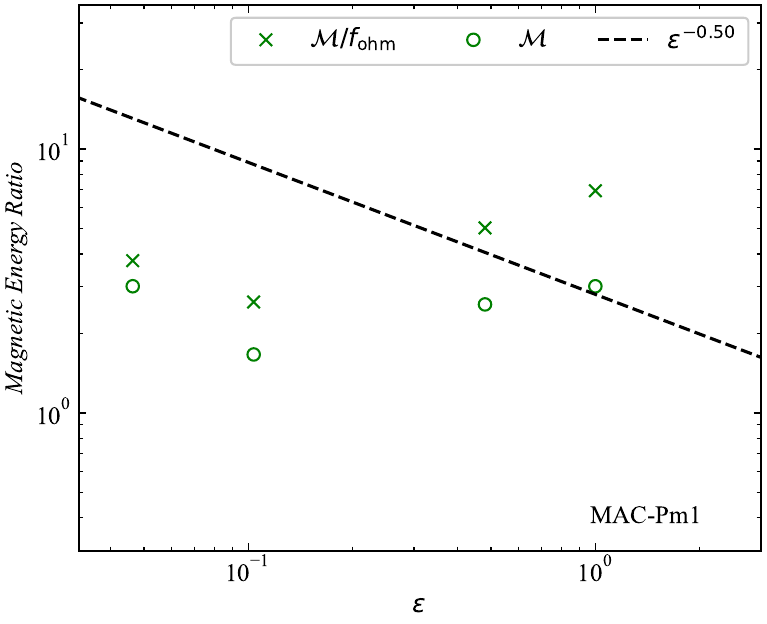} 
    \includegraphics[height=4.5cm]{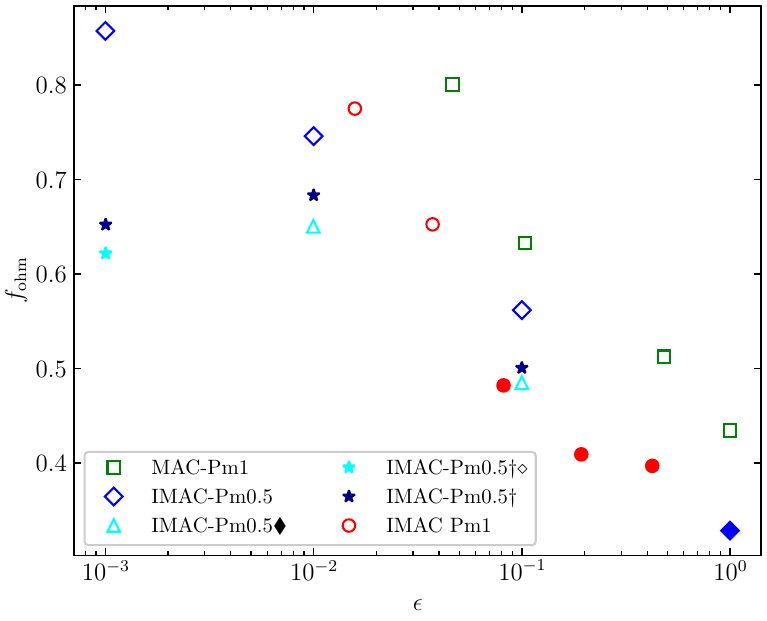}
    \caption{Magnetic to Kinetic energy ratio showing the effect of normalising with $\fohm$ for the IMAC-Pm0.5 path (top left), IMAC-Pm1 (top right) and MAC-Pm1 (bottom left). Purple points on IMAC-Pm0.5 plot show off-path high Ra simulations. Ohmic dissipation fraction (bottom right) for all paths. }
    \label{fig:M_fohm}
\end{figure*}

The scaling behaviour of the 3 paths (IMAC-Pm1, IMAC-Pm0.5 and MAC-Pm1) are summarised in Table~\ref{tab:path_predictions_eps} and the results are plotted in figures~\ref{fig:Rm_scaling}-\ref{fig:M_fohm}.  The caveat noted above about the limited range of $\epsilon$ that can be tested still applies, but we can check the expected $\Ek$-dependence over a broader range of values. 

The variation of the input parameters $\Ek$, $\Pm$ and $\Ra$ are trivially obeyed along the paths; however, the diagnostic parameters that depend on properties of the field and flow emerge from the calculations and are not bound to follow the theoretical predictions. \citet{aubert_spherical_2017} found good agreement between predicted and calculated diagnostics for their path, which encoded a MAC balance with the scaling $\Pm \sim \epsilon^{1/2}$, and so we might anticipate a similar result. However, the present paths are designed to straddle the dipole-multipole transition, and it is well known that diagnostics can vary strongly across this transition \citep[e.g.][]{tassin_geomagnetic_2021}. Simulations along our individual paths cross the dipole-multipole transition, which contributes to deviations from the predicted diagnostic scalings. 
\begin{table*}
\begin{minipage}{160mm}
\centering
\caption{Predicted scalings with epsilon compared to best fit to data for different scaling rules. Path predictions for the 4 different unidimensional parameter paths considered in this study. The starting point for IMAC paths is the simulation LEDT002, which has $\Rm_0 = 1185$, $\Ek_0 = 10^{-3}$, $\Pm_0 = 35.3$, $\lehnert_0 = 7.158 \times 10^{-2} \times \sqrt(0.33)$, $\Raf_0 = 4.8 \times 10^{-4}$, $\Mratio_0 = 5.0$. Starting point for MAC-Pm1 path is sim 15 of this work, with $\Rm_0=1417.5$, $\Ek_0 = 3\times 10^{-4}$, $\Pm_0 = 14.96$, $\lehnert = 4.39\times 10^{-2}$, $\Ra_{F,0}=2.031\times10^{-4}$, $\Mratio=2.388$. }
\label{tab:path_predictions_eps}
\begin{tabular}{lllllllll}
 & IMAC-Pm1 &  &  & IMAC-Pm0.5 &  &  & MAC-Pm1 &  \\
 \hline
Quantity & Prediction & Best fit &  & Prediction & Best fit &  & Prediction & Best fit \\
\hline
$\Pm$          & 1     & 0.9999 &  & 0.5 & 0.5 &  & 1 & 1.0001 \\
$\Ro$          & 0.4   & 0.3659 &  & 0.4 & 0.4033&  & 0.5 & 0.3311 \\
$\Ek$          & 1.4   & 1.4    &  & 0.9 & 0.9 &  & 1.5 & 1.5 \\
$\Rey$          & -1    & -1.0304  &  &-0.5 & -0.4967 &  & -1 & -1.1689 \\
$\Rm$          & 0     & -0.0305  &  & 0.0 & 0.0032 &  & 0 & -0.1688 \\
$\dupol$    & 0.2   & 0.1819 &  & 0.2 & 0.1693 &  & 0 & 0.0657 \\
$\Rol$      & 0.2   & 0.1877 &  & 0.2 & 0.2340 &  & 0.5 & 0.2653 \\
\hline
$\dbpol$    & $0.1$ & 0.0498  & &$0.1$  & 0.1021      &  & 0.0  & 0.0852\\
$\dbmin$    & $0$   & 0.1437  & &$0$    & 0.0644      &  & 0.0  & 0.2263 \\
$\lehnert$  & $0.3$ & 0.3213  & & $0.3$   & 0.2698      &  & 0.25 & 0.3577 \\
$\elsasser$ & $0.2$ & 0.2442  & & $0.2$   & 0.1421      &  & 0.0  & 0.2182 \\
$\Mratio$   & $-0.2$& -0.0958 & & $-0.2$  &-0.2651      &  & -0.5 & 0.0546 \\
\hline
\end{tabular}
\end{minipage}
\end{table*}

Figure~\ref{fig:Rm_scaling} shows that all 3 paths display a small variation of $\Rm$ with $\epsilon$. The transition from dipolar to multipolar solutions corresponds to an overall reduction in $\Rm$ as a greater fraction of the total energy is partitioned into the magnetic field. On the IMAC-Pm0.5 path, which provides the best coverage in $\epsilon$, $\Rm$ scales empirically as $\epsilon^{0.003}$. This weak dependence on $\epsilon$ arises in part because most simulations along the IMAC-Pm0.5 path are in the dipole-dominated regime. For the other two paths there is a clear change in behaviour at the dipole-multipole transition, which, combined with the relatively narrow range of $\epsilon$ explored, means that we do not place too much emphasis on the overall fit to the data. Instead, we note that $\Rm$ varies little along the sampled regions of these paths (less than a factor of 2 for the MAC-Pm1 path and a factor of 1.3 for the IMAC-Pm1 path) and is consistent with the theoretical prediction to the extent that the comparison can be made. Figure~\ref{fig:Rm_scaling} shows that these points also apply to the variation of $\Ro$ along the IMAC-Pm0.5 path. However, for the other two paths there is substantial curvature in the calculated values of $\Ro$. This, combined with the narrow range of $\epsilon$, shows that the fitted exponents along the Pm1 paths should be treated with some caution.  

Figure~\ref{fig:lengthscales} shows the characteristic dimensionless lengthscales of the flow and field, denoted here simply as $\dupol$ and $\dbpol$, and the magnetic dissipation scale (denoted as $\dbmin$) for each path. The general agreement between calculated and theoretical values for $\dupol$ and $\dbpol$ is quite good along all paths. For the Pm1 paths $\dupol$ and $\dbpol$ do not vary monotonically and so the fitted exponents only represent the general trend over a very limited range of $\epsilon$. Discrepancies between calculated and predicted values of $\dbmin$ are larger, particularly for the Pm1 paths, though $\dbmin$ varies by less than a factor of 2 and might be trending toward a more constant value at the lowest $\epsilon$ considered. The relative variations of flow and field lengthscales are also broadly consistent with the theory: compared to $\dbpol$, $\dupol$ exhibits greater variations along the IMAC paths and comparable variation along the MAC path. 

Figure~\ref{fig:Le_sqrt_fohm} shows the scaling of $\lehnert$ for the 3 paths. The IMAC-Pm0.5 path again provides the best match to the theoretical prediction, though at the lowest values of $\epsilon$ the trend does appear to start flattening out towards a scaling that is closer to the $\epsilon^{1/4}$ predicted by the MAC theory. This is explained by the analysis of the dynamical balances in section~\ref{sec:forces}, which shows a gradual transiton towards a MAC balance along the IMAC-Pm0.5 path. It is also interesting that the IMAC-Pm0.5 simulations more closely follow the theoretical prediction when the $\fohm$ correction is omitted from the fitting rather than when it is included. The Pm1 paths display $\lehnert$ scalings that are ostensibly similar to the theoretical predictions; however, they also show distinct trends at high and low $\epsilon$ that are not captured by the simple linear fit, regardless of the inclusion of the $\fohm$ factor. This behaviour is slightly surprising given that the analysis of dynamical balances in section~\ref{sec:forces} indicates that the secondary IMAC balance is quite robust along these paths, though that analysis also shows that the ratio of terms in the vorticity balance used to derive the paths are not entirely borne out by the simulations. 

Figure~\ref{fig:M_fohm} shows the scaling of the magnetic/kinetic energy ratio $\Mratio$ for the three paths. The IMAC-Pm0.5 path is relatively close to the theoretical prediction when the $\fohm$ factor is omitted. For the Pm1 paths $\Mratio$ is strongly curved and does not follow the predicted scaling. Also shown is the ohmic dissipation fraction $\fohm$. In all cases $\fohm$ increases with $\epsilon$, reaching values of $0.7-0.8$, consistent with the assumptions used to derive the path scaling laws and expectations for Earth's core. 

Our results indicate that the effect of inertia starts to diminish as the different paths are traversed to lower $\epsilon$ (Figure~\ref{fig:force_curl_int_avg}). We have also found that $\Mratio$ and $\fohm$ increase along all paths. These are all properties of the MAC balance, which seems to emerge naturally despite the IMAC paths being designed to preserve a different large-scale balance. This lends support to the results of \citet{yadav_approaching_2016} and \citet{schwaiger_force_2019}, who argued that the MAC balance prevails over a wide range of parameter space at low $\Ek$ and $\Pm$. 

\subsection{Ekman number dependence}
\label{sec:ekman}

\begin{figure*}
    \centering
    \includegraphics[height=5cm]{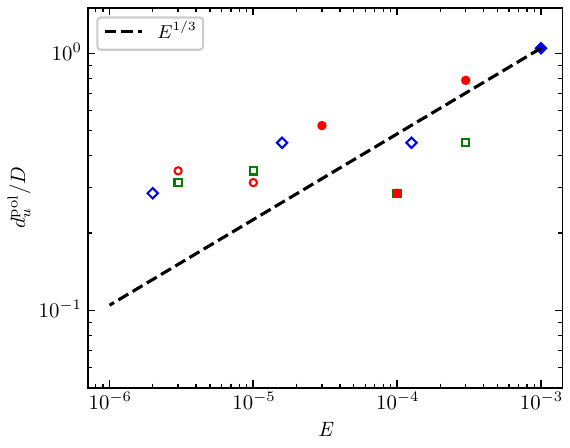}
    \includegraphics[height=5cm]{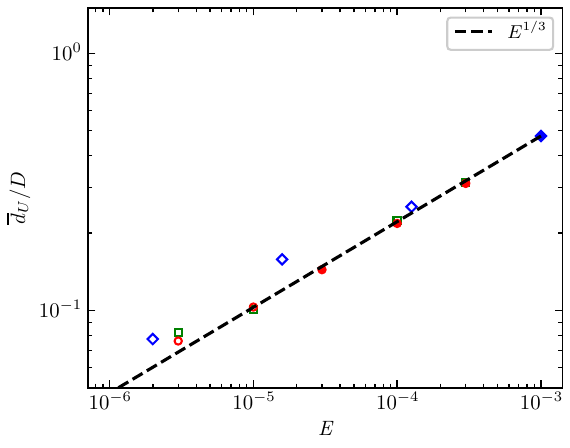} \\
    \includegraphics[height=5cm]{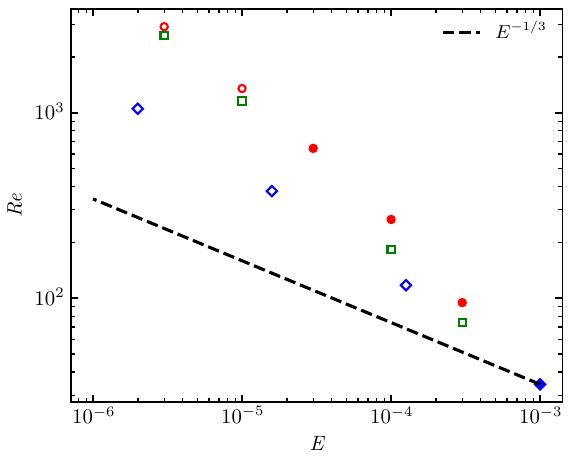}
    \includegraphics[height=5cm]{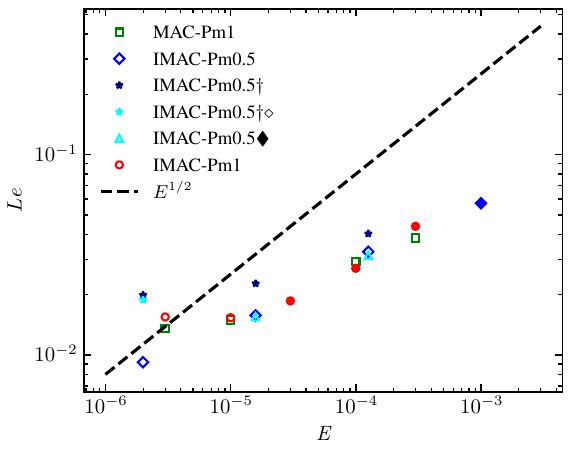}
    \caption{Comparison of characteristic flow lengthscales calculated as peak of poloidal flow $\dupol$ (top left), and weighted kinetic energy spectrum $\duchrs$ \citep{christensen_convection-driven_2006}  (top right) against Ekman number $\Ek$. Bottom panel show Reynolds $\Rey$ (bottom left), and Lehnert $\lehnert$ (bottom right) numbers.}
    \label{fig:Ekman_2}
\end{figure*}

In this section we briefly consider the $\Ek$-dependence of our path simulations. Our simulations cover a greater range of $\Ek$ than $\epsilon$ and hence any dependence on $\Ek$ should be more robustly resolved. Furthermore, some studies have suggested that the dynamics of strongly forced convection-driven dynamos maintain an asymptotic dependence on $\Ek$ \citep[e.g.][]{king_flow_2013, calkins_large-scale_2021}. Figure~\ref{fig:Ekman_2} shows the $\Ek$-dependence of $\dupol$, $\Rey$, and $\lehnert$. Linear theory shows, at convective onset,  that the lengthscale $\du$ scales as $\Ek^{1/3}$ \citep{chandrasekhar_hydrodynamic_1961}, while asymptotic QG theory predicts that the fluctuating velocity scales as $\Ek^{-1/3}$ and both mean and fluctuating magnetic field scale as $\Ek^{1/2}$ in our units \citep[e.g.][]{calkins_large-scale_2021}. The simulation data evidently shows an $\Ek$-dependence, which is 
not suggested by the force balance analysis but is suggested by the vorticity balance (Figure~\ref{fig:force_curl_int_avg}). The simulation data are also not quite in line with the theoretical predictions, perhaps because the simulations do not operate at asymptotically small $\Ek$.  However, it is interesting to note that the flow lengthscale calculated from the weighted energy spectrum $\duchrs$ \citep{christensen_convection-driven_2006} follows the $\Ek^{1/3}$ scaling very closely, while the $\dupol$ scale does not. Since the weighted energy spectrum scale tends to reflect behaviour at higher wavenumbers than the peak, this may indicate distinct scaling behaviour in the low and high wavenumber regions of the kinetic energy spectrum, as has been suggested by \citet{nicoski_asymptotic_2024} for non-magnetic rapidly rotating spherical shell convection. 

\subsection{Reversal Behaviour and `Off-Path' Simulations}
\label{sec:reversals}

Figures~\ref{fig:Rm_scaling}--\ref{fig:M_fohm} show that simulated fields transition from multipolar-dominated to dipole-dominated on all 3 paths as $\epsilon$ decreases despite the large-scale force balance varying only weakly along the path. This result is consistent with the fact that the dipole-multipole transition occurs abruptly \citep{christensen_convection-driven_2006, oruba_predictive_2014, tassin_geomagnetic_2021} and is associated with small changes in the state of the dynamo, at least in the parameter regime studied.  Previous studies have suggested various quantities that determine the transition from dipolar to multipolar solutions, including $\Mratio \sim 1$ and $\Rol \sim 0.1$. As $\epsilon$ decreases, Figure~\ref{fig:M_fohm} shows that $\Mratio$ grows above $1$ while Table~\ref{tab:sim_outputs} shows that $\Rol$ falls below $0.1$. This suggests that one way to reach the dipole-multipole transition is to change the parameters from their path values in order to reduce $\Mratio$ and increase $\Rol$. 

We demonstrate the effect of increasing $\Ra$ from the theoretical path value for the IMAC-Pm0.5 path (parameters are given in the ``IMAC-Pm0.5$\dagger$'' rows in Table~\ref{tab:sim_outputs}). These ``off path” simulations access the dipole-multipole transition down to an Ekman number $\Ek = 2 \times 10^{-6}$. Increasing $\Ra$ increases $\Rm$ to values of $O(10^4)$ at the lowest $\Ek$, while $\Ro$ and $\lehnert$ remain approximately constant (Figure~\ref{fig:Rm_scaling}). $\Mratio$ decreases from $1.4$ to $0.7$ (Figure~\ref{fig:M_fohm}) while $\Rol$, calculated using equation~(\ref{eq:Rol}), increases from $0.12$ to $0.27$ with decreasing $\epsilon$ and so these simulations are consistent with both proposed diagnostics of the dipole-multipole transition. 

\section{Discussion and Conclusions} \label{sec:discussion}

We have developed and analysed three uni-dimensional parameter paths that are designed to access the dipole-multipole transition in rapidly rotating spherical shell dynamos with an Earth-like magnetic Reynolds number, $\Rm\sim 1000$. The three paths differ in the assumed force balance and the variation of the magnetic Prandtl number $\Pm$ with the path parameter $\epsilon$: one path aims to preserve a Magneto-Archimedian-Coriolis (MAC) balance and assumes $\Pm \sim \epsilon^{1/2}$ while the other paths aim to preserve an Inertia-MAC (IMAC) balance and assume $\Pm \sim \epsilon^{1/2}$ and $\Pm \sim \epsilon^{1}$ respectively. We find that simulations transition from multipole-dominated to dipole-dominated along each path and therefore it is necessary to conduct addition simulations with parameters that deviate from those of the path. Using ``path'' simulations to select appropriate ``off path'' simulations, we have managed to access the dipole-multipole transition down to $\Ek = 2\times 10^{-6}$, which is one of the lowest values obtained to date. 

The challenge of tracking the dipole-multipole transition using a path theory is immediately apparent from the simple fact that it represents a regime transition rather than a regime of its own. It is, therefore, inevitable that ``off path'' simulations, employing conditions deviating from those specified by the path theory, are needed to access the dipole-multipole transition. However, by using path theory to access conditions that are close to the dipole-multipole transition, it is possible to make substantial computational savings compared to a conventional systematic sampling of parameter space. The path theory can also be used to construct paths that attempt to preserve a quantity hypothesised to distinguish the transition, such as the magnetic/kinetic energy $\Mratio$ or the local Rossby number $\Rol$. 

We have found, as did \citet{aubert_spherical_2017}, that a scale-dependent hyperdiffusion applied to the velocity and temperature fields produces large-scale simulated behaviour that is similar to that obtained from Direct Numerical Simulation (DNS) at the same physical conditions. However, our numerical setup, together with the properties of the IMAC paths, has meant that we were not able to reach such extreme physical conditions as \citet{aubert_spherical_2017}. In particular, our use of no-slip (rather than stress-free) velocity boundary conditions requires greater radial resolution, while the use of a larger HD cutoff (here $\lh > 50$) reduces the overall effect of the HD. Our `off path'' simulations further require higher $\Ra$ than the path simulations, which necessitates increasing $\lh$. Therefore, for the configuration we have studied, the computational gains are reduced compared to previous studies. Recently \citet{frasson2024geomagnetic} have applied HD to the magnetic field as well as the velocity and temperature fields. Since the magnetic field often limits the spatio-temporal resolution of dynamo simulations \citep[e.g.][]{davies2011scalability}, this approach may afford further computational savings that allow future simulations to reach lower values of $\epsilon$. 

These computational issues limit our capacity to test the proposed scalings as a function of $\epsilon$, despite reaching values of $\Ek = 2 \times 10^{-6}$ that are close to many state-of-the-art dynamo simulations \citep{sheyko_magnetic_2016, schaeffer_turbulent_2017, aubert_spherical_2017, mound_longitudinal_2023, frasson2024geomagnetic}. One issue is that dynamos transition from multipole-dominated to dipole-dominated as $\epsilon$ decreases, which does little to $\Rm$ and $\Ro$ (Figure~\ref{fig:Rm_scaling}) and $\fohm$ (Figure~\ref{fig:M_fohm}) but does affect the scaling of $\lehnert$ (Figure~\ref{fig:Le_sqrt_fohm}) and $\Mratio$ (Figure~\ref{fig:M_fohm}) and the lengthscales (Figure~\ref{fig:lengthscales}). A second issue is that the relationship between $\epsilon$ and $\Ek$ changes between paths: for the IMAC-Pm0.5 path $\Ek = 2 \times 10^{-6}$ corresponds to $\epsilon = 10^{-3}$ while for the Pm1 paths $\Ek = 2 \times 10^{-6}$ corresponds to $\epsilon > 10^{-2}$. Since $\Ek$ sets the computational requirements of the simulation, it will clearly remain challenging to test the predicted behaviour of the Pm1 paths. 

Due to the relatively large values of $\epsilon$ achieved along some paths (particularly those with $\Pm \sim \epsilon^{1}$) and the fact that all paths have similar starting conditions, distinguishing differences in behaviour requires care. Simulations by \citet{aubert_spherical_2017} and \citet{schwaiger_force_2019} indicate that $\Ek \sim 10^{-6}$ is required to see a 1 order of magnitude difference between Lorentz and interial terms in the force balance spectra and our simulations are broadly compatible with this. $\Rm$ is designed to be constant along all paths, while the scaling behaviour of $\Ro$ is well-known to be similar for MAC and IAC theories \citep[e.g.][]{king_flow_2013}. $\Mratio$ clearly depends on $\Pm$ and is up to an order of magnitude larger for the $\Pm \sim \epsilon^{1/2}$ path than for the $\Pm \sim \epsilon^{1}$ paths at the same $\Ek$. At our lowest $\Ek$, values of $\lehnert$ also differ by about a factor of 2 along the different paths. Therefore, we believe it is possible to observe distinct dynamo behaviour along different parameter paths within the computationally accessible parameter range. 

We analysed both scale-dependent and volume-integrated dynamical balances to test compatibility between the simulation outputs and assumed balances used to derive the path theory. The force analysis revealed large-scale dynamics that are broadly consistent with theoretical predictions, while the integrated balances showed a somewhat different picture, with a leading order magnetostrophic balance. The curled force analysis, which is more dynamically relevant than the force balance, revealed a complex picture with no obvious asymptotic behaviour and a non-negligible role of viscosity, though this could be confined to scales smaller than the peak of the kinetic energy spectrum. We also found that the lengthscale dependence of inertial and Lorentz terms is best represented in our simulations by the dissipation scales, $\dumin$ and $\dbmin$ respectively. \citet{schwaiger_relating_2021} have recently argued that the dynamically relevant lengthscales can be determined from cross-over points in the force spectra. Our approach is different because it focuses on the lengthscales in real space and seeks to identify the appropriate scale for each term, similar to the study of \citet{cox_penetration_2019} for boundary-driven rotating convection. More broadly, our results emphasise the need to consider both force and vorticity balances in the analyses of the dynamics, as has been advocated by \citet{naskar_2025} for non-magnetic rotating spherical shell convection. 

The starting condition for the IMAC paths is a simulation from \citet{nakagawa_combined_2022} with $\Rm = 1185$ and $\Mratio = 5$. We considered this an appropriate starting condition because it already encodes an Earth-like $\Rm$, has a subdominant viscous term, and comparable inertial and Lorentz terms. We cannot rule out that using different starting conditions for different paths (e.g. ones tuned to the expected balance along the path) would produce cleaner scaling behaviour and enable a clearer distinction among them. Nevertheless, as $\epsilon$ decreases the simulations all show signs of dynamically selecting an increasing $\Mratio$, $\fohm$ and ratio of Coriolis/Inertial forces, all of which are consistent with a MAC balance even when the path theory attempts to impose an IMAC balance. This result complements the analysis of \citet{schwaiger_force_2019}, who found that the QG-MAC balance is observed over an widening range of parameter space as rotation rate increases. 

The choice to investigate IMAC paths is based on the findings of previous numerical dynamo simulations in order to test our methodology and is not intended to link to the conditions of the present geodynamo. Previous path studies \citep{aubert_spherical_2017, aubert_approaching_2019} have found that $\fdip$ mildly increases (or at least does not decrease towards values generally associated with the dipole-multipole transition) as $\epsilon$ decreases, which generally does not favour reversing behaviour. However, since the dipole-multipole transition is sensitive to the buoyancy distribution and boundary conditions of the dynamo \citep[e.g.][]{kutzner_stable_2002} it might be possible to change the setup of these simulations to promote reversals. Another possibility is suggested by the recent work of \citet{jones2025low}, who induced reversals with $\Mratio\sim 10$ by increasing $\Pra$ and $\Pm$, thus reducing the non-linear momentum advection and promoting nonlinear thermal advection. The present path theories assume $\Pra = 1 = \mathrm{constant}$, which is lower than the $\Pra = 40$ used by \citet{jones2025low}. Incorporating variable $\Pra$ in the path theory depends on whether thermal or chemical buoyancy dominates. In the former case, $\Pra \sim 10^{-2}$ in Earth's core \citep{pozzo_transport_2013}, suggesting $\Pra \sim \epsilon^{2/5}$, while in the latter case $\Pra$ is replaced by the Schmidt number $\Sc =\kinvisc /\compdiff \sim 100$ where $\compdiff$ is the compositional diffusivity, suggesting $\Sc \approx \mathrm{constant}$. The path setup can be easily modified to account for either case. 

Potential geomagnetic applications of IMAC paths may however have arisen in Earth's history. During the Devonian (419.2--358.9 Ma) the geomagnetic field is thought to have been very weak (comparable to the limit of paleointensity detection) and potentially multipolar \citep{van2022persistent, shcherbakova2017devonian}, while during the Ediacaran/late Cambrian ($\sim$600-500 Ma) the geomagnetic field was weak, reversed frequently, and showed a high rate of secular variation \citep{li2023late, lloyd2024weak}, all characteristics of a multipolar state. The study of \citet{driscoll_simulating_2016} based on dynamo simulations also predicted a multipolar field around 2-1.7 Ga. The Ediacaran has been suggested as a time corresponding to the formation of the solid inner core \citep{bono2019young, lloyd2021first, zhou2022early, davies_dynamo_2022}, which would make the inner core relatively young during the Devonian. IMAC paths for these periods would therefore require a smaller inner core than what we have considered. The buoyancy source distribution may also need to be modified, though this is highly uncertain back in the past \citep{davies_buoyancy_2011}. 

We have used our path simulations to compare two previously proposed diagnostics of the dipole-multipole transition: $\Mratio \sim 1$ and $\Rol \approx 0.12$. In the parameter range studied, we have found that both diagnostics describe our results and hence our simulations cannot distinguish the two possibilities. Therefore we do not observe a change in the character of the dipole-multipole transition at the sampled conditions. Recently \citet{frasson2024geomagnetic} have suggested a new diagnostic based on the ratio of the zonal antisymmetric Elsasser number and the zonal antisymmetric magnetic Reynolds number. It would be interesting to assess this diagnostic using the path simulations in a future study. Finally, we note that the path theories developed here can in principle be used to study dynamos with heterogeneous outer boundary forcing \citep[e.g.][]{aubert_spherical_2017, mound_longitudinal_2023}. It would be interesting to test the ideas of \citet{terra2024regionally} that polarity reversals are triggered by regional dynamical variations induced by heterogeneous outer boundary heat flow.
We have demonstrated that the path approach is an efficient method for seeking the dipole-multipole transition in rapidly rotating dynamos. However, along IMAC paths, the conditions under which we access the dipole-multipole transition become increasingly unrealistic because $\Rm$ rises above any plausible bounds inferred from geophysical observations.
Therefore, our results support the idea that inertially-driven reversals are not relevant for the geomagnetic field, a conclusion obtained by previous studies along somewhat different lines of reasoning \citep{davidson_scaling_2013, tassin_geomagnetic_2021}. We believe that a more promising direction for future research lies in building path theories based on low inertia dynamos \citep{jones2025low}. It is also worth considering different distributions of buoyancy, in particular the strongly bottom-driven chemical driving, which tends to produce simulated fields that are more similar to the paleomagnetic field than thermally-driven dynamos \citep{meduri_numerical_2021}. 
\begin{acknowledgments}
We gratefully acknowledge support from Natural Environment Research Council grants NE/Y003500/1, NE/V010867/1 and NE/W005247/1. Calculations were performed on the UK National supercomputing service ARCHER2.
\end{acknowledgments}

\FloatBarrier
\bibliography{new_refs}

\begin{thebibliography}{88}
\expandafter\ifx\csname natexlab\endcsname\relax\def\natexlab#1{#1}\fi

\bibitem[Aubert(2019)]{aubert_approaching_2019}
Aubert, J., 2019.
\newblock Approaching {Earth}’s core conditions in high-resolution geodynamo simulations, {\it Geophysical Journal International\/}, {\bf 219}(Supplement\_1), S137--S151.

\bibitem[Aubert(2023)]{aubert_state_2023}
Aubert, J., 2023.
\newblock State and evolution of the geodynamo from numerical models reaching the physical conditions of {Earth}’s core, {\it Geophysical Journal International\/}, {\bf 235}(1), 468--487.

\bibitem[Aubert et~al.(2001)Aubert, Brito, Nataf, Cardin, \& Masson]{aubert2001systematic}
Aubert, J., Brito, D., Nataf, H.-C., Cardin, P., \& Masson, J.-P., 2001.
\newblock A systematic experimental study of rapidly rotating spherical convection in water and liquid gallium, {\it Physics of the Earth and Planetary Interiors\/}, {\bf 128}(1-4), 51--74.

\bibitem[Aubert et~al.(2017)Aubert, Gastine, \& Fournier]{aubert_spherical_2017}
Aubert, J., Gastine, T., \& Fournier, A., 2017.
\newblock Spherical convective dynamos in the rapidly rotating asymptotic regime, {\it Journal of Fluid Mechanics\/}, {\bf 813}, 558--593.

\bibitem[Aurnou \& King(2017)]{aurnou_cross-over_2017}
Aurnou, J.~M. \& King, E.~M., 2017.
\newblock The cross-over to magnetostrophic convection in planetary dynamo systems, {\it Proceedings of the Royal Society A: Mathematical, Physical and Engineering Sciences\/}, {\bf 473}(2199), 20160731.

\bibitem[Biggin et~al.(2012)Biggin, Steinberger, Aubert, Suttie, Holme, Torsvik, van~der Meer, \& van Hinsbergen]{biggin_possible_2012}
Biggin, A.~J., Steinberger, B., Aubert, J., Suttie, N., Holme, R., Torsvik, T.~H., van~der Meer, D.~G., \& van Hinsbergen, D. J.~J., 2012.
\newblock Possible links between long-term geomagnetic variations and whole-mantle convection processes, {\it Nature Geoscience\/}, {\bf 5}(8), 526--533.

\bibitem[Bono et~al.(2019)Bono, Tarduno, Nimmo, \& Cottrell]{bono2019young}
Bono, R.~K., Tarduno, J.~A., Nimmo, F., \& Cottrell, R.~D., 2019.
\newblock Young inner core inferred from ediacaran ultra-low geomagnetic field intensity, {\it Nature Geoscience\/}, {\bf 12}(2), 143--147.

\bibitem[Buffett(2010)]{buffett2010tidal}
Buffett, B.~A., 2010.
\newblock Tidal dissipation and the strength of the earth’s internal magnetic field, {\it Nature\/}, {\bf 468}(7326), 952--954.

\bibitem[Calkins(2018)]{calkins2018quasi}
Calkins, M.~A., 2018.
\newblock Quasi-geostrophic dynamo theory, {\it Physics of the Earth and Planetary Interiors\/}, {\bf 276}, 182--189.

\bibitem[Calkins et~al.(2021)Calkins, Orvedahl, \& Featherstone]{calkins_large-scale_2021}
Calkins, M.~A., Orvedahl, R.~J., \& Featherstone, N.~A., 2021.
\newblock Large-scale balances and asymptotic scaling behaviour in spherical dynamos, {\it Geophysical Journal International\/}, {\bf 227}(2), 1228--1245.

\bibitem[Chandrasekhar(1961)]{chandrasekhar_hydrodynamic_1961}
Chandrasekhar, S., 1961.
\newblock {\it Hydrodynamic and {Hydromagnetic} {Stability}\/}, Courier Corporation.

\bibitem[Christensen \& Aubert(2006)]{christensen_scaling_2006}
Christensen, U.~R. \& Aubert, J., 2006.
\newblock Scaling properties of convection-driven dynamos in rotating spherical shells and application to planetary magnetic fields, {\it Geophysical Journal International\/}, {\bf 166}(1), 97--114.

\bibitem[Christensen \& Tilgner(2004)]{christensen_power_2004}
Christensen, U.~R. \& Tilgner, A., 2004.
\newblock Power requirement of the geodynamo from ohmic losses in numerical and laboratory dynamos, {\it Nature\/}, {\bf 429}(6988), 169--171.

\bibitem[Christensen et~al.(2001)Christensen, Aubert, Cardin, Dormy, Gibbons, Glatzmaier, Grote, Honkura, Jones, Kono, Matsushima, Sakuraba, Takahashi, Tilgner, Wicht, \& Zhang]{christensen_numerical_2001}
Christensen, U.~R., Aubert, J., Cardin, P., Dormy, E., Gibbons, S., Glatzmaier, G.~A., Grote, E., Honkura, Y., Jones, C., Kono, M., Matsushima, M., Sakuraba, A., Takahashi, F., Tilgner, A., Wicht, J., \& Zhang, K., 2001.
\newblock A numerical dynamo benchmark, {\it Physics of the Earth and Planetary Interiors\/}, {\bf 128}(1), 25--34.

\bibitem[Christensen et~al.(2006)Christensen, Aubert, \& Olson]{christensen_convection-driven_2006}
Christensen, U.~R., Aubert, J., \& Olson, P., 2006.
\newblock Convection-driven planetary dynamos, {\it Proceedings of the International Astronomical Union\/}, {\bf 2}(S239), 188--195.

\bibitem[Cox et~al.(2019)Cox, Davies, Livermore, \& Singleton]{cox_penetration_2019}
Cox, G.~A., Davies, C.~J., Livermore, P.~W., \& Singleton, J., 2019.
\newblock Penetration of boundary-driven flows into a rotating spherical thermally stratified fluid, {\it Journal of Fluid Mechanics\/}, {\bf 864}, 519--553.

\bibitem[Davidson(2013)]{davidson_scaling_2013}
Davidson, P.~A., 2013.
\newblock Scaling laws for planetary dynamos, {\it Geophysical Journal International\/}, {\bf 195}(1), 67--74.

\bibitem[Davies \& Gubbins(2011)]{davies_buoyancy_2011}
Davies, C. \& Gubbins, D., 2011.
\newblock A buoyancy profile for the {Earth's} core, {\it Geophysical Journal International\/}, {\bf 187}(2), 549--563.

\bibitem[Davies et~al.(2015)Davies, Pozzo, Gubbins, \& Alfè]{davies_constraints_2015}
Davies, C., Pozzo, M., Gubbins, D., \& Alfè, D., 2015.
\newblock Constraints from material properties on the dynamics and evolution of {Earth}'s core, {\it Nature Geoscience\/}, {\bf 8}(9), 678--685.

\bibitem[Davies et~al.(2011)Davies, Gubbins, \& Jimack]{davies2011scalability}
Davies, C.~J., Gubbins, D., \& Jimack, P.~K., 2011.
\newblock Scalability of pseudospectral methods for geodynamo simulations, {\it Concurrency and Computation: Practice and Experience\/}, {\bf 23}(1), 38--56.

\bibitem[Davies et~al.(2013)Davies, Silva, \& Mound]{davies_influence_2013}
Davies, C.~J., Silva, L., \& Mound, J., 2013.
\newblock On the influence of a translating inner core in models of outer core convection, {\it Physics of the Earth and Planetary Interiors\/}, {\bf 214}, 104--114.

\bibitem[Davies et~al.(2022)Davies, Bono, Meduri, Aubert, Greenwood, \& Biggin]{davies_dynamo_2022}
Davies, C.~J., Bono, R.~K., Meduri, D.~G., Aubert, J., Greenwood, S., \& Biggin, A.~J., 2022.
\newblock Dynamo constraints on the long-term evolution of {Earth}’s magnetic field strength, {\it Geophysical Journal International\/}, {\bf 228}(1), 316--336.

\bibitem[Dormy(2016)]{dormy_strong-field_2016}
Dormy, E., 2016.
\newblock Strong-field spherical dynamos, {\it Journal of Fluid Mechanics\/}, {\bf 789}, 500--513.

\bibitem[Dormy et~al.(2018)Dormy, Oruba, \& Petitdemange]{dormy2018three}
Dormy, E., Oruba, L., \& Petitdemange, L., 2018.
\newblock Three branches of dynamo action, {\it Fluid Dynamics Research\/}, {\bf 50}(1), 011415.

\bibitem[Driscoll \& Olson(2009{\natexlab{a}})]{driscoll2009polarity}
Driscoll, P. \& Olson, P., 2009{\natexlab{a}}.
\newblock Polarity reversals in geodynamo models with core evolution, {\it Earth and Planetary Science Letters\/}, {\bf 282}(1-4), 24--33.

\bibitem[Driscoll \& Olson(2009{\natexlab{b}})]{driscoll_effects_2009}
Driscoll, P. \& Olson, P., 2009{\natexlab{b}}.
\newblock Effects of buoyancy and rotation on the polarity reversal frequency of gravitationally driven numerical dynamos, {\it Geophysical Journal International\/}, {\bf 178}(3), 1337--1350.

\bibitem[Driscoll(2016)]{driscoll_simulating_2016}
Driscoll, P.~E., 2016.
\newblock Simulating 2 {Ga} of geodynamo history, {\it Geophysical Research Letters\/}, {\bf 43}(11), 5680--5687.

\bibitem[Driscoll \& Evans(2016)]{driscoll2016frequency}
Driscoll, P.~E. \& Evans, D.~A., 2016.
\newblock Frequency of proterozoic geomagnetic superchrons, {\it Earth and Planetary Science Letters\/}, {\bf 437}, 9--14.

\bibitem[Driscoll \& Wilson(2018)]{driscoll2018paleomagnetic}
Driscoll, P.~E. \& Wilson, C., 2018.
\newblock Paleomagnetic biases inferred from numerical dynamos and the search for geodynamo evolution, {\it Frontiers in Earth Science\/}, {\bf 6}, 113.

\bibitem[Dziewonski \& Anderson(1981)]{dziewonski1981preliminary}
Dziewonski, A.~M. \& Anderson, D.~L., 1981.
\newblock Preliminary reference {Earth} model, {\it Physics of the earth and planetary interiors\/}, {\bf 25}(4), 297--356.

\bibitem[Finlay \& Amit(2011)]{finlay2011flow}
Finlay, C.~C. \& Amit, H., 2011.
\newblock On flow magnitude and field-flow alignment at {Earth's} core surface, {\it Geophysical Journal International\/}, {\bf 186}(1), 175--192.

\bibitem[Frasson et~al.(2024)Frasson, Schaeffer, Nataf, \& Labrosse]{frasson2024geomagnetic}
Frasson, T., Schaeffer, N., Nataf, H.-C., \& Labrosse, S., 2024.
\newblock Geomagnetic dipole stability and zonal flows controlled by mantle heat flux heterogeneities, {\it Geophysical Journal International\/}, {\bf 240}(3), 1481--1504.

\bibitem[Garcia et~al.(2017)Garcia, Oruba, \& Dormy]{garcia2017equatorial}
Garcia, F., Oruba, L., \& Dormy, E., 2017.
\newblock Equatorial symmetry breaking and the loss of dipolarity in rapidly rotating dynamos, {\it Geophysical \& Astrophysical Fluid Dynamics\/}, {\bf 111}(5), 380--393.

\bibitem[Gillet et~al.(2010)Gillet, Jault, Canet, \& Fournier]{gillet2010fast}
Gillet, N., Jault, D., Canet, E., \& Fournier, A., 2010.
\newblock Fast torsional waves and strong magnetic field within the earth’s core, {\it Nature\/}, {\bf 465}(7294), 74--77.

\bibitem[Glatzmaiers \& Roberts(1995)]{glatzmaiers_three-dimensional_1995}
Glatzmaiers, G.~A. \& Roberts, P.~H., 1995.
\newblock A three-dimensional self-consistent computer simulation of a geomagnetic field reversal, {\it Nature\/}, {\bf 377}(6546), 203--209.

\bibitem[Gross(2007)]{gross2007earth}
Gross, R.~S., 2007.
\newblock Earth rotation variations-long period, {\it Treatise on geophysics\/}, {\bf 3}(821), 239--294.

\bibitem[Gwirtz et~al.(2022)Gwirtz, Davis, Morzfeld, Constable, Fournier, \& Hulot]{gwirtz2022can}
Gwirtz, K., Davis, T., Morzfeld, M., Constable, C., Fournier, A., \& Hulot, G., 2022.
\newblock Can machine learning reveal precursors of reversals of the geomagnetic axial dipole field?, {\it Geophysical Journal International\/}, {\bf 231}(1), 520--535.

\bibitem[Heimpel \& Evans(2013)]{heimpel_testing_2013}
Heimpel, M.~H. \& Evans, M.~E., 2013.
\newblock Testing the geomagnetic dipole and reversing dynamo models over {Earth}’s cooling history, {\it Physics of the Earth and Planetary Interiors\/}, {\bf 224}, 124--131.

\bibitem[Holme et~al.(2015)Holme, Olson, \& Schubert]{holme2015large}
Holme, R., Olson, P., \& Schubert, G., 2015.
\newblock Large-scale flow in the core, {\it Treatise on geophysics\/}, {\bf 8}, 107--130.

\bibitem[Jones(2015)]{jones_805_2015}
Jones, C.~A., 2015.
\newblock 8.05 - {Thermal} and {Compositional} {Convection} in the {Outer} {Core}, in {\em Treatise on {Geophysics} ({Second} {Edition})\/}, pp. 115--159, ed. Schubert, G., Elsevier, Oxford.

\bibitem[Jones \& Tsang(2025)]{jones2025low}
Jones, C.~A. \& Tsang, Y.-K., 2025.
\newblock Low inertia reversing geodynamos, {\it Physics of the Earth and Planetary Interiors\/}, {\bf 360}, 107303.

\bibitem[King \& Buffett(2013)]{king_flow_2013}
King, E.~M. \& Buffett, B.~A., 2013.
\newblock Flow speeds and length scales in geodynamo models: {The} role of viscosity, {\it Earth and Planetary Science Letters\/}, {\bf 371-372}, 156--162.

\bibitem[Kutzner \& Christensen(2002)]{kutzner_stable_2002}
Kutzner, C. \& Christensen, U.~R., 2002.
\newblock From stable dipolar towards reversing numerical dynamos, {\it Physics of the Earth and Planetary Interiors\/}, {\bf 131}(1), 29--45.

\bibitem[Labrosse(2015)]{labrosse2015thermal}
Labrosse, S., 2015.
\newblock Thermal evolution of the core with a high thermal conductivity, {\it Physics of the Earth and Planetary Interiors\/}, {\bf 247}, 36--55.

\bibitem[Li et~al.(2023)Li, Tarduno, Jiao, Liu, Peng, Xu, Yang, \& Yang]{li2023late}
Li, Y.-X., Tarduno, J.~A., Jiao, W., Liu, X., Peng, S., Xu, S., Yang, A., \& Yang, Z., 2023.
\newblock Late cambrian geomagnetic instability after the onset of inner core nucleation, {\it nature communications\/}, {\bf 14}(1), 4596.

\bibitem[Lister \& Buffett(1995)]{lister1995strength}
Lister, J.~R. \& Buffett, B.~A., 1995.
\newblock The strength and efficiency of thermal and compositional convection in the geodynamo, {\it Physics of the Earth and Planetary Interiors\/}, {\bf 91}(1-3), 17--30.

\bibitem[Lloyd et~al.(2021)Lloyd, Biggin, Halls, \& Hill]{lloyd2021first}
Lloyd, S.~J., Biggin, A.~J., Halls, H., \& Hill, M.~J., 2021.
\newblock First palaeointensity data from the cryogenian and their potential implications for inner core nucleation age, {\it Geophysical Journal International\/}, {\bf 226}(1), 66--77.

\bibitem[Lloyd et~al.(2024)Lloyd, Biggin, Halls, \& Denyszyn]{lloyd2024weak}
Lloyd, S.~J., Biggin, A.~J., Halls, H., \& Denyszyn, S., 2024.
\newblock Weak paleointensities from 1.6 ga {Greenland} dykes: Further evidence for a billion-year period of paleomagnetic dipole low during the {Paleoproterozoic}, {\it Earth and Planetary Science Letters\/}, {\bf 648}, 119110.

\bibitem[Majumder et~al.(2024)Majumder, Sreenivasan, \& Maurya]{majumder_self-similarity_2024}
Majumder, D., Sreenivasan, B., \& Maurya, G., 2024.
\newblock Self-similarity of the dipole–multipole transition in rapidly rotating dynamos, {\it Journal of Fluid Mechanics\/}, {\bf 980}, A30.

\bibitem[Matsui et~al.(2016)Matsui, Heien, Aubert, Aurnou, Avery, Brown, Buffett, Busse, Christensen, Davies, Featherstone, Gastine, Glatzmaier, Gubbins, Guermond, Hayashi, Hollerbach, Hwang, Jackson, Jones, Jiang, Kellogg, Kuang, Landeau, Marti, Olson, Ribeiro, Sasaki, Schaeffer, Simitev, Sheyko, Silva, Stanley, Takahashi, Takehiro, Wicht, \& Willis]{matsui_performance_2016}
Matsui, H., Heien, E., Aubert, J., Aurnou, J.~M., Avery, M., Brown, B., Buffett, B.~A., Busse, F., Christensen, U.~R., Davies, C.~J., Featherstone, N., Gastine, T., Glatzmaier, G.~A., Gubbins, D., Guermond, J.-L., Hayashi, Y.-Y., Hollerbach, R., Hwang, L.~J., Jackson, A., Jones, C.~A., Jiang, W., Kellogg, L.~H., Kuang, W., Landeau, M., Marti, P., Olson, P., Ribeiro, A., Sasaki, Y., Schaeffer, N., Simitev, R.~D., Sheyko, A., Silva, L., Stanley, S., Takahashi, F., Takehiro, S.-i., Wicht, J., \& Willis, A.~P., 2016.
\newblock Performance benchmarks for a next generation numerical dynamo model, {\it Geochemistry, Geophysics, Geosystems\/}, {\bf 17}(5), 1586--1607.

\bibitem[McDermott \& Davidson(2019)]{mcdermott2019physical}
McDermott, B. \& Davidson, P., 2019.
\newblock A physical conjecture for the dipolar--multipolar dynamo transition, {\it Journal of Fluid Mechanics\/}, {\bf 874}, 995--1020.

\bibitem[Meduri et~al.(2021)Meduri, Biggin, Davies, Bono, Sprain, \& Wicht]{meduri_numerical_2021}
Meduri, D.~G., Biggin, A.~J., Davies, C.~J., Bono, R.~K., Sprain, C.~J., \& Wicht, J., 2021.
\newblock Numerical dynamo simulations reproduce paleomagnetic field behavior, {\it Geophysical Research Letters\/}, {\bf 48}(5), e2020GL090544.

\bibitem[Menu et~al.(2020)Menu, Petitdemange, \& Galtier]{menu_magnetic_2020}
Menu, M.~D., Petitdemange, L., \& Galtier, S., 2020.
\newblock Magnetic effects on fields morphologies and reversals in geodynamo simulations, {\it Physics of the Earth and Planetary Interiors\/}, {\bf 307}, 106542.

\bibitem[Moffatt(1978)]{moffatt_magnetic_1978}
Moffatt, H.~K., 1978.
\newblock {\it Magnetic field generation in electrically conducting fluids\/}, Cambridge monographs on mechanics and applied mathematics, Cambridge Univ. Pr, Cambridge, 1st edn.

\bibitem[Mound \& Davies(2017)]{mound_heat_2017}
Mound, J.~E. \& Davies, C.~J., 2017.
\newblock Heat transfer in rapidly rotating convection with heterogeneous thermal boundary conditions, {\it Journal of Fluid Mechanics\/}, {\bf 828}, 601--629.

\bibitem[Mound \& Davies(2023)]{mound_longitudinal_2023}
Mound, J.~E. \& Davies, C.~J., 2023.
\newblock Longitudinal structure of {Earth}’s magnetic field controlled by lower mantle heat flow, {\it Nature Geoscience\/}, {\bf 16}(4), 380--385.

\bibitem[Nakagawa \& Davies(2022)]{nakagawa_combined_2022}
Nakagawa, T. \& Davies, C.~J., 2022.
\newblock Combined dynamical and morphological characterisation of geodynamo simulations, {\it Earth and Planetary Science Letters\/}, {\bf 594}, 117752.

\bibitem[Naskar et~al.(2025)Naskar, Davies, Mound, \& Clarke]{naskar_2025}
Naskar, S., Davies, C., Mound, J., \& Clarke, A., 2025.
\newblock Force balances in spherical shell rotating convection, {\it Journal of Fluid Mechanics\/}, {\bf 1009}, A70.

\bibitem[Nataf \& Schaeffer(2015)]{nataf_806_2015}
Nataf, H.~C. \& Schaeffer, N., 2015.
\newblock 8.06 - {Turbulence} in the {Core}, in {\em Treatise on {Geophysics} ({Second} {Edition})\/}, pp. 161--181, ed. Schubert, G., Elsevier, Oxford.

\bibitem[Nataf \& Schaeffer(2024)]{nataf_dynamic_2024}
Nataf, H.-C. \& Schaeffer, N., 2024.
\newblock Dynamic regimes in planetary cores: $\tau ${\textendash}$\ell $ diagrams, {\it Comptes Rendus. G\'eoscience\/}, {\bf 356}, 1--30.

\bibitem[Nicoski et~al.(2024)Nicoski, O’Connor, \& Calkins]{nicoski_asymptotic_2024}
Nicoski, J.~A., O’Connor, A.~R., \& Calkins, M.~A., 2024.
\newblock Asymptotic scaling relations for rotating spherical convection with strong zonal flows, {\it Journal of Fluid Mechanics\/}, {\bf 981}, A22.

\bibitem[Ogg(2020)]{ogg2020geomagnetic}
Ogg, J., 2020.
\newblock Geomagnetic polarity time scale, in {\em Geologic time scale 2020\/}, pp. 159--192, Elsevier.

\bibitem[Olson \& Amit(2014)]{olson_magnetic_2014}
Olson, P. \& Amit, H., 2014.
\newblock Magnetic reversal frequency scaling in dynamos with thermochemical convection, {\it Physics of the Earth and Planetary Interiors\/}, {\bf 229}, 122--133.

\bibitem[Olson \& Christensen(2006)]{olson_dipole_2006}
Olson, P. \& Christensen, U.~R., 2006.
\newblock Dipole moment scaling for convection-driven planetary dynamos, {\it Earth and Planetary Science Letters\/}, {\bf 250}(3), 561--571.

\bibitem[Olson et~al.(2011)Olson, Glatzmaier, \& Coe]{olson_complex_2011}
Olson, P.~L., Glatzmaier, G.~A., \& Coe, R.~S., 2011.
\newblock Complex polarity reversals in a geodynamo model, {\it Earth and Planetary Science Letters\/}, {\bf 304}(1), 168--179.

\bibitem[Oruba \& Dormy(2014{\natexlab{a}})]{oruba2014transition}
Oruba, L. \& Dormy, E., 2014{\natexlab{a}}.
\newblock Transition between viscous dipolar and inertial multipolar dynamos, {\it Geophysical Research Letters\/}, {\bf 41}(20), 7115--7120.

\bibitem[Oruba \& Dormy(2014{\natexlab{b}})]{oruba_predictive_2014}
Oruba, L. \& Dormy, E., 2014{\natexlab{b}}.
\newblock Predictive scaling laws for spherical rotating dynamos, {\it Geophysical Journal International\/}, {\bf 198}(2), 828--847.

\bibitem[Pozzo et~al.(2013)Pozzo, Davies, Gubbins, \& Alfè]{pozzo_transport_2013}
Pozzo, M., Davies, C., Gubbins, D., \& Alfè, D., 2013.
\newblock Transport properties for liquid silicon-oxygen-iron mixtures at {Earth}'s core conditions, {\it Physical Review B\/}, {\bf 87}(1), 014110.

\bibitem[Ranjan et~al.(2018)Ranjan, Davidson, Christensen, \& Wicht]{ranjan2018internally}
Ranjan, A., Davidson, P., Christensen, U.~R., \& Wicht, J., 2018.
\newblock Internally driven inertial waves in geodynamo simulations, {\it Geophysical Journal International\/}, {\bf 213}(2), 1281--1295.

\bibitem[Schaeffer et~al.(2017)Schaeffer, Jault, Nataf, \& Fournier]{schaeffer_turbulent_2017}
Schaeffer, N., Jault, D., Nataf, H.-C., \& Fournier, A., 2017.
\newblock Turbulent geodynamo simulations: a leap towards {Earth}’s core, {\it Geophysical Journal International\/}, {\bf 211}(1), 1--29.

\bibitem[Schwaiger et~al.(2019)Schwaiger, Gastine, \& Aubert]{schwaiger_force_2019}
Schwaiger, T., Gastine, T., \& Aubert, J., 2019.
\newblock Force balance in numerical geodynamo simulations: a systematic study, {\it Geophysical Journal International\/}, {\bf 219}(Supplement\_1), S101--S114.

\bibitem[Schwaiger et~al.(2021)Schwaiger, Gastine, \& Aubert]{schwaiger_relating_2021}
Schwaiger, T., Gastine, T., \& Aubert, J., 2021.
\newblock Relating force balances and flow length scales in geodynamo simulations, {\it Geophysical Journal International\/}, {\bf 224}(3), 1890--1904.

\bibitem[Shcherbakova et~al.(2017)Shcherbakova, Biggin, Veselovskiy, Shatsillo, Hawkins, Shcherbakov, \& Zhidkov]{shcherbakova2017devonian}
Shcherbakova, V., Biggin, A., Veselovskiy, R., Shatsillo, A., Hawkins, L., Shcherbakov, V., \& Zhidkov, G., 2017.
\newblock Was the {Devonian} geomagnetic field dipolar or multipolar? {Palaeointensity} studies of {Devonian} igneous rocks from the {Minusa} {Basin} ({Siberia}) and the {Kola} {Peninsula} dykes, {Russia}, {\it Geophysical Journal International\/}, {\bf 209}(2), 1265--1286.

\bibitem[Sheyko et~al.(2016)Sheyko, Finlay, \& Jackson]{sheyko_magnetic_2016}
Sheyko, A., Finlay, C.~C., \& Jackson, A., 2016.
\newblock Magnetic reversals from planetary dynamo waves, {\it Nature\/}, {\bf 539}(7630), 551--554.

\bibitem[Soderlund et~al.(2012)Soderlund, King, \& Aurnou]{soderlund_influence_2012}
Soderlund, K.~M., King, E.~M., \& Aurnou, J.~M., 2012.
\newblock The influence of magnetic fields in planetary dynamo models, {\it Earth and Planetary Science Letters\/}, {\bf 333-334}, 9--20.

\bibitem[Sprain et~al.(2019)Sprain, Biggin, Davies, Bono, \& Meduri]{sprain_assessment_2019}
Sprain, C.~J., Biggin, A.~J., Davies, C.~J., Bono, R.~K., \& Meduri, D.~G., 2019.
\newblock An assessment of long duration geodynamo simulations using new paleomagnetic modeling criteria ({QPM}), {\it Earth and Planetary Science Letters\/}, {\bf 526}, 115758.

\bibitem[Sreenivasan \& Jones(2006)]{sreenivasan_role_2006}
Sreenivasan, B. \& Jones, C.~A., 2006.
\newblock The role of inertia in the evolution of spherical dynamos, {\it Geophysical Journal International\/}, {\bf 164}(2), 467--476.

\bibitem[Strik et~al.(2003)Strik, Blake, Zegers, White, \& Langereis]{strik_palaeomagnetism_2003}
Strik, G., Blake, T.~S., Zegers, T.~E., White, S.~H., \& Langereis, C.~G., 2003.
\newblock Palaeomagnetism of flood basalts in the {Pilbara} {Craton}, {Western} {Australia}: {Late} {Archaean} continental drift and the oldest known reversal of the geomagnetic field, {\it Journal of Geophysical Research: Solid Earth\/}, {\bf 108}(B12).

\bibitem[Tassin et~al.(2021)Tassin, Gastine, \& Fournier]{tassin_geomagnetic_2021}
Tassin, T., Gastine, T., \& Fournier, A., 2021.
\newblock Geomagnetic semblance and dipolar–multipolar transition in top-heavy double-diffusive geodynamo models, {\it Geophysical Journal International\/}, {\bf 226}(3), 1897--1919.

\bibitem[Teed \& Dormy(2023)]{Teed_Dormy_2023}
Teed, R.~J. \& Dormy, E., 2023.
\newblock Solenoidal force balances in numerical dynamos, {\it Journal of Fluid Mechanics\/}, {\bf 964}, A26.

\bibitem[Terra-Nova \& Amit(2024)]{terra2024regionally}
Terra-Nova, F. \& Amit, H., 2024.
\newblock Regionally-triggered geomagnetic reversals, {\it Scientific Reports\/}, {\bf 14}(1), 9639.

\bibitem[van~der Boon et~al.(2022)van~der Boon, Biggin, Thallner, Hounslow, Bono, Nawrocki, W{\'o}jcik, Paszkowski, K{\"o}nigshof, de~Backer, et~al.]{van2022persistent}
van~der Boon, A., Biggin, A.~J., Thallner, D., Hounslow, M.~W., Bono, R., Nawrocki, J., W{\'o}jcik, K., Paszkowski, M., K{\"o}nigshof, P., de~Backer, T., et~al., 2022.
\newblock A persistent non-uniformitarian paleomagnetic field in the devonian?, {\it Earth-Science Reviews\/}, {\bf 231}, 104073.

\bibitem[Wicht(2002)]{wicht_inner-core_2002}
Wicht, J., 2002.
\newblock Inner-core conductivity in numerical dynamo simulations, {\it Physics of the Earth and Planetary Interiors\/}, {\bf 132}(4), 281--302.

\bibitem[Willis et~al.(2007)Willis, Sreenivasan, \& Gubbins]{willis_thermal_2007}
Willis, A.~P., Sreenivasan, B., \& Gubbins, D., 2007.
\newblock Thermal core–mantle interaction: {Exploring} regimes for ‘locked’ dynamo action, {\it Physics of the Earth and Planetary Interiors\/}, {\bf 165}(1-2), 83--92.

\bibitem[Wilson et~al.(2022)Wilson, Pozzo, Alf{\`e}, Walker, Greenwood, Pommier, \& Davies]{wilson2022powering}
Wilson, A.~J., Pozzo, M., Alf{\`e}, D., Walker, A.~M., Greenwood, S., Pommier, A., \& Davies, C.~J., 2022.
\newblock Powering {Earth's} ancient dynamo with silicon precipitation, {\it Geophysical Research Letters\/}, {\bf 49}(22), e2022GL100692.

\bibitem[Yadav et~al.(2016{\natexlab{a}})Yadav, Gastine, Christensen, Duarte, \& Reiners]{yadav_effect_2016}
Yadav, R., Gastine, T., Christensen, U., Duarte, L., \& Reiners, A., 2016{\natexlab{a}}.
\newblock Effect of shear and magnetic field on the heat-transfer efficiency of convection in rotating spherical shells, {\it Geophysical Journal International\/}, {\bf 204}(2), 1120--1133.

\bibitem[Yadav et~al.(2016{\natexlab{b}})Yadav, Gastine, Christensen, Wolk, \& Poppenhaeger]{yadav_approaching_2016}
Yadav, R.~K., Gastine, T., Christensen, U.~R., Wolk, S.~J., \& Poppenhaeger, K., 2016{\natexlab{b}}.
\newblock Approaching a realistic force balance in geodynamo simulations, {\it Proceedings of the National Academy of Sciences\/}, {\bf 113}(43), 12065--12070.

\bibitem[Zhou et~al.(2022)Zhou, Tarduno, Nimmo, Cottrell, Bono, Ibanez-Mejia, Huang, Hamilton, Kodama, Smirnov, et~al.]{zhou2022early}
Zhou, T., Tarduno, J.~A., Nimmo, F., Cottrell, R.~D., Bono, R.~K., Ibanez-Mejia, M., Huang, W., Hamilton, M., Kodama, K., Smirnov, A.~V., et~al., 2022.
\newblock Early cambrian renewal of the geodynamo and the origin of inner core structure, {\it Nature communications\/}, {\bf 13}(1), 4161.

\end{thebibliography}

\appendix
\section{Table of Results} \label{sec:appendix}
\vspace{5em}
\hspace{-2em}
\begin{minipage}{170mm}
\captionof{table}{Summary of time-averaged diagnostics for the simulations reported in this work. The scaling rules are explained in the text. \(\blacklozenge\) indicates DNS simulation, \(\dagger\) indicates an off-path simulation with increased Rayleigh number, \(\diamond\) indicates increased resolution simulation. Simulations 9, 10, and 19 are duplicates of 6, 7, and 16, respectively, with higher resolution to investigate model convergence. }
\begin{tabular}{llllllllllllll}
\hline
No & rule & $\Raf$ & $\Ek$ & $\Pm$ & $\lh$ & $\lmax$ & $\Nr$ & $\Ro$ & $\Rm$ & $\elsasser$ & $\lehnert$ & $\Mratio$ & $\fdip$ \\ \hline
1 & MAC-Pm1 & 5.817e-06 & 3.000e-06 & 0.539 & 150 & 256 & 512 & 7.834e-03 & 1406.6 & 16.59 & 1.360e-02 & 3.016 & 0.790 \\
2 & MAC-Pm1 & 1.298e-05 & 1.000e-05 & 1.201 & 140 & 192 & 288 & 1.158e-02 & 1391.1 & 13.38 & 1.493e-02 & 1.663 & 0.705 \\
3 & MAC-Pm1 & 6.025e-05 & 1.000e-04 & 5.579 & 96 & 160 & 160 & 1.829e-02 & 1020.2 & 24.45 & 2.936e-02 & 2.574 & 0.534 \\
4 & MAC-Pm1 & 1.253e-04 & 3.000e-04 & 11.604 & 128 & 160 & 160 & 2.217e-02 & 857.6 & 28.44 & 3.835e-02 & 3.015 & 0.421 \\
5 & IMAC-Pm0.5 & 4.797e-07 & 1.995e-06 & 1.118 & 192 & 256 & 384 & 2.100e-03 & 1176.7 & 23.72 & 9.200e-03 & 19.187 & 0.763 \\
6 & IMAC-Pm0.5 & 4.777e-06 & 1.585e-05 & 3.536 & 50 & 192 & 256 & 6.004e-03 & 1339.6 & 27.71 & 1.576e-02 & 6.888 & 0.625 \\
7 & IMAC-Pm0.5 & 4.789e-05 & 1.259e-04 & 11.180 & 50 & 160 & 160 & 1.480e-02 & 1314.4 & 47.61 & 3.275e-02 & 4.898 & 0.475 \\
8 & IMAC-Pm0.5 & 4.800e-04 & 1.000e-03 & 35.350 & 64 & 96 & 96 & 3.434e-02 & 1213.8 & 57.79 & 5.718e-02 & 2.809 & 0.212 \\
9 & IMAC-Pm0.5\(\blacklozenge\) & 4.788e-05 & 1.259e-04 & 11.182 & - & 160 & 160 & 1.511e-02 & 1342.5 & 43.84 & 3.142e-02 & 4.321 & 0.489 \\
10 & IMAC-Pm0.5\(\blacklozenge\) & 4.733e-06 & 1.580e-05 & 3.540 & - & 256 & 256 & 6.051e-03 & 1355.8 & 26.92 & 1.550e-02 & 6.544 & 0.695 \\
11 & IMAC-Pm1 & 7.571e-06 & 3.000e-06 & 0.558 & 150 & 192 & 768 & 8.747e-03 & 1625.9 & 22.27 & 1.548e-02 & 3.133 & 0.799 \\
12 & IMAC-Pm1 & 1.800e-05 & 1.000e-05 & 1.318 & 128 & 256 & 320 & 1.353e-02 & 1783.3 & 15.52 & 1.535e-02 & 1.291 & 0.542 \\
13 & IMAC-Pm1 & 3.915e-05 & 3.000e-05 & 2.888 & 128 & 192 & 256 & 1.931e-02 & 1859.1 & 16.67 & 1.861e-02 & 0.950 & 0.247 \\
14 & IMAC-Pm1 & 1.000e-04 & 1.000e-04 & 6.820 & 128 & 192 & 192 & 2.662e-02 & 1815.4 & 24.99 & 2.707e-02 & 1.038 & 0.161 \\
15 & IMAC-Pm1 & 2.031e-04 & 3.000e-04 & 14.959 & 118 & 128 & 192 & 2.843e-02 & 1417.5 & 47.96 & 4.386e-02 & 2.388 & 0.339 \\
16 & IMAC-Pm0.5\(\dagger\)& 4.800e-05 & 2.000e-06 & 1.000 & 160 & 256 & 768 & 2.416e-02 & 12079.3 & 98.93 & 1.989e-02 & 0.677 & 0.299 \\
17 & IMAC-Pm0.5\(\dagger\)& 4.777e-05 & 1.585e-05 & 3.536 & 50 & 256 & 320 & 2.111e-02 & 4709.9 & 57.64 & 2.273e-02 & 1.159 & 0.145 \\
18 & IMAC-Pm0.5\(\dagger\)& 1.996e-04 & 1.259e-04 & 11.180 & 50 & 256 & 256 & 3.434e-02 & 3049.2 & 72.15 & 4.031e-02 & 1.379 & 0.153 \\
19 & IMAC-Pm0.5\(\dagger\)\(\diamond\)& 4.800e-05 & 2.000e-06 & 1.000 & 450 & 512 & 768 & 2.461e-02 & 12303.7 & 88.73 & 1.884e-02 & 0.586 & 0.112 \\ \hline
\end{tabular}
\label{tab:sim_outputs}
\end{minipage}

\newpage
\onecolumn
\makeatletter
\textbf{\hspace{-2em}\huge Supplemental Information:\\ Accessing the dipole-multipole transition in rapidly rotating spherical shell dynamos}
\FloatBarrier
\setcounter{equation}{0}
\setcounter{figure}{0}
\setcounter{table}{0}
\setcounter{page}{1}
\renewcommand{\theequation}{S\arabic{equation}}
\renewcommand{\thefigure}{S\arabic{figure}}
\vspace{5em}
    \centering
        \includegraphics[width=0.33\linewidth]{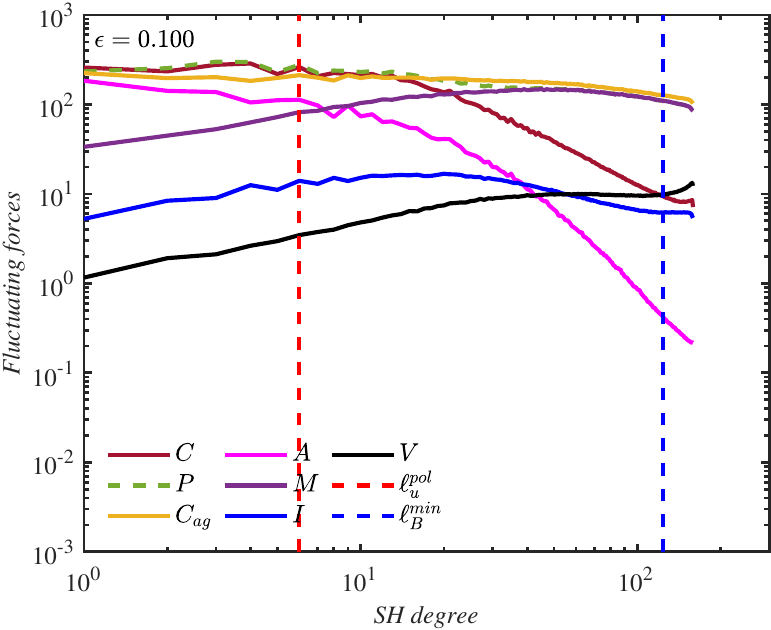} 
        \includegraphics[width=0.33\linewidth]{figs/fluc_force_spec_dyn_IMAC_PM0p5_eps=0.1.pdf}     
    \captionof{figure}{Effect of HD on energy spectra (top row) and force spectra (bottom row), for a simulation on the IMAC-Pm0.5 path with Ekman number $1.26\times 10^{-4}$. The DNS simulation is shown on the left and the HD simulation on the right. Fluctuating forces $C$ (brown), $P$ (dashed green), $C_{ag}$ (yellow), $A$ (pink), $M$ (purple), $I$ (solid blue) and $V$ (black) are defined in Section \ref{subsec:equations}. The red and blue dashed lines show the spherical harmonic degree corresponding to the peak in poloidal kinetic energy spectra $\lupol$ and the ohmic dissipation $\lbmin$, respectively.}
    \label{fig:dns_vs_les_ek_1e-4}
\vspace{10em}
    \includegraphics[width=0.33\linewidth]{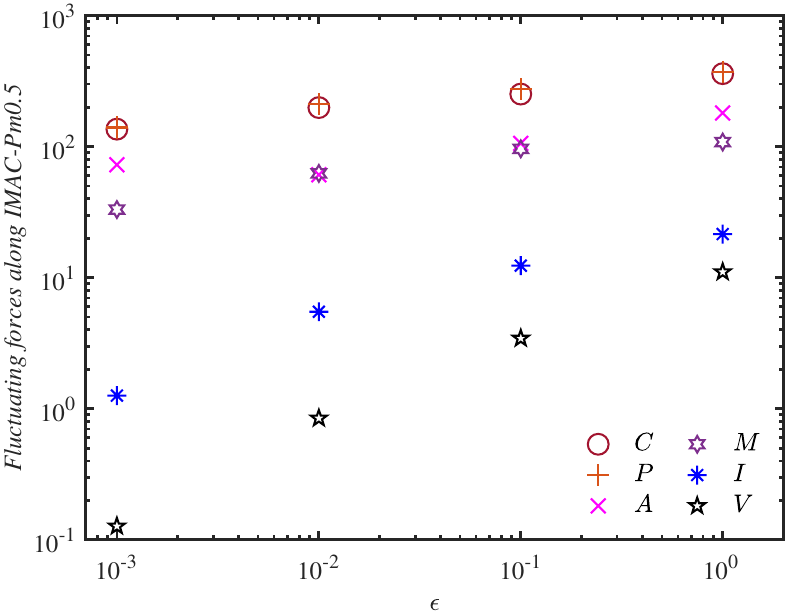}%
    \includegraphics[width=0.33\linewidth]{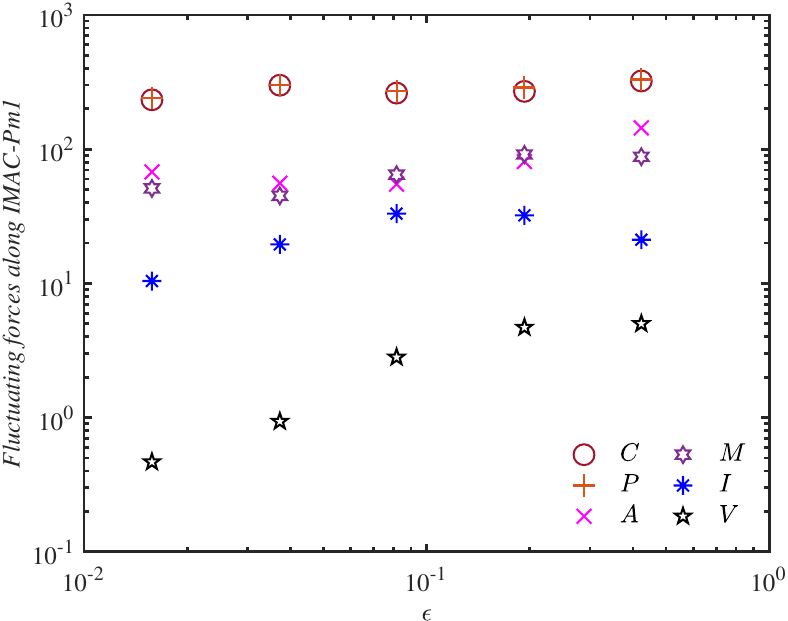}
    \includegraphics[width=0.33\linewidth]{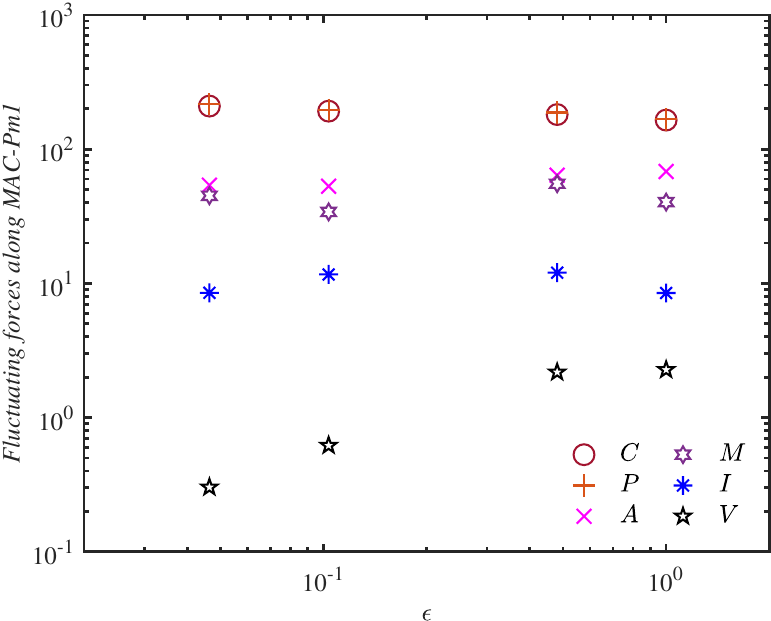}
    \captionof{figure}{Volume-integrated forces (top row) along the IMAC-Pm0.5 (left), IMAC-Pm1 (middle) and MAC-Pm1 (right) paths. The forces are evaluated at the spherical harmonic degree $\lupol$.}
    \label{fig:force_int_lpol}
    
\end{document}